\documentclass[pdflatex,sn-nature]{sn-jnl}

\usepackage{amsmath,amssymb,amsfonts}
\usepackage{graphicx}
\usepackage{xcolor}
\usepackage{booktabs}
\usepackage{array}
\usepackage{colortbl}
\usepackage{adjustbox}
\usepackage{longtable}
\usepackage[title]{appendix}
\definecolor{bestcol}{HTML}{0B6E4F}
\definecolor{ctrlcol}{HTML}{9B2C2C}
\definecolor{hdrcol}{HTML}{EAF1F7}
\definecolor{oursbg}{HTML}{E8F6EF}
\definecolor{ctrlbg}{HTML}{FCEDEB}
\definecolor{basebar}{HTML}{BFC8D4}
\definecolor{ctrlbar}{HTML}{EE9B91}
\definecolor{ctrlbarlight}{HTML}{9EA9C7}
\definecolor{gradbar}{HTML}{F1C98F}
\definecolor{trustbar}{HTML}{C9DDE5}
\definecolor{gridbar}{HTML}{E7E9EF}
\usepackage{amsthm}
\usepackage{tikz}
\usepackage{pgfplots}
\usetikzlibrary{arrows.meta,positioning,shapes.geometric,calc,patterns}
\pgfplotsset{compat=1.18}

\newcommand{\ket}[1]{|#1\rangle}
\newcommand{\bra}[1]{\langle #1|}
\newcommand{\K}{\mathcal{K}}
\newcommand{\G}{\mathcal{G}}
\newcommand{\Hterms}{\mathcal{H}}
\newcommand{\M}{\mathcal{M}}
\newcommand{\xyyx}{XY/YX}
\newcommand{\xxyy}{XX/YY}
\newcommand{\yzzy}{YZ/ZY}

\makeatletter
\let\table\tableorg
\let\endtable\endtableorg
\makeatother

\begin{document}

\title[Commutator-guided shallow-circuit repair]{Commutator-guided repair and classical-accessibility boundaries for shallow quantum circuits}

\author[1]{\fnm{Yifeng} \sur{Peng}}
\author[1]{\fnm{Sunpreet Singh} \sur{Arora}}
\author[1]{\fnm{Harshal} \sur{Shah}}
\author[1]{\fnm{Srinivasan} \sur{Raghuraman}}
\author[1]{\fnm{Kim} \sur{Wagner}}
\author[1]{\fnm{Bin} \sur{Yan}}
\affil[1]{\orgname{Visa}, \orgaddress{\city{Foster City}, \state{CA}, \country{USA}}}

\abstract{
Trainability does not guarantee low-energy arrival: a shallow quantum circuit can be optimizable yet fail, under fixed depth and evaluation budgets, to reach Hamiltonian-selected states. We introduce a commutator-guided repair for this operational failure mode. Low-order host--Hamiltonian commutators and fixed same-support companion axes define a restricted pool of local Pauli rotations, from which a compact completion layer is selected before full reoptimization. Across 189 paired six-qubit trials, unrepaired circuits achieve $0/189$ strict successes, a non-deployable per-trial control envelope achieves $109/189$, and two predeclared selectors achieve $133/189$ and $127/189$. The repair loses to matched controls for transverse-field Ising, is weak for pure-$ZZ$ MaxCut, and remains size-limited for weighted MIS. Under fixed-depth geometric locality, repaired losses and gradients are classically computable, while corresponding ground-state targets admit low-bond-dimension tensor-network approximations through the tested sizes. The method reveals a trainability--arrival--classical-accessibility tradeoff rather than quantum advantage.
}


\maketitle
\section{Introduction}

Near-term quantum algorithms begin with a practical compromise.  Deep circuits can represent many states, but they are expensive to run on noisy hardware and difficult to optimize.  Shallow circuits, by contrast, use only a small number of layers.  They reduce noise exposure, keep control overhead modest, and can avoid some barren-plateau mechanisms---optimization regimes in which gradients become too small to guide training---that affect deeper variational circuits \cite{mcclean2018barren,cerezo2021cost,marrero2021entanglement,holmes2022connecting}.  This is why shallow parameterized circuits are widely used in variational quantum eigensolvers (VQE), which minimize energy over a circuit family to approximate a ground state \cite{peruzzo2014vqe,cerezo2021vqa,tilly2022vqe}, the quantum approximate optimization algorithm (QAOA), which alternates cost and mixer operations for discrete optimization problems \cite{farhi2014qaoa,hadfield2019from}, and other near-term ansatz designs.  Throughout the paper, an \emph{ansatz} means the chosen parameterized circuit family being optimized.

The usual question is whether such a circuit can be trained.  Here ``trainable'' means that changing the circuit parameters can reliably lower the objective, for example because the gradients are not exponentially small.  But training is only half of the story.  The circuit must also arrive at a useful low-energy state under the depth and optimization budget available in practice.  In a ground-state or optimization problem, the Hamiltonian is the operator that assigns an energy to each quantum state, and the algorithm is trying to prepare states near its lowest energy.  A shallow ansatz can remain locally optimizable under the tested protocol and still fail the low-energy window because the finite-depth circuit lacks useful Hamiltonian-selected directions, or because the available optimization budget cannot exploit them.  Related reachability deficits have been observed in shallow QAOA settings \cite{akshay2020reachability}.  Complementary input-state-design approaches address this trainability--reachability tension by changing the initial state for a fixed ansatz \cite{wu2026enhancing}; here we keep the input state and host circuit fixed and instead repair the host by adding a restricted commutator-guided completion block.  We therefore use reachability operationally in this paper: as finite-budget low-energy arrival for a fixed host, optimizer, and evaluation cap, not as a proof that the global variational minimum of the host is separated from the ground state in every instance.

This paper asks whether such a finite-depth low-energy-arrival failure can be repaired without abandoning the shallow circuit.  The shallow setting is deliberate.  If arbitrary depth is allowed, missing directions can be hidden by generic expressivity growth; at fixed shallow depth, the design problem is sharper.  We must decide which few Hamiltonian-selected directions should be added first.  The repair we study is therefore not an unrestricted all-to-all ansatz.  In the local benchmarks, the added layer is restricted to one- and two-qubit Pauli rotations supported by the circuit model and inserted under a fixed gate budget; the scheduled depth of the inserted block is audited rather than assumed.

The guiding idea is to compare two pieces of information.  The first is the \emph{host ansatz}: the shallow circuit we already have, with its available generators, meaning the elementary rotations whose angles are trainable.  The second is the \emph{Hamiltonian}: the target operator whose low-energy states we want.  When a host generator and a Hamiltonian term fail to commute, their commutator $[A,B]=AB-BA$ points to a direction that is not simply another copy of either one.  In familiar Pauli language, where $X$, $Y$, and $Z$ are the three standard single-qubit rotation axes, a local mixer $X_i$ and a local field $Z_i$ generate a missing $Y_i$ direction through $[X_i,Z_i]$; an $X_i$ mixer and a coupling $Z_iZ_j$ generate edge directions such as $Y_iZ_j$ through $[X_i,Z_iZ_j]$.  These axes are natural repair candidates because they are tied to the mismatch between the chosen host and the chosen problem, not to a generic pool of extra gates.

This viewpoint also clarifies how the method differs from problem-aware ansatz designs.  A problem-aware ansatz such as QAOA or a Hamiltonian variational ansatz (HVA) inserts Hamiltonian terms, mixer terms, or known problem structure directly into the circuit \cite{wecker2015progress,wiersema2020hva,lucas2014ising}, whereas hardware-efficient designs prioritize native local rotations and entanglers without problem structure \cite{kandala2017hardware}.  Commutator completion asks a different question: after such a host has been chosen, what residual directions are still missing?  The answer can be an axis absent from the Hamiltonian itself, such as $Y_i$ or $Y_iZ_j/Z_iY_j$.  The method can therefore be used as a repair layer on top of a problem-aware host rather than only as a competing ansatz family.

Commutators are not new objects in quantum algorithms.  Adaptive ansatz methods such as ADAPT-VQE, qubit-ADAPT, QEB-ADAPT, and ADAPT-QAOA rank operators from a pool and reoptimize after each insertion \cite{grimsley2019adaptive,tang2021qubitadapt,yordanov2021qebadapt,zhu2022adaptive}.  Batched and TETRIS-ADAPT-VQE variants of adaptive ansatz construction also reduce selection rounds or circuit depth by adding several gradient-ranked operators per stage \cite{sapova2022variational,anastasiou2024tetris}.  Our repair is different in scope: it constructs a host--Hamiltonian-restricted pool before selection and inserts one budgeted completion block, rather than iteratively rebuilding a generic-pool ansatz.  ADAPT-QAOA is especially close because it can discover useful $Y$-type mixer directions on QAOA hosts; here those axes are derived from the host--Hamiltonian commutator closure before full reoptimization and then tested under strict-window and matched-control diagnostics.  QITE replaces imaginary-time steps by local unitary approximations \cite{motta2020qite}.  Counterdiabatic and adiabatic-gauge-potential (AGP) methods use related algebraic structure to follow a path or suppress diabatic transitions, and CD-QAOA, feedback, and reinforcement-learning variants add counterdiabatic or nested-commutator-inspired terms to QAOA-style dynamics \cite{sels2017minimizing,kolodrubetz2017geometry,wurtz2022counterdiabaticity,chandarana2022digitized,yao2021reinforcement,magann2022feedback}.  Lie-algebraic and expressibility analyses diagnose what circuit generators can produce or train in principle \cite{sim2019expressibility,larocca2023overparam,fontana2024adjoint,larocca2022diagnosing}.  Our contribution is more specific: a front-loaded, budgeted completion rule that builds a restricted local repair space from the chosen host and Hamiltonian, ranks it once, inserts a compact block before full reoptimization, and tests it against same-depth, wrong-axis, Hamiltonian-term, optimizer, adaptive-pool, generic one-shot-pool, and classical-accessibility controls.  The reported selector uses the low-order commutator closure together with a fixed same-support companion-axis set; the contribution is the commutator-guided restricted pool and one-shot repair protocol, not the invention of commutators themselves.  Supplementary Table S1 summarizes this positioning.

The resulting picture is a diagnostic map, not a claim that one ansatz family follows a clean monotone scaling law.  We test when Hamiltonian-aware shallow repair works and when it fails.  First, a paired benchmark over seven shallow hosts and nine Hamiltonian settings asks whether commutator-guided completion improves strict low-energy arrival under matched inserted-gate budgets and final reoptimization caps, with selector costs tracked separately.  Second, axis-specific controls ask whether the gain follows the predicted commutator directions rather than generic added depth.  Third, application quadratic unconstrained binary optimization (QUBO) problems and tensor-network probes ask whether the repaired circuits remain size-limited or classically accessible.  Here QUBO denotes an optimization problem with binary variables and a quadratic cost function, while tensor networks are classical compressed representations of quantum states used to test whether the states remain easy to simulate.  Boundary cases are kept in the narrative because they are part of the result: they show where the first low-order closure is insufficient, where a matched control is already enough, or where a repaired regime is not evidence of quantum advantage.  In this sense the paper is about a trainability--low-energy-arrival--classical-accessibility tradeoff: commutator-guided repair can make a shallow circuit reach low-energy states it previously missed under a fixed protocol, but the resulting success must still be tested against classical accessibility \cite{white1992dmrg,schollwock2011density,cerezo2025does}.

\begin{figure}[!t]
\centering
\includegraphics[width=0.98\textwidth]{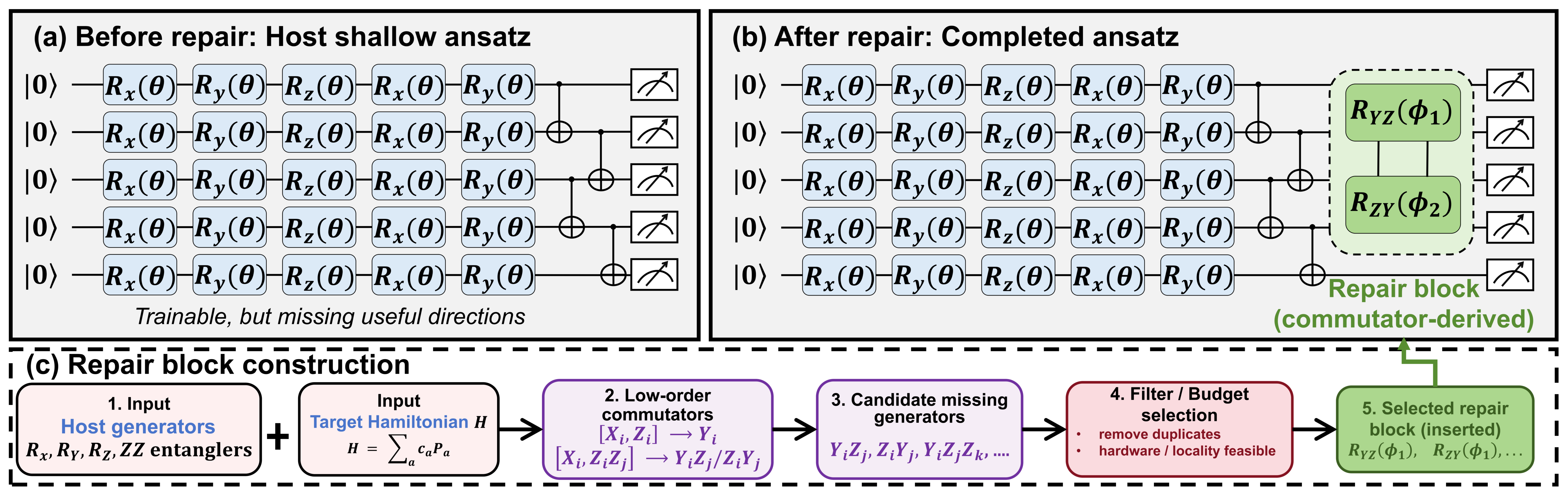}
\caption{\label{fig:mechanism}Local optimizability is not low-energy arrival, and commutators provide a Hamiltonian-aware repair.  A shallow host that remains locally optimizable under the tested protocol can still miss useful low-energy directions under a fixed depth and evaluation budget.  The repair algorithm takes the host generators and target Hamiltonian as inputs, computes low-order stripped commutators, filters the resulting Pauli axes by locality and budget, and appends the selected axes as a compact completion block after the host before full reoptimization.  The displayed example highlights two first-shell motifs used throughout the paper: local $Y_i$ from $[X_i,Z_i]$ and edge $Y_iZ_j/Z_iY_j$ from $[X_i,Z_iZ_j]$.  Here $YZ/ZY$ denotes the two oriented Pauli rotations on the same edge, $Y_iZ_j$ and $Z_iY_j$, generated from the two local mixers $X_i$ and $X_j$ acting on a $Z_iZ_j$ coupling.  Later results bound the interpretation by testing when loss evaluations or targets remain classically accessible.}
\end{figure}

\section{Methods}

Our approach combines an operational low-energy-arrival metric that makes the shallow-VQE failure mode measurable, a Hamiltonian-aware completion that supplies candidate missing directions, and a paired benchmark protocol that isolates the repair from generic alternatives.

\subsection*{Reachability metrics}

We track low-energy arrival through
\begin{equation}
  P(n,d,\mathrm{it})
  =
  \Pr\!\left[
    E_{\rm best}(n,d,\mathrm{it})-E_0
    \le
    \epsilon(n)
  \right],
  \label{eq:arrivalprob}
\end{equation}
where $n$ is the number of qubits, $d$ is the depth within a fixed ansatz family, $\mathrm{it}$ is the function-evaluation budget, $E_{\rm best}$ is the lowest energy found within that budget, $E_0$ is the exact ground energy, and $\epsilon(n)$ is the target low-energy window.  This is an operational arrival probability: it depends on the ansatz, optimizer, initialization, and evaluation cap, and should not be read as a separate proof of the exact variational optimum of the ansatz.  A run is counted as a strict success only if its best energy error falls inside this window.  In the exact-diagonalization experiments we use the strict low-energy window $\epsilon(n)=\max(10^{-6},0.1\Delta_{\rm gap})$ unless otherwise stated, where $\Delta_{\rm gap}$ is the spectral gap, i.e., the energy separation between the ground-state manifold and the first excited level.  For diagonal QUBO problems with degenerate optima, we also report the optimal-subspace sampling probability
\begin{equation}
  P_{\rm opt}
  =
  \sum_{x\in\arg\min H}|\langle x|\psi(\theta^\star)\rangle|^2,
  \label{eq:popt}
\end{equation}
which measures how much optimized probability mass lies on optimal bitstrings, even when the expectation value misses the strict window.  The term ``subspace'' is used because a QUBO can have several optimal bitstrings with the same minimum cost.

We use the resulting arrival frontiers as finite-size diagnostics, not as asymptotic scaling laws or pure representation-gap certificates.  They test whether Hamiltonian-aware completion shifts the empirical boundary in $n$, circuit depth, and optimization budget relative to base and matched-control ansatz families.

\subsection*{Low-order commutator completion}

Intuitively, a nonzero ansatz--Hamiltonian commutator marks a direction selected by the Hamiltonian but not generated directly by the shallow host. Collecting the lowest-order such directions, filtered by locality and budget, yields a compact repair layer. The locality filter is part of the algorithm rather than an implementation detail: in the geometrically local benchmarks the retained axes are one- and two-qubit rotations of bounded lattice range, scheduled into sublayers of non-overlapping rotations; the resulting sublayer count is bounded by the gate budget and audited in Supplementary Table S37.  Here ``geometrically local'' means that gates and Hamiltonian terms connect only nearby sites on the lattice.  Dense or non-geometric QUBO instances, whose graph couplings need not be local in this geometric sense, are treated separately and are not used for the local light-cone theorem.

Let $\ket{\psi(\theta)}=U(\theta)\ket{0}$ be a shallow ansatz generated by Pauli strings $\G_0=\{P_a\}$, where a Pauli string is a tensor product of single-qubit $I$, $X$, $Y$, and $Z$ operators.  Its tangent vectors can be written as
\begin{equation}
  \ket{t_a(\theta)}
  =
  \partial_{\theta_a}\ket{\psi(\theta)}
  =
  -\frac{i}{2}\widetilde P_a(\theta)\ket{\psi(\theta)},
  \label{eq:tangent}
\end{equation}
where $\widetilde P_a(\theta)$ is the generator dressed by the suffix of the circuit, meaning conjugated by the gates that follow it.  The energy gradient is the projection of the residual onto this tangent span,
\begin{equation}
  \partial_{\theta_a}E
  =
  2\,{\rm Re}\bra{t_a(\theta)}(H-E)\ket{\psi(\theta)}.
  \label{eq:gradient}
\end{equation}
Thus a shallow circuit can be trainable but incomplete: $(H-E)\ket{\psi}$ may have important components outside $T_\psi\M_0$, the local tangent space of states reachable by infinitesimal parameter changes.

Now add a candidate generator $A$.  To first order, it rotates an existing generator $P$ by
\begin{equation}
  e^{-i\alpha A/2}Pe^{i\alpha A/2}
  =
  P-\frac{i\alpha}{2}[A,P]+O(\alpha^2).
  \label{eq:rotation}
\end{equation}
A completion layer therefore does more than add isolated tangent vectors; it changes the dressed tangent geometry through commutators.  The lowest Hamiltonian-aware candidate set is the first stripped shell
\begin{equation}
  \K_1 =
  {\rm strip}\{i[P,Q]\neq 0:P\in\G_0,\;Q\in\Hterms\},
  \label{eq:k1}
\end{equation}
where $\Hterms$ is the set of Hamiltonian Pauli terms and ${\rm strip}$ removes identities, duplicates, base generators, and Hamiltonian terms already present; this leaves only genuinely new candidate axes.
This first-order argument is a construction principle for candidate axes, not by itself a predictor of strict-window success.  The finite-depth success of an inserted block is assessed only after full reoptimization and against matched controls; in the results below, static tangent-space diagnostics are explicitly treated as incomplete.

For a QAOA-style host with local $X$ mixers and a diagonal Ising/QUBO cost
$H_C=\sum_{(i,j)}J_{ij}Z_iZ_j+\sum_i h_iZ_i$, the first shell is explicit:
\begin{equation}
  \begin{aligned}
  i[X_i,Z_iZ_j]&\propto Y_iZ_j,\\
  i[X_j,Z_iZ_j]&\propto Z_iY_j,\\
  i[X_i,Z_i]&\propto Y_i .
  \end{aligned}
  \label{eq:qaoa}
\end{equation}
Thus pure-$ZZ$ objectives predict an edge $\yzzy$ completion, while QUBOs with local fields ($h_i\neq0$; ``fielded'' QUBOs below) additionally predict a local-$Y$ sublayer; for J1-J2 and XXZ targets, the same rule adds diagonal and mixed-shell terms when they appear in $\Hterms$. The elementary Pauli derivation is given in Supplementary Section 1. We use \emph{commutator completion} as the umbrella term: its concrete instances below are the closure selectors in the paired benchmark, the fixed $\yzzy+Y$ axis in the QAOA2D mechanism test, and the local-$Y$ completion in fielded QUBOs.
Figure~\ref{fig:mechanism} summarizes the construction as a circuit-design rule: compute the lowest commutator shell, insert the resulting axes as a compact layer, and compare the repaired ansatz against matched-resource alternatives.  Supplementary Fig. S2 gives the corresponding motif dictionary and the control logic for interpreting a positive result.

This construction is the lowest-order nontrivial closure of the ansatz--Hamiltonian Lie algebra under locality and budget constraints.  In this context, ``Lie algebra'' simply refers to the set of directions generated by repeatedly taking commutators of available circuit and Hamiltonian operators.  The first-shell rule is the cheapest instance.  The full algorithm used in the expanded host test constructs
\begin{equation}
  \mathcal C_{\le 2}(\G_0,H)= {\rm impl}\!\left(\K_1\cup \K_2\cup \mathcal Q(\G_0,H)\right),
\end{equation}
where $\K_1={\rm strip}\,i[\G_0,\Hterms]$, $\K_2={\rm strip}\,i[\K_1,\G_0\cup\Hterms]$, and ${\rm impl}$ keeps Pauli rotations of weight at most two on supports available to the host, Hamiltonian, or low-order closure.  The companion set $\mathcal Q$ is fixed exhaustively by the same supports in the reported benchmark: local $Y_i$ axes are included on sites with local-field commutators, edge $Y_iZ_j/Z_iY_j$ axes are included on $ZZ$ supports exposed by $X$ mixers, and edge $X_iZ_j/Z_iX_j$ axes are included on the corresponding brickwork/hardware-efficient-ansatz (HEA) companion supports; no other Pauli axes are added to $\mathcal Q$.  We select a budgeted sublayer from $\mathcal C_{\le2}$ \emph{before} the reported full reoptimization.  The default selector scores each candidate at the optimized base with all new repair angles set to zero, then ranks candidates by the absolute parameter-shift derivative with respect to that candidate angle.  Ties are broken deterministically by the fixed candidate ordering generated after duplicate and sign stripping.  The trust-region selector uses a one-dimensional finite-difference gradient and curvature for each candidate to estimate the single-axis energy decrease allowed by a small trust radius, then ranks candidates by that predicted decrease.  In both cases the highest-ranked candidates are inserted up to the gate budget, new angles are initialized by the selector rule, and all base and repair parameters are reoptimized together.  Unless otherwise stated, reported post-base evaluation caps apply to the final full reoptimization after operator selection.  Candidate-scoring evaluations are tracked separately as selector cost; end-to-end accounting is $N_{\rm total}=N_{\rm selection}+N_{\rm reoptimization}$ after the optimized base is fixed.  This differs from a gradient-ranked ADAPT pool in two ways: the pool is not a generic global Pauli list, and the selected block is inserted at once and fully reoptimized rather than grown by sequential best-of insertion.  Additional implementation details, full cell-level settings, and audit tables are provided in the Supplementary Information.

The rule is intentionally \emph{ansatz-dependent}.  It is not a Hamiltonian-only prescription and does not propose one universal layer.  Changing the base generator set $\G_0$ changes the closure: local $X$ mixers with $ZZ$ costs expose $\yzzy$; brickwork hosts with local $Y/Z$ rotations emphasize $XZ/ZX$ in the first shell, while the closure selector can also recover useful companion axes; HEA hosts change the ranking through their local-axis content; and HVA leaves a harder mixed Heisenberg shell.  Supplementary Figs. S1 and S3 and Supplementary Tables S11--S13 report the original host-scope matrix and the expanded seven-host closure-selector matrix.  We treat host dependence as a scope condition, not a loophole: low-order completion repairs a mismatch between a chosen shallow ansatz and a Hamiltonian, so the host, commutator order, and selector are part of the algorithm.

\subsection*{Benchmarks and protocol}

We use exact statevector simulation and dense exact diagonalization for lattice Hamiltonians and diagonal application-QUBO instances with six to fifteen qubits in the main benchmarks.  Statevector simulation evolves the full $2^n$-dimensional quantum state without sampling noise, while exact diagonalization computes the true ground energy by diagonalizing the Hamiltonian matrix.  All reported optimizations use bounded L-BFGS-B with parameter bounds $[-\pi,\pi]$, tolerances \texttt{ftol=1e-10} and \texttt{gtol=1e-8}, shared seeds within each paired cell, and a fixed objective-evaluation cap for the final post-selection reoptimization; censored cost to target is set equal to that cap when the strict target is not reached.  Completion arms use seeded multistart initializations when available, with the inherited optimized-base parameters as one start and seeded perturbations as additional starts; the evaluation cap is shared across starts rather than reset.  Strict success means
\begin{equation}
  E(\theta)-E_0 \leq \max(10^{-6},0.1\Delta_{\rm gap}),
\end{equation}
where $\Delta_{\rm gap}$ is the spectral gap.  This gap-relative window is intentionally stringent but heterogeneous; therefore we also report energy margins, fidelities or optimal-subspace overlaps.  Fidelity is the squared overlap between the optimized state and the target ground state when the target is nondegenerate; for degenerate diagonal QUBOs we instead use optimal-subspace overlap.  All completion arms append Pauli rotations to an optimized base circuit and then reoptimize all parameters.  The full cell-level breakdown, initialization rules, selector accounting, and locality audit are reported in the Supplementary Information.

\section{Results}

The benchmarks below are not designed to produce a monotone scaling law for a single ansatz family. They are designed as a diagnostic map: where commutator-guided repair improves finite-budget low-energy arrival for shallow hosts, where matched controls already suffice or win, and where any repaired success remains classically accessible. We therefore move from the broadest paired benchmark to motif-specific application tests and finally to the sharpest classical-accessibility limit, retaining positive and boundary cases side by side. Supplementary Table S3 collects a representative cross-reference map of these regimes.

\subsection{Hamiltonian-aware completion improves shallow low-energy arrival across paired benchmarks}

We first test whether a front-loaded commutator-closure selector improves strict low-energy arrival across many host ansatz families and Hamiltonian families, under resources matched to generic alternatives. The paired lattice benchmark spans seven shallow host ansatz families and nine $2\times3$ Hamiltonian settings---two transverse-field Ising (TFIM) fields, three J1-J2 Ising couplings, two Heisenberg J1-J2 couplings, and two XXZ J1-J2 anisotropies---with three shared seeds and a matched $56$-gate insertion budget, for $7\times9\times3=189$ paired six-qubit trials. TFIM is an Ising model with a transverse single-qubit field; J1-J2 models include both nearest-neighbor and next-nearest-neighbor couplings; Heisenberg models use symmetric $XX$, $YY$, and $ZZ$ spin couplings; and XXZ models are anisotropic variants in which the $ZZ$ coupling has a different strength from the $XX/YY$ terms. For each host and Hamiltonian we optimize the base circuit, rank the implementable low-order closure $\mathcal{C}_{\le2}$ once at the optimized base, insert the selected block under the shared gate cap, and reoptimize all parameters. The two reported selectors---a gradient-ranked selector and a trust-region energy-descent selector---are predeclared, not chosen after seeing the outcomes.

The seven hosts span different amounts of structure.  The product host contains independent one-qubit rotations and no entangling layer, so it tests whether local basis freedom alone can help.  QAOA2D is the most problem-structured Ising host: it alternates a two-dimensional $ZZ$ cost layer with an $X$ mixer using shared angles.  The brickwork host uses local $R_Y/R_Z$ rotations plus a checkerboard pattern of CZ entanglers, representing a hardware-efficient but less Hamiltonian-specific circuit.  The three HEA variants keep the same CZ entangling scaffold but change the local axes available to the host ($X/Y$, $X/Z$, or $X/Y/Z$).  The HVA-Heisenberg host is a problem-aware Hamiltonian variational ansatz with shared $XX$, $YY$, and $ZZ$ rotations on edges and diagonals.

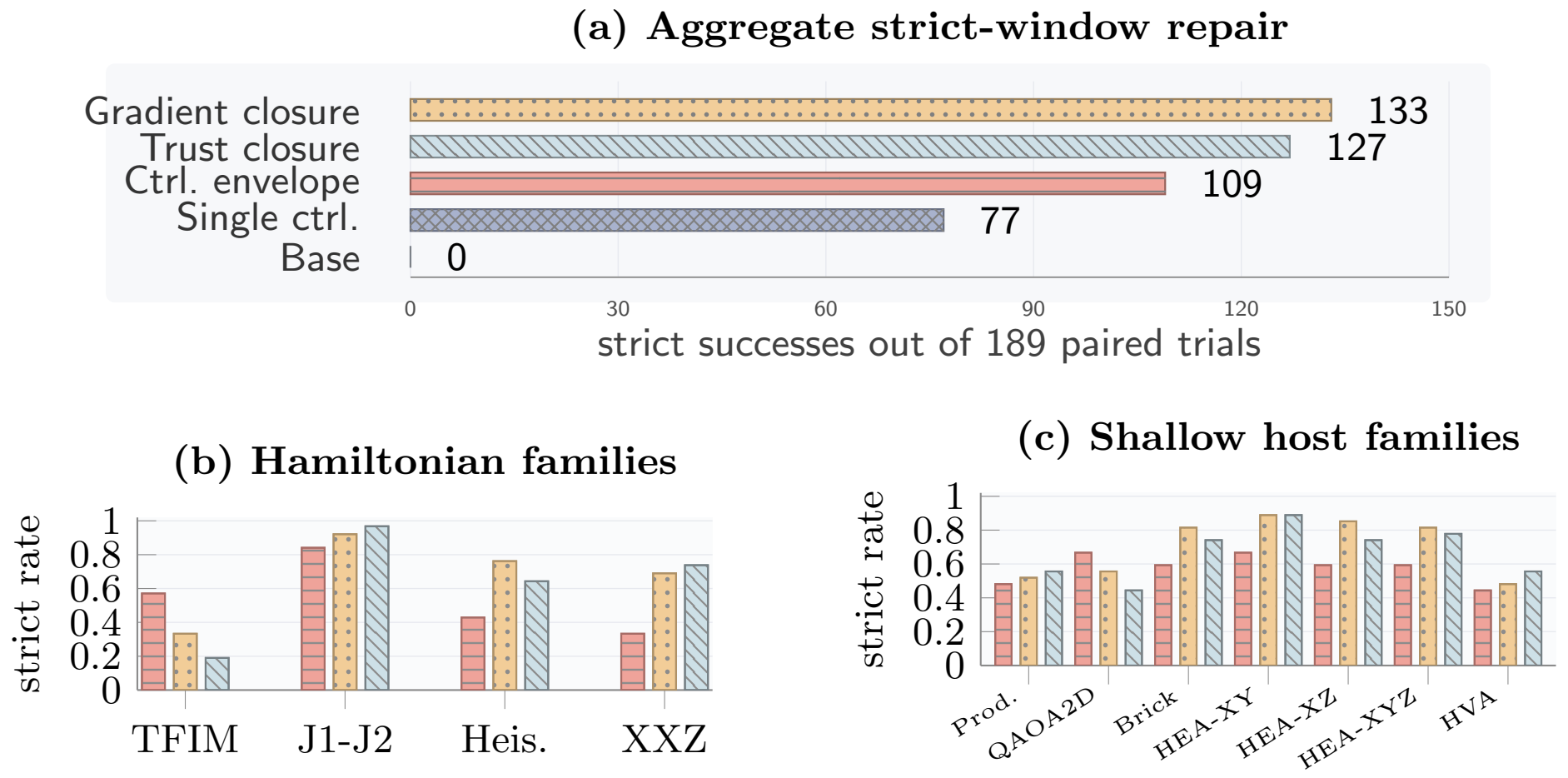
\begin{figure}[!t]
\centering
\begin{adjustbox}{max width=0.98\textwidth}
\begin{minipage}{0.98\textwidth}
\centering
\begin{tikzpicture}[x=0.055cm,y=0.54cm,font=\sffamily\scriptsize]
\fill[gridbar!32,rounded corners=2pt] (-44,0.03) rectangle (156,3.54);
\node[font=\rmfamily\small\bfseries] at (75,4.05) {(a) Aggregate strict-window repair};
\foreach \x in {0,30,60,90,120,150} {
  \draw[gridbar,line width=0.25pt] (\x,0.40) -- (\x,3.28);
  \node[anchor=north,font=\sffamily\tiny,text=black!70] at (\x,0.28) {\x};
}
\draw[black!45,line width=0.35pt] (0,0.40) -- (150,0.40);
\def\aggbar#1#2#3#4#5{%
  \node[anchor=east,text=black!80] at (-5,#4) {#1};
  \draw[fill=#3!90,draw=#3!65!black,line width=0.42pt] (0,#4-0.16) rectangle (#2,#4+0.16);
  \path[#5, pattern color=black!50] (0,#4-0.16) rectangle (#2,#4+0.16);
  \pgfmathsetmacro{\labelx}{max(#2+3,3)}
  \node[anchor=west,text=black] at (\labelx,#4) {#2};
}
\aggbar{Base}{0}{basebar}{0.70}{pattern=north east lines}
\aggbar{Single ctrl.}{77}{ctrlbarlight}{1.24}{pattern=crosshatch}
\aggbar{Ctrl. envelope}{109}{ctrlbar}{1.78}{pattern=horizontal lines}
\aggbar{Trust closure}{127}{trustbar}{2.32}{pattern=north west lines}
\aggbar{Gradient closure}{133}{gradbar}{2.86}{pattern=dots}
\node[anchor=north,font=\sffamily\scriptsize,text=black!75] at (75,-0.12) {strict successes out of 189 paired trials};
\end{tikzpicture}

\vspace{5pt}
\begin{minipage}{0.48\linewidth}
\centering
\begin{tikzpicture}
\begin{axis}[
  width=\linewidth,
  height=0.48\linewidth,
  ybar,
  ymin=0,ymax=1.02,
  ylabel={strict rate},
  symbolic x coords={TFIM,J1J2,Heis,XXZ},
  xticklabels={TFIM,J1-J2,Heis.,XXZ},
  xtick=data,
  x tick label style={font=\scriptsize},
  y tick label style={font=\scriptsize},
  label style={font=\scriptsize},
  ymajorgrids=true,
  grid style={gridbar, line width=0.25pt},
  axis x line*=bottom,
  axis y line*=left,
  axis line style={black!45},
  tick style={black!45},
  axis background/.style={fill=gridbar!20},
  bar width=5.2pt,
  title={(b) Hamiltonian families},
  title style={font=\small\bfseries, yshift=-2pt},
]
\addplot+[fill=ctrlbar!92, draw=ctrlbar!70!black,
  postaction={pattern=horizontal lines, pattern color=black!45}]
  coordinates {(TFIM,0.571) (J1J2,0.841) (Heis,0.429) (XXZ,0.333)};
\addplot+[fill=gradbar!92, draw=gradbar!72!black,
  postaction={pattern=dots, pattern color=black!45}]
  coordinates {(TFIM,0.333) (J1J2,0.921) (Heis,0.762) (XXZ,0.690)};
\addplot+[fill=trustbar!92, draw=trustbar!58!black,
  postaction={pattern=north west lines, pattern color=black!45}]
  coordinates {(TFIM,0.190) (J1J2,0.968) (Heis,0.643) (XXZ,0.738)};
\end{axis}
\end{tikzpicture}
\end{minipage}
\hfill
\begin{minipage}{0.48\linewidth}
\centering
\begin{tikzpicture}
\begin{axis}[
  width=\linewidth,
  height=0.48\linewidth,
  ybar,
  ymin=0,ymax=1.02,
  ylabel={strict rate},
  symbolic x coords={Prod,QAOA,Brick,HEAXY,HEAXZ,HEAXYZ,HVA},
  xticklabels={Prod.,QAOA2D,Brick,HEA-XY,HEA-XZ,HEA-XYZ,HVA},
  xtick=data,
  x tick label style={font=\tiny,rotate=32,anchor=east},
  y tick label style={font=\scriptsize},
  label style={font=\scriptsize},
  ymajorgrids=true,
  grid style={gridbar, line width=0.25pt},
  axis x line*=bottom,
  axis y line*=left,
  axis line style={black!45},
  tick style={black!45},
  axis background/.style={fill=gridbar!20},
  bar width=3.7pt,
  title={(c) Shallow host families},
  title style={font=\small\bfseries, yshift=-2pt},
]
\addplot+[fill=ctrlbar!92, draw=ctrlbar!70!black,
  postaction={pattern=horizontal lines, pattern color=black!45}]
  coordinates {(Prod,0.481) (QAOA,0.667) (Brick,0.593) (HEAXY,0.667) (HEAXZ,0.593) (HEAXYZ,0.593) (HVA,0.444)};
\addplot+[fill=gradbar!92, draw=gradbar!72!black,
  postaction={pattern=dots, pattern color=black!45}]
  coordinates {(Prod,0.519) (QAOA,0.556) (Brick,0.815) (HEAXY,0.889) (HEAXZ,0.852) (HEAXYZ,0.815) (HVA,0.481)};
\addplot+[fill=trustbar!92, draw=trustbar!58!black,
  postaction={pattern=north west lines, pattern color=black!45}]
  coordinates {(Prod,0.556) (QAOA,0.444) (Brick,0.741) (HEAXY,0.889) (HEAXZ,0.741) (HEAXYZ,0.778) (HVA,0.556)};
\end{axis}
\end{tikzpicture}
\end{minipage}
\end{minipage}
\end{adjustbox}
\caption{\label{fig:closureselectorvisual}Commutator-guided repair improves strict low-energy arrival relative to a conservative matched-control envelope.  Panel (a) compares strict-window successes over 189 paired six-qubit trials: optimized base $0/189$, the strongest single matched noncommutator control $77/189$, a non-deployable per-trial control envelope $109/189$, and the two predeclared closure selectors $133/189$ and $127/189$.  The $109/189$ comparator is an envelope over several noncommutator controls, not one deployable baseline: for each host--Hamiltonian cell and seed it credits the best noncommutator control arm before aggregation.  It is therefore the conservative comparison emphasized in the text, and it intentionally gives the noncommutator controls a per-trial choice that the reported selectors do not receive. Panels (b) and (c) show that the gain is structured by Hamiltonian family and host.  Product denotes unentangled one-qubit rotations, QAOA2D denotes a two-dimensional $ZZ$-cost/$X$-mixer host, Brick denotes a brickwork $R_Y/R_Z$-CZ circuit, HEA variants share a CZ scaffold with different local axes, and HVA denotes a Heisenberg Hamiltonian variational ansatz. The optimized base is zero in every aggregate and is omitted from these two panels for legibility.  Bar colors and hatch patterns in (b) and (c) match the corresponding arms in (a): control envelope (salmon, horizontal hatching), gradient closure (orange, dotted hatching), and trust closure (light blue, diagonal hatching).  The selectors outperform the control envelope on J1-J2 Ising, Heisenberg, XXZ, and most non-QAOA hosts, while TFIM is a control-win family and the QAOA2D aggregate remains a mixed boundary case. Wilson 95\% intervals for the envelope and the two selector rates are $0.51$--$0.64$, $0.64$--$0.76$, and $0.60$--$0.73$, respectively; because all methods are evaluated on the same host--Hamiltonian cells and seeds, paired cell-level comparisons are reported in the text.}
\end{figure}

Across the full paired matrix the optimized base never reaches the strict window ($0/189$), the non-deployable per-trial control envelope reaches $109/189$, and the two commutator-closure selectors reach $133/189$ (gradient) and $127/189$ (trust) (Fig.~\ref{fig:closureselectorvisual}).  The envelope is intentionally non-deployable: for each host--Hamiltonian cell and seed, it credits the strongest noncommutator arm among same-family parameter-matched depth, Hamiltonian-term layers, local-$Y$-only layers, wrong-axis layers, and random edge-axis layers before aggregation.  The strongest single deployable noncommutator control is the random edge-axis layer, a seeded same-count layer that samples local $Y_i$ axes and two-qubit edge Pauli axes on available local/edge supports under the same 56-rotation and 6000-evaluation caps; it reaches $77/189$.  We therefore treat $133/189$ versus $109/189$, rather than $133/189$ versus $77/189$, as the main aggregate comparison.  This framing deliberately favors the controls because the envelope may choose a different noncommutator arm in each trial, whereas the gradient and trust closure columns are fixed predeclared selectors.

The raw fixed-shell arms are weaker: a fixed $\K_1+Y$ arm reaches $62/189$, and a fixed $\K_1+\K_2$ arm reaches $81/189$ (Supplementary Table S13).  A stricter selector ablation that removes the low-order companion axes and ranks only $\K_1\cup\K_2$ reaches $104/189$ with gradient ranking and $98/189$ with trust ranking, placing strict commutator closure just below the per-trial envelope but well above the strongest single control and the fixed unranked $\K_1+\K_2$ block.  Adding the fixed same-support companion-axis pool and reranking accounts for the remainder of the lift from $104/189$ to $133/189$.  Thus the headline result should be read as the performance of a predeclared selector over the budgeted low-order closure and companion-axis pool, not as the performance of one unranked first-shell block.  The Supplementary Information gives the full cell-level matrix and per-host selector breakdown (Supplementary Tables S4 and S13), along with the protocol checklist, control-arm audit, $\K_1\cup\K_2$-only selector ablation, and locality audit (Supplementary Tables S34--S37).

The resulting gain is $24$ additional strict successes over the non-deployable control envelope, or about $13$ percentage points on the paired grid. At the cell level, where each host--Hamiltonian cell contains three shared seeds, the gradient selector has a higher strict-success count than the per-trial control envelope in 22 cells, ties it in 33, and loses in 8 (two-sided sign test over non-tied cells, $p=0.016$).  The trust selector gives 22 wins, 31 ties, and 10 losses ($p=0.050$), so we regard it as a near-boundary robustness check rather than an independent statistically decisive win.  A diagnostic selector envelope, reported only in the Supplementary Information as headroom rather than as the default method, wins 25 cells, ties 31, and loses 7.

On representative control subsets, additional tests address explanations of the aggregate gain that do not involve missing directions, such as optimizer choice, initialization, and budget. Alternative initializations, aggressive reoptimization, and larger optimizer budgets do not rescue the tested base failures, same-family depth increases do not reproduce the repair, and matched full-pool controls require substantially larger selection cost to become competitive. Here an ADAPT-style control means a sequential procedure that repeatedly selects one operator from a larger pool and reoptimizes after each insertion. Under the same post-base evaluation cap and final gate budget, the full-pool sequential control reaches 1/24 and 3/18 strict successes on the application-QUBO and QAOA2D stress subsets, against 14/24 and 12/18 for the fixed first-shell rule; with a five-fold cap and six extra insertions it recovers to 13/24 and 9/18 but at roughly $7.2\times$ and $9.1\times$ the censored cost to target. We also ran a direct one-shot full local Pauli-pool control on the same 42 stress-test trials (24 application-QUBO and 18 QAOA2D trials). This control uses the same optimized bases, inserted-gate budget, block-insertion protocol, and final reoptimization cap as the fixed first-shell arm; unlike fixed $\K_1$, it additionally performs one-shot gradient screening over the full implementable local/edge Pauli pool. It reaches 0/24 and 3/18 strict successes on the two subsets; fixed $\K_1$ wins paired margins in 39/42 trials, with mean margin advantage $1.058$ and median advantage $0.599$ (Supplementary Table S33). These controls do not cover every cell of the 189-trial matrix; instead, they are finite-budget falsification gates for the most direct alternative explanations. A full 189-trial one-shot local-pool sweep, and a sequential search restricted to exactly the same $\mathcal C_{\le2}$ pool, remain open controls. The contribution is therefore not adaptive search itself, but the observation that low-order ansatz--Hamiltonian commutator closure already supplies a coherent first adaptive layer. The full optimizer, resource, full-pool, and ADAPT-style controls are reported in Supplementary Tables S6--S9 and S32--S33 and in the Supplementary Information section ``Negative and ADAPT-Style Controls''.

The boundaries are also informative. On the transverse-field Ising family and on the QAOA2D host the matched controls are already competitive: TFIM is a control-win family, and QAOA2D is a mixed host whose TFIM rows make its aggregate ambiguous. The claim is correspondingly scoped---commutator closure supplies a structured, Hamiltonian-aware repair space that is more reliable than generic added depth on this benchmark, not a universal win over every resource-matched alternative on every host and Hamiltonian.

Conversely, the problem-aware HVA host makes the residual framing concrete: unrepaired it reaches $0/27$ strict windows, and the closure selectors reach $13/27$ and $15/27$ against $12/27$ for the per-trial control envelope (Fig.~\ref{fig:closureselectorvisual}c).  Encoding problem structure in the host therefore does not by itself close the finite-depth arrival gap.

\subsection{The repair is axis-specific, not merely deeper}

To separate algebraic mechanism from additional depth, we isolate the repair on the QAOA2D Ising host. In this setting the first commutator shell predicts a single edge motif, $[X_i,Z_iZ_j]\rightarrow Y_iZ_j/Z_iY_j$, together with a local $Y_i$ from $[X_i,Z_i]$.

Repeated insertion of the predicted $\yzzy+Y$ block reaches strict success on all paired trials at both tested insertion budgets ($B=40$ and $B=56$ gates). The strongest matched-parameter QAOA depth control reaches only two thirds of the $B=40$ cells and one third of the hardest $B=56$ cells, at much larger cost, while repeated wrong-axis layers ($\xyyx+Y$, $\xxyy+Y$) reach zero strict successes. The predicted axis thus carries the repair: adding parameters, repeating an arbitrary structured layer, or using generic edge expressibility does not substitute for it. A counterdiabatic/AGP completion, budget-matched to a single fixed commutator-axis block (both arms use 14 inserted gates and 16 parameters on 18 held-out QAOA2D trials; Supplementary Table S18), improves the base but reaches only $6/18$ strict windows (mean fidelity $0.946$) against $12/18$ for the single commutator-axis block (mean fidelity $0.994$). Low-order commutator completion is therefore a sparse, axis-specific repair with counterdiabatic motivation but a reachability objective rather than schedule approximation. Supplementary Tables S17--S21 and Supplementary Fig. S4 give the fixed-axis controls and stop/repeat sparsification analysis.

The motif structure transfers beyond the cell in which it was identified. Freezing the full first shell $\K_1+Y$ and applying it to two-dimensional Heisenberg J1-J2 and to held-out $2\times3$ XXZ J1-J2 reaches $8/9$ strict successes in each family, against $3/9$ and $2/9$ for budget-matched random-greedy selection and $0/9$ for noncommutator controls (Supplementary Fig. S11 and Supplementary Table S22). The boundary is equally clear: on the larger $2\times4$ XXZ cells the same frozen rule lifts mean ground-state fidelity from $0.002$ to $0.917$ but does not enter the strict window. On the depth-tested QAOA2D TFIM and J1-J2 cells, the analogous single-block boundary is a finite-depth threshold rather than a structural span failure: static tangent-space and commutator-shell coverage do not separate the cells that succeed at depth one from those that need a second block, and repeating the first-shell block enters the strict window with at most two inserted blocks for every tested depth-threshold cell (Supplementary Tables S14--S15). We therefore do not claim a universal solver: the first shell is a strong direction finder, while strict success on larger or harder cells can require repeated shells or a higher-order complement.

\subsection{Fielded QUBOs give a size-limited optimal-sampling diagnostic}

Application-motivated QUBOs provide a complementary diagnostic because commutator completion predicts a different repair axis. After the binary-to-Ising map, which rewrites binary variables as Pauli-$Z$ operators, weighted maximum independent set (MIS) and minimum vertex cover (MVC) produce both $ZZ$ penalties and local $Z$ fields, whereas weighted maximum cut (MaxCut) is a pure-$ZZ$ baseline. We call MIS and MVC ``fielded'' QUBOs because the Ising form contains these local $Z_i$ field terms. The local-field instance therefore makes a falsifiable prediction: fielded QUBOs should expose a local-$Y$ repair axis from $[X_i,Z_i]$, while pure MaxCut should not.

The graph families used here---sparse and denser Erd\H{o}s--R\'enyi graphs, random geometric graphs, and an irregular star-tail family for MIS/MVC, with a ring-plus-chords family replacing the irregular family for MaxCut (Supplementary Information section ``Application-QUBO Reachability, Screening, and Scaling Detail'')---probe sparse-to-moderately-dense structured optimization regimes rather than arbitrary toy Hamiltonians. QUBO and Ising formulations cover many combinatorial optimization models \cite{kochenberger2014unconstrained,glover2022quantum,lucas2014ising}, and sparse or bounded-degree interaction graphs arise naturally when each decision variable participates in only a limited set of constraints. Such sparsity does not make the problems trivial: classical graph-optimization problems remain nontrivial under strong degree restrictions \cite{alimonti2000some}, and sparse large-girth regular MaxCut is a standard setting for analyzing QAOA and local algorithms \cite{basso2021quantum,barak2021classical}. Related MIS/MWIS encodings on structured graphs have also become central benchmarks for neutral-atom quantum optimization \cite{ebadi2022quantum,nguyen2023quantum}. This sparse-graph regime is therefore aligned with the shallow-repair question. For the first-shell completions deployed in this section, the commutator terms are supported on the instance's sites and edges, so when the cost graph has $O(n)$ edges the candidate pool scales with the edge count rather than automatically becoming all-to-all; within this regime, fielded MIS/MVC instances test the predicted local-$Y$ repair, while pure-$ZZ$ MaxCut provides a boundary case in which that local-field motif is absent.

\begin{figure}[!t]
\centering
\includegraphics[width=0.98\textwidth]{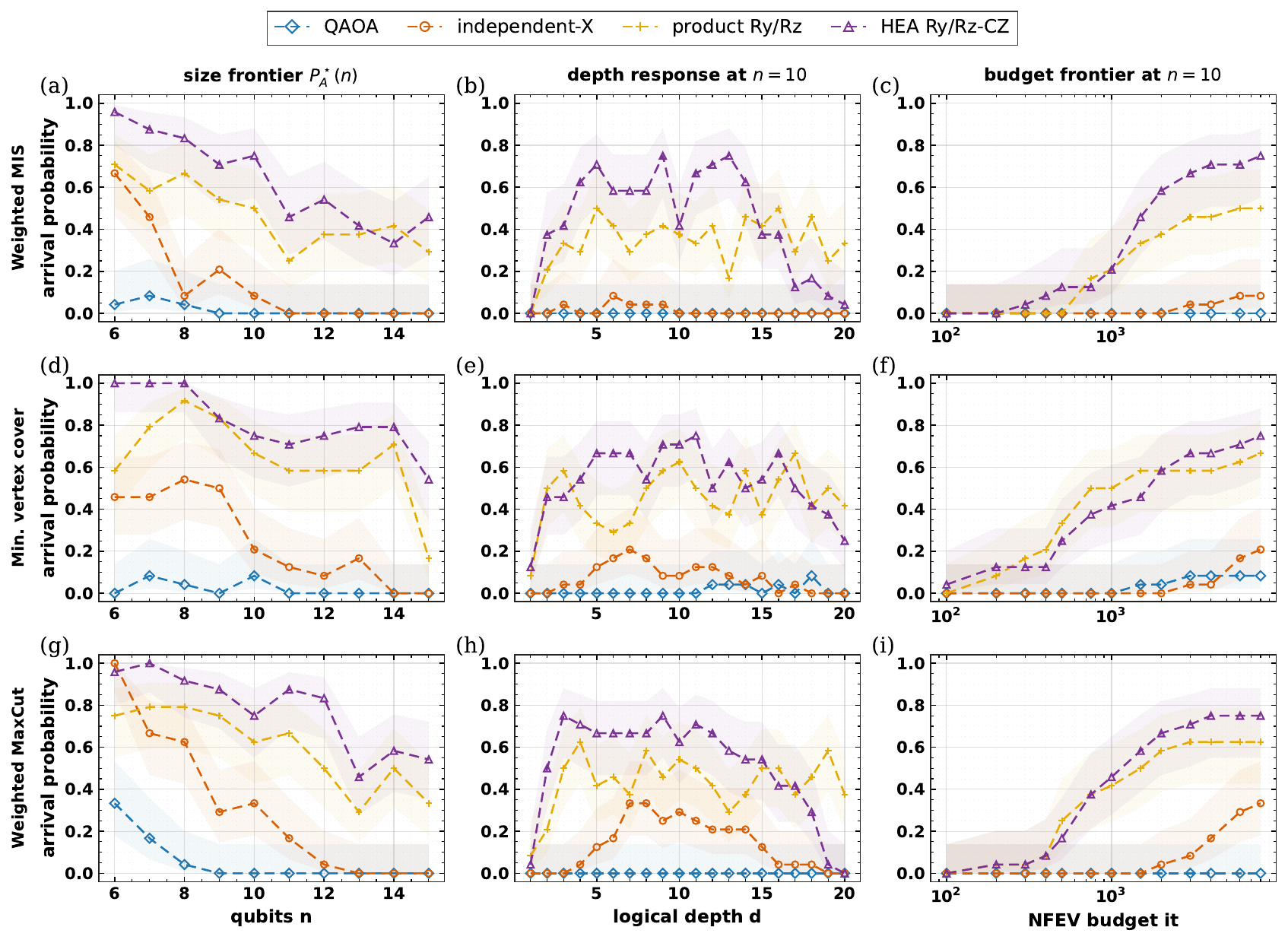}
\caption{\label{fig:shallow_phenomenology}Shallow-host reachability before commutator completion.  The measured quantity $P_A(n,d,\mathrm{it})$ is the probability that ansatz family $A$ reaches the strict low-energy window after a fixed training budget, and $P_A^\star(n)$ is the maximum of $P_A(n,d,\mathrm{it})$ over the tested depths and budgets for ansatz family $A$ at size $n$.  QAOA uses shared cost and mixer angles; independent-$X$ keeps the cost layer but gives each qubit its own $X$-mixer angle; product $R_Y/R_Z$ uses unentangled one-qubit rotations; and HEA $R_Y/R_Z$-CZ adds a shallow CZ entangling pattern. Rows are weighted MIS, minimum vertex cover, and weighted MaxCut; columns show the best size frontier over $n=6,\ldots,15$, the dense depth response at $n=10$, and the training-budget frontier at $n=10$.  NFEV denotes the number of objective-function evaluations.  Transparent bands are Wilson 95\% binomial intervals over the 24 paired trials contributing to each plotted success probability.  Product and HEA circuits are much stronger than QAOA, yet low-energy arrival still depends sharply on Hamiltonian family, size, depth, and budget.}
\end{figure}

Before any completion is added, we compare four ordinary shallow hosts in the ansatz-zoo sweep of Fig.~\ref{fig:shallow_phenomenology}.  QAOA uses a global cost angle and a global $X$-mixer angle in each layer, so all qubits share the same mixer strength.  The independent-$X$ variant, a mixer-side local-angle variant related to multi-angle QAOA \cite{herrman2022multi}, keeps the same diagonal cost layer but gives each qubit its own $X$-mixer angle; it tests whether local mixer freedom alone fixes the arrival gap.  The product $R_Y/R_Z$ host uses independent single-qubit rotations with no entanglers, testing how far separable states can go on these diagonal objectives.  The HEA $R_Y/R_Z$-CZ host adds a shallow CZ entangling pattern to the same local rotations, testing the additional value of short-range entanglement.  This sweep shows that the low-energy-arrival failure is not merely a weak-QAOA artifact: hardware-efficient and product hosts raise the arrival frontier substantially above QAOA, yet low-energy arrival still depends sharply on family, size, depth, and budget, with even the strongest shallow host falling to partial success by $n=15$.

This pre-completion sweep establishes the baseline landscape and motivates a Hamiltonian-aware repair rather than a stronger generic baseline alone.  The decline in the first-column frontiers is expected under fixed shallow-depth and fixed-budget conditions: as $n$ grows, the strict low-energy manifold becomes more selective, while the numbers of variational parameters and optimization trials do not scale comparably.  These curves therefore measure a finite-resource reachability boundary rather than an intrinsic loss of trainability.

\begin{figure}[!t]
\centering
\includegraphics[width=\textwidth]{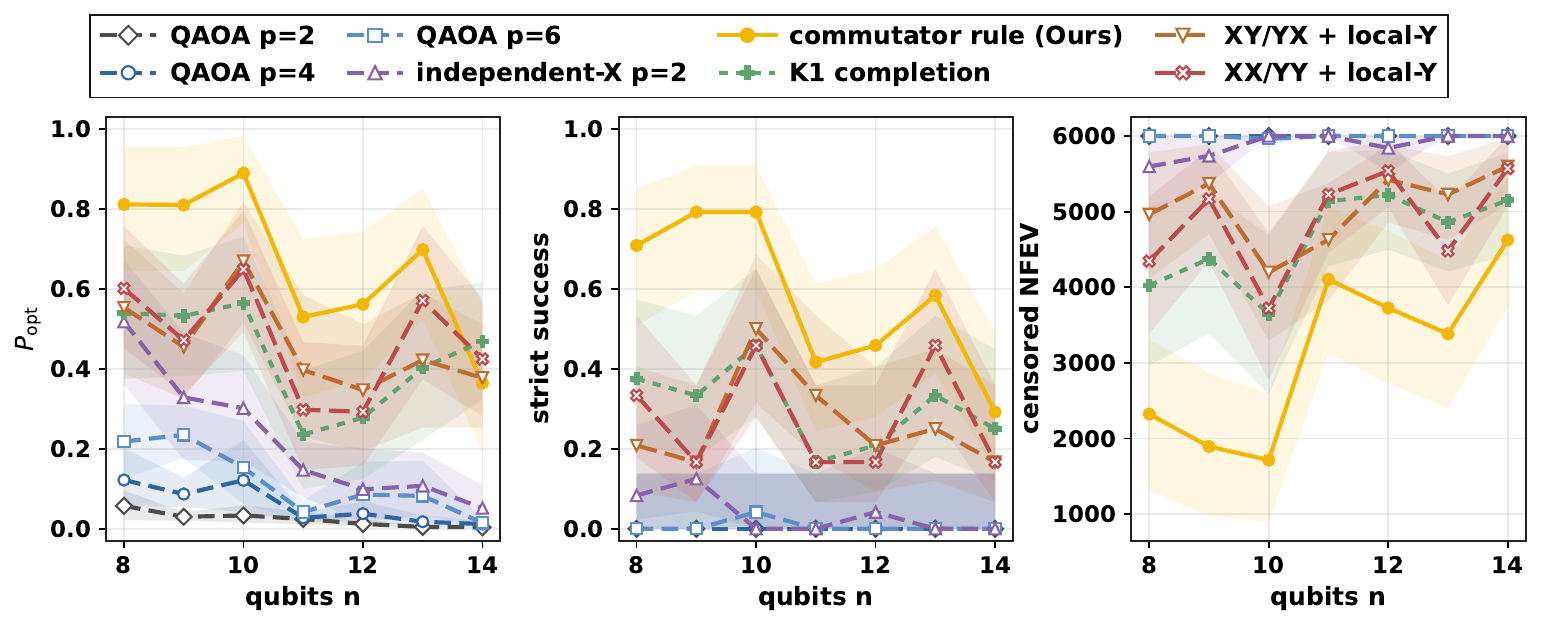}
\caption{\label{fig:mis_powered_scaling}Weighted-MIS repair scaling as an optimal-sampling diagnostic. Each point uses six graph instances and four shared seeds, exact diagonal spectra, full statevector optimization, and a shared $6000$-evaluation cap. The three panels report optimal-subspace probability $P_{\rm opt}$, strict-window success rate, and censored number of function evaluations (NFEV) to the strict target.  Shaded bands for $P_{\rm opt}$ and NFEV are bootstrap 95\% confidence intervals for the mean over the 24 paired trials; shaded bands for strict-window success are Wilson 95\% binomial intervals. The local-$Y$ completion derived from $[X,Z]$ (plotted as ``commutator rule (Ours)'') separates strongly from QAOA $p=2$--$6$ and independent-$X$ $p=2$.  The two edge-axis controls that also include local $Y$ are shared-axis controls, not pure wrong-axis negatives; they show that the local-field motif carries much of the gain, although shared-axis and full-first-shell variants can match or exceed its $P_{\rm opt}$ at the largest size. ``$\K_1$ completion'' denotes the full first-shell $\K_1$ arm of Eq.~(6).  The local-$Y$ completion remains the strict-success and target-cost leader, while the middle panel maps the finite-size boundary where a single repair block stops sufficing for strict-window success.}
\end{figure}

On a powered weighted-MIS scaling sweep---six graph instances and four shared seeds at each $n=8,\ldots,14$, a common $6000$-evaluation cap, and QAOA controls to $p=6$, where $p$ is the number of QAOA cost-mixer layers---the front-loaded local-$Y$ completion predicted by $[X_i,Z_i]$ remains the strict-success winner among the QAOA-style structured-family arms at every tested size.  It also maintains a large optimal-subspace sampling advantage in $P_{\rm opt}$ (Eq.~\ref{eq:popt}) over QAOA up to $p=6$ and independent-$X$ (Fig.~\ref{fig:mis_powered_scaling}). Two nearest-axis controls include the same predicted local-$Y$ sublayer but pair it with noncommutator edge axes; they therefore test whether local-$Y$ alone explains the gain rather than serving as purely wrong-axis controls. Those shared-axis controls narrow or overtake the local-$Y$ completion in $P_{\rm opt}$ at the largest size, and the full first-shell arm (``$\K_1$ completion'' in Fig.~\ref{fig:mis_powered_scaling}; Supplementary Tables S25 and S30) also exceeds it in $P_{\rm opt}$ at $n=14$; the local-$Y$ completion nevertheless remains best in strict success and target cost. The negative half of the prediction also holds: on pure-$ZZ$ weighted MaxCut the local-$Y$ axis gives no strict successes ($0/18$; Supplementary Table S3), and only the edge $\yzzy$ axis contributes weakly ($3/18$; Supplementary Table S29), confirming the axis-specificity of the local-field rule while marking pure-$ZZ$ objectives as a boundary.

The finite-size boundary is explicit and central to the claim. The optimal-sampling advantage over the standard QAOA-style baselines persists across the continuous $n=8,\ldots,14$ sweep, while strict success traces where a single repair block stops sufficing: it is high at the smallest sizes ($17/24$ and $19/24$ at $n=8,10$) and falls to $7/24$ by $n=14$. An earlier small-instance screen ($11/12$ strict successes at $n=8$) also falls to $10/18$ when the instance set is redrawn in a balanced way, and on that redraw the bootstrap confidence interval for the per-instance energy-margin difference between the commutator rule and the best control crosses zero. The robust application evidence is therefore an optimal-sampling diagnostic, in the style of Zhou \emph{et al.}~\cite{zhou2020qaoa}, not a large-$n$ solver claim: the commutator-predicted local axis substantially increases the probability of sampling an optimal bitstring at small and intermediate sizes, while strict-window reachability remains size-sensitive and shared-axis controls identify local $Y$ as the dominant motif. Supplementary Figs. S5--S10 and S12--S14 and Supplementary Tables S23--S25 and S27--S31 collect the full QUBO screens, dense depth slices, replayed budget curves, and companion spatial diagnostics; Supplementary Table S26 gives the expanded host/Hamiltonian aggregate behind Fig.~\ref{fig:closureselectorvisual}.

\begin{figure}[!t]
\centering
\includegraphics[width=\textwidth]{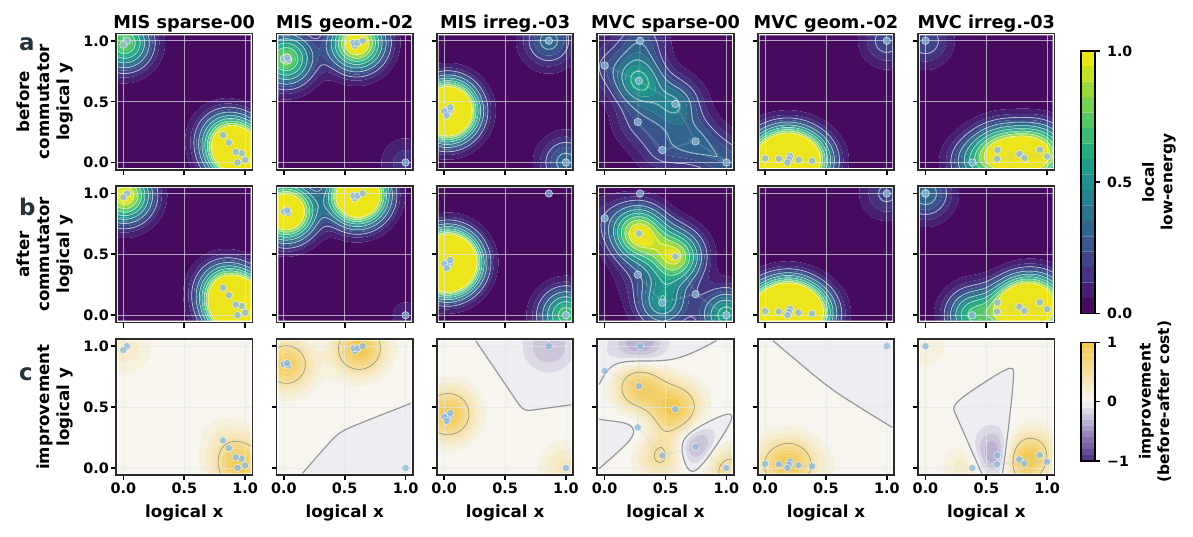}
\caption{\label{fig:qubo_spatial_improvement}Spatial diagnostic for the local-field commutator repair.  Six representative $n=8$ fielded-QUBO instances are shown on fixed graph layouts.  Points denote QUBO variables/qubits; the logical $x,y$ axes are layout coordinates, not circuit parameters, and the contours are interpolated from the eight discrete site-allocated values in each panel as a guide to the eye.  \textbf{a}, Base QAOA local-energy maps after optimization.  \textbf{b}, The same maps after adding and reoptimizing the local-$Y_i$ commutator layer predicted by $[X_i,Z_i]$.  In rows \textbf{a} and \textbf{b}, color is normalized within each instance and uses an inverted convention so that low-cost basins are visually bright: brighter color denotes lower allocated local Hamiltonian contribution, not higher energy.  \textbf{c}, Local improvement map, computed as the site-allocated contribution before repair minus that after repair; yellow marks regions where the commutator layer reduces local cost and purple marks regions that worsen.  This figure is an interpretability diagnostic only: it visualizes where the repair changes local contributions in selected instances and does not enter the success statistics.}
\end{figure}

The same application cells also give a spatial diagnostic of what the repair changes.  For each optimized state, we allocate the diagonal Hamiltonian expectation to sites by assigning local $Z_i$ contributions to site $i$ and splitting each $Z_iZ_j$ contribution equally across its endpoints.  Figure~\ref{fig:qubo_spatial_improvement} visualizes six representative fielded-QUBO instances before and after the local-$Y$ commutator layer, together with the corresponding local improvement map.  Each point is a QUBO variable/qubit, and the logical $x,y$ axes are fixed graph-layout coordinates rather than variational parameters; the contours are visual interpolations of these discrete site-allocated values across that layout.  Lower allocated contribution means that the optimized state places less expected Hamiltonian cost on that variable neighborhood, and the improvement row directly plots the local cost reduction induced by the repair. The plot is not used as an additional success count or scaling benchmark.  Its role is narrower: to show that the local-field repair changes identifiable variable neighborhoods, with both improved and worsened local contributions visible in individual instances.

\subsection{Classical accessibility bounds the reachability claim}
\label{sec:limitation}

The repaired circuits must also be tested against classical accessibility. We separate the expressivity-repair claim from any quantum-advantage claim, and distinguish what is proven from what remains open. For geometrically local hosts---the Ising strips on which the repair is most coherent---any bounded-order local commutator-closure completion scheduled into fixed depth has a bounded light cone.  The light cone is the finite neighborhood of qubits that can influence a local energy term after a shallow local circuit. Under these fixed-order, fixed-depth, bounded-degree locality assumptions, the loss and its parameter-shift gradients are classically computable in time linear in the number of qubits, with a prefactor that grows exponentially with the fixed light-cone size set by depth, range, and commutator order; the formal statement and proof are given as Supplementary Theorem 1. This is a scoping result for loss evaluation, not a full dequantization of all possible optimization trajectories, and it does not by itself decide the classical hardness of sampling from the repaired circuit.  An independent low-weight Pauli back-propagation confirms the same boundary numerically: this calculation propagates the measured Pauli operators backward through the circuit and truncates terms above a chosen Pauli weight, i.e., above a chosen number of non-identity single-qubit factors.  The Pauli weight needed for fixed accuracy saturates and the retained-term count grows only polynomially through $n=20$ (Supplementary Table S16). For dense or high-degree QUBOs the light cone can grow with system size, and for bounded-degree but non-geometric graphs the geometric ball-growth bound used in Supplementary Theorem 1 is no longer the right statement; whether such instances remain classically easy requires separate graph-specific or low-weight Pauli-propagation analysis and is left open.

To probe the finite-size reachability boundary at sizes beyond exact diagonalization, a certified width-2 weighted-MIS ladder serves as a scaling microscope. Because its width is fixed, a transfer dynamic program---a classical dynamic-programming algorithm that sweeps across the ladder while keeping only a constant-width boundary state---certifies the optimal independent-set weight and bitstrings in time linear in $n$.  This certificate lets us measure the exact optimal-subspace probability $P_{\rm opt}$ for shallow QAOA and the local-field completion at $n=20$--$50$ without heuristic ground-energy estimation. Across the completed grid---depths $d=1,2$ through $n=50$ and the more expensive $d=4$ pass through $n=30$, with the interrupted $d=4,n\ge40$ segments excluded---both QAOA and QAOA plus the local-field completion have zero strict energy successes, while median $P_{\rm opt}$ decays rapidly with size (Supplementary Fig. S15). In several completed large-$n$ points the completion is neutral or worse than the unrepaired QAOA host. These certified-ladder data are therefore presented as a limitation, not as a positive scaling result: fixed-depth local repair can be trainable and locally meaningful yet still fail to maintain a useful probability of sampling the exact optimum as the certified system grows.

The cleanest repaired Ising-strip targets are independently classically accessible. A tensor-network analysis using TeNPy~\cite{hauschild2018tenpy} converges to the ground-energy reference at $\chi^\star=4$ for TFIM and $\chi^\star=4$--$8$ for J1-J2 Ising, with no monotone growth through $n=50$ at an energy-error target of $10^{-2}$. The parameter $\chi^\star$ is the required matrix-product-state bond dimension, a standard measure of how much entanglement the tensor-network representation must keep. This is a ground-energy accessibility statement rather than a full dequantization of the optimization loop, and the $10^{-2}$ target is looser than the strict-window criterion used for the VQE runs. The corresponding certified-ladder and density-matrix renormalization group (DMRG) diagnostics are kept in the Supplementary Information together with the full tensor-network summary (Supplementary Table S2), because their role is to bound the interpretation of the repair rather than to provide an additional main-text success figure.

\section{Discussion}

The central result is a systematic repair for a concrete shallow-VQE failure mode: a shallow host ansatz can remain locally optimizable under the tested protocol yet fail to arrive at the low-energy states selected by a Hamiltonian under a fixed depth and evaluation budget, and useful missing directions can be inferred from the ansatz--Hamiltonian commutator algebra before opening a large adaptive pool. The method is a Hamiltonian-aware shallow-host repair rather than an optimizer improvement---it changes the local orbit available to the circuit, and full reoptimization then exploits the installed directions. Although commutators of ansatz generators and Hamiltonian terms are standard objects, the contribution is to use their low-order closure, together with a fixed same-support companion set, as a front-loaded and budgeted repair space ranked once and inserted before reoptimization rather than grown by sequential best-of-pool search. In this sense the method is not a universal arbitrary-depth ansatz builder; it is a diagnostic design rule for deciding what to add when depth is expensive and the shallow host is the object being repaired.

The evidence supports a scoped algorithmic claim. Across seven hosts and nine fixed $2\times3$ Hamiltonian settings, the optimized base reaches $0/189$ strict windows, a non-deployable per-trial control envelope reaches $109/189$, and the default closure-gradient selector reaches $133/189$; the strongest single matched noncommutator control reaches $77/189$ but is not the primary comparator. At the cell level the gradient selector beats this conservative envelope more often than it loses, while the trust selector provides a near-boundary robustness check rather than an independent decisive win. The fixed-axis studies expose the underlying motifs, with $[X,ZZ]$ giving $\yzzy$ for Ising couplings and $[X,Z]$ giving local $Y$ for fielded QUBOs. The useful object is not an arbitrary extra layer, but a low-order algebraic repair space tied to the chosen host and Hamiltonian. The one-shot full local-pool control sharpens this point: the larger pool contains the commutator axes, yet its locally largest gradients can crowd out the coordinated first-shell motif, consistent with algebraic pool restriction acting as a structural inductive bias rather than only as a measurement-saving device.

These boundaries are themselves the result. Pure MaxCut lacks the local-field motif that local-$Y$ completion exploits; $2\times4$ XXZ reaches a high-fidelity basin just outside the strict window; weighted-MIS success traces a finite-size boundary; and the QAOA2D Ising single-block misses are finite-depth thresholds, not span failures. The unifying lesson is diagnostic: static tests such as tangent-span coverage and commutator-shell coverage do not by themselves predict finite-depth low-energy arrival, because the measured property is reoptimized and protocol-limited; the ansatz--Hamiltonian commutator algebra, by contrast, names the missing directions constructively and cheaply.

The present study is also limited by its noiseless setting.  Exact statevector simulation makes the paired control tests and strict energy windows clean, but it does not model finite-shot estimation, hardware noise, compilation overhead, or the cost of estimating many candidate gradients on a device.  In the geometrically local regimes covered by the light-cone theorem, the same loss and gradient queries are classically computable, while in dense or non-geometric QUBO regimes the theorem does not apply and shot complexity remains a separate question.  Hardware-level noise and sampling costs are therefore outside the scope of the present evidence.

The broader point is that shallow quantum algorithms should not be judged by trainability alone. Recent work has connected provable trainability conditions to classical simulability \cite{cerezo2025does}; the present results add the operational low-energy-arrival axis to that discussion. This work separates three notions that are often conflated: whether a circuit can be optimized, whether its finite-depth orbit and optimization protocol can reach the Hamiltonian-selected low-energy window, and whether success in that repaired regime could evade classical analysis. Read together with the classical-accessibility bound, the picture is a scoped trainability--arrival--classical-accessibility tradeoff: a shallow circuit can be trainable, repaired by a low-order commutator-guided block, and still have classically computable losses or low-bond targets in the geometrically local regimes where the repair is cleanest. Dense or non-geometric application QUBOs are outside that proof, so their quantum-advantage status is not settled here. We expect this lens--repair the finite-depth low-energy-arrival gap, then ask whether the repaired regime is classically hard--to be a useful default for designing and stress-testing shallow quantum algorithms. A useful reporting standard is therefore to give three companion quantities for any shallow-repair claim: strict-window success against matched controls, the classical cost of evaluating the repaired loss when locality allows such a bound, and a tensor-network or related classical-accessibility check on the repaired targets.

\backmatter

\section*{Acknowledgements}
The authors thank colleagues at Visa for helpful discussions.




\clearpage
\bibliography{references}

\end{document}


\title[Supplementary information]{Supplementary Information: Commutator-guided repair and classical-accessibility boundaries for shallow quantum circuits}

\author[1]{\fnm{Yifeng} \sur{Peng}}
\author[1]{\fnm{Sunpreet Singh} \sur{Arora}}
\author[1]{\fnm{Harshal} \sur{Shah}}
\author[1]{\fnm{Srinivasan} \sur{Raghuraman}}
\author[1]{\fnm{Kim} \sur{Wagner}}
\author[1]{\fnm{Bin} \sur{Yan}}
\affil[1]{\orgname{Visa}, \orgaddress{\city{Foster City}, \state{CA}, \country{USA}}}

\maketitle

\setcounter{equation}{0}
\renewcommand{\theequation}{S\arabic{equation}}
\renewcommand{\theHequation}{S\arabic{equation}}
\setcounter{figure}{0}
\renewcommand{\thefigure}{S\arabic{figure}}
\renewcommand{\theHfigure}{S\arabic{figure}}
\setcounter{table}{0}
\renewcommand{\thetable}{S\arabic{table}}
\renewcommand{\theHtable}{S\arabic{table}}

\noindent\textbf{Terminology note.}\par\noindent
This Supplementary Information uses the same terminology as the main text.  An \emph{ansatz} is a parameterized circuit family; a \emph{host} is the ansatz before adding the commutator repair layer; a \emph{Pauli string} is a tensor product of $I,X,Y,Z$ operators; and a \emph{commutator} is $[A,B]=AB-BA$.  The first-shell closure $\K_1$ contains new Pauli axes obtained from commutators between host generators and Hamiltonian terms, while $\K_2$ repeats the operation once more.  A strict success means that the optimized energy lies inside the gap-relative low-energy window defined in the main text.  The reported reachability metrics are operational low-energy-arrival metrics: they depend on the ansatz family, optimizer, initialization, and evaluation budget, and are not standalone certificates of the exact variational optimum.  For QUBO experiments, $P_{\rm opt}$ is the probability of sampling any optimal bitstring, and for local classical-accessibility results a light cone is the finite set of qubits that can influence a local observable through a fixed-depth local circuit.

\section{Supplementary Derivations and Protocol Audits}
\label{app:methodcontrols}

This section provides additional derivations, complete implementation settings, and audit tables supporting the Methods and Results in the main text: the elementary Pauli derivation of the first-shell axes, the closure-selector protocol and reproducibility checklist, the prior-work positioning table, and the tensor-network capstone table.

\begin{proposition}[First-shell axes for fielded Ising/QUBO costs]
Consider a QAOA-style base containing local mixer generators $X_i$ and a diagonal cost Hamiltonian
$H_C=\sum_{(i,j)}J_{ij}Z_iZ_j+\sum_i h_iZ_i$.
The nonzero first stripped commutator shell generated by the mixer and the cost contains, up to signs and nonzero scalar factors,
\[
  \{Y_iZ_j,Z_iY_j:(i,j)\in E,\;J_{ij}\neq0\}
  \cup
  \{Y_i:h_i\neq0\}.
\]
Thus pure $ZZ$ costs expose edge $\yzzy$ axes, while fielded QUBOs additionally expose local-$Y$ axes.
\end{proposition}

\begin{proof}
Pauli strings on different qubits commute, so only terms sharing a site with the mixer generator contribute.  On one site, $XZ=-ZX$ and $[X,Z]=-2iY$ under the standard Pauli convention.  Hence $i[X_i,Z_i]$ is proportional to $Y_i$, and
$i[X_i,Z_iZ_j]=i[X_i,Z_i]Z_j$ is proportional to $Y_iZ_j$; similarly $i[X_j,Z_iZ_j]$ is proportional to $Z_iY_j$.  All remaining $X_k$ with $k\notin\{i,j\}$ commute with $Z_iZ_j$, and stripping removes identities, duplicates, and axes already present in the base or cost.  The sign is immaterial because a Pauli rotation angle can absorb it.
\end{proof}

\subsection*{Closure-selector protocol}

The expanded 189-trial matrix uses a front-loaded selector rather than sequential adaptive growth.  For each host--Hamiltonian cell and seed, the base circuit is first optimized.  The candidate pool is then built from implementable $\K_1$, implementable $\K_2$, and the fixed same-support companion set $\mathcal Q$, after stripping scalar factors, signs, duplicate Pauli strings, and axes already present in the host or Hamiltonian layer.  Here ``implementable'' means Pauli rotations of weight at most two on supports available to the host, Hamiltonian, or low-order closure.  In the reported benchmark, $\mathcal Q$ consists exhaustively of local $Y_i$ axes on local-field supports, edge $Y_iZ_j/Z_iY_j$ axes on $ZZ$ supports exposed by $X$ mixers, and edge $X_iZ_j/Z_iX_j$ axes on the corresponding brickwork/HEA companion supports; no other Pauli axes are added to $\mathcal Q$.

For the gradient selector, each candidate Pauli axis $A$ is appended to the optimized base with a new angle initialized at zero.  Its score is the magnitude of the parameter-shift derivative
\[
  s_{\rm grad}(A)
  =
  \left|
  \frac{E_A(+\pi/2)-E_A(-\pi/2)}{2}
  \right|,
\]
where $E_A(\phi)$ is the energy after adding only the candidate rotation $e^{-i\phi A/2}$ to the optimized base.  Candidates are sorted by this score; ties inherit the deterministic stripped-pool order, namely the canonical Pauli-label/support order after stripping duplicate signs and scalar factors.  The highest-ranked candidates are inserted until the gate cap is reached, with new angles initialized at zero, and then all base and repair parameters are reoptimized together.

For the trust-region selector, each candidate is again evaluated one axis at a time at the optimized base.  A centered finite difference with step $\delta=0.08$ estimates a one-dimensional gradient $g_A$ and curvature $h_A$.  The candidate's provisional angle is
\[
  \theta_A
  =
  {\rm clip}\!\left(
    -\frac{g_A}{|h_A|+10^{-6}},
    -0.5,0.5
  \right),
\]
with $\theta_A=0$ when $|h_A|<10^{-9}$.  The score is the predicted one-axis energy decrease,
\[
  s_{\rm trust}(A)
  =
  -g_A\theta_A-\frac{1}{2}h_A\theta_A^2 .
\]
The highest-scoring axes are inserted until the gate cap is reached.  Their provisional angles are projected to an $\ell_2$ trust radius of $0.8$ before final initialization, and then the full circuit is reoptimized.  In the main matrix the insertion budget is 56 Pauli rotations for both selectors and for the matched noncommutator controls.

\noindent\textbf{Reproducibility checklist for the 189-trial matrix.}\par\noindent
The aggregate benchmark in the main text uses the archived run named
\texttt{commutator\_closure\_selector\_expanded\_b56\_allhosts\_3seed}.
The following checklist collects the protocol details that are otherwise distributed across table captions and run metadata.

The detailed checklist is reported as Table~\ref{tab:selector_repro_checklist}.

The checklist is intended to make the finite-budget comparison reproducible from the text.  It does not remove the remaining algorithmic confound that a sequential ADAPT-style search restricted to exactly the same $\mathcal C_{\le2}$ pool was not run; the ADAPT controls below instead compare against a larger full-Pauli-pool sequential baseline under matched and generous evaluation caps.

\begin{table}[!t]
\caption{\label{tab:priorpositioning}Positioning relative to nearby commutator, adaptive, and algebraic approaches.  The comparison is meant to delimit the novelty claim rather than to rank methods.  We do not claim that commutators are new; the contribution is a front-loaded, budgeted, host--Hamiltonian completion rule for shallow-circuit low-energy arrival, evaluated together with matched controls and classical-accessibility boundaries.}
\footnotesize
\setlength{\tabcolsep}{4.5pt}
\renewcommand{\arraystretch}{1.16}
\adjustbox{max width=\textwidth}{%
\begin{tabular}{p{0.18\textwidth}p{0.34\textwidth}p{0.40\textwidth}}
\toprule
\rowcolor{hdrcol}
\textbf{Approach} & \textbf{Nearby role} & \textbf{Distinction from this work}\\
\midrule
ADAPT-VQE / qubit-ADAPT / QEB-ADAPT / ADAPT-QAOA \cite{grimsley2019adaptive,tang2021qubitadapt,yordanov2021qebadapt,zhu2022adaptive} & Sequential, gradient-ranked ansatz growth from a prescribed operator pool, with reoptimization after each insertion & Our repair is front-loaded: a compact block is selected once from the host--Hamiltonian low-order closure and then fully reoptimized under a fixed budget. ADAPT-QAOA can discover related $Y$-type mixer directions sequentially; here those directions are derived before optimization from the host--Hamiltonian closure and evaluated as a reachability repair.\\
\addlinespace[2pt]
Multi-operator or depth-aware ADAPT variants, including TETRIS-ADAPT-VQE \cite{sapova2022variational,anastasiou2024tetris} & Add several gradient-selected operators per stage to reduce selection rounds or circuit depth & These variants are closer to our block insertion than one-operator ADAPT, but they still start from a prescribed adaptive pool and iteratively rebuild the ansatz.  Our pool is restricted before ranking by the chosen host and Hamiltonian, and the selected repair is inserted as one front-loaded completion block.\\
\addlinespace[2pt]
Problem-aware ansatz designs / HVA / QAOA \cite{wecker2015progress,wiersema2020hva,hadfield2019from,lucas2014ising} & Encode Hamiltonian terms or mixer--cost structure directly into a shallow circuit family & Our repair is residual and host-dependent: after a problem-aware or hardware-efficient host is chosen, the commutator closure identifies missing companion directions such as $Y_i$ or $Y_iZ_j/Z_iY_j$ that need not appear as Hamiltonian terms.\\
\addlinespace[2pt]
Input-state reachability design \cite{wu2026enhancing} & Modify the initial state to enlarge or alter what a fixed ansatz can reach & Our repair keeps the input state and host circuit fixed and instead adds a bounded host--Hamiltonian completion block; the benchmark is operational low-energy arrival under explicit depth and evaluation budgets.\\
\addlinespace[2pt]
CD-QAOA / feedback variants \cite{wurtz2022counterdiabaticity,chandarana2022digitized,yao2021reinforcement,magann2022feedback} & Add counterdiabatic, nested-commutator-inspired, feedback, or learned terms to improve adiabatic or QAOA-style dynamics & Our target is not schedule following; it is finite-depth low-energy-arrival repair for a chosen shallow host, with wrong-axis, depth, and control-envelope comparisons.\\
\addlinespace[2pt]
AGP / counterdiabatic ansatz designs \cite{sels2017minimizing,kolodrubetz2017geometry} & Approximate gauge-potential generators along a path or parameter manifold to suppress diabatic transitions & We use low-order commutator closure as a design space for missing reachable directions, not as an approximation to a continuous gauge potential.\\
\addlinespace[2pt]
Lie-algebra diagnostics \cite{sim2019expressibility,larocca2022diagnosing,larocca2023overparam,fontana2024adjoint} & Analyze generator algebras, controllability, expressibility, or trainability properties & Our method is an implemented completion protocol with finite-budget benchmarks; algebraic closure is used to construct repair layers, not only to diagnose expressibility.\\
\addlinespace[2pt]
Classical-accessibility and simulability analyses \cite{fontana2023noisy,rudolph2023lowesa,angrisani2025classically,cerezo2025does} & Bound or estimate when restricted circuits, trainable regimes, or low-weight observable propagation remain classically tractable & We use these ideas as a boundary test: repaired reachability is reported together with light-cone, Pauli-path, and tensor-network checks rather than as a quantum-advantage claim.\\
\addlinespace[2pt]
This work & Predeclared low-order closure of the selected host generators and Hamiltonian terms, plus matched low-order companion axes & The claim is deliberately bounded: trainability is separated from reachability, and repaired fixed-depth local regimes are not presented as quantum-advantage evidence.\\
\bottomrule
\end{tabular}}
\end{table}

\begin{table}[!t]
\caption{\label{tab:tncapstone}Tensor-network capstone for the two-row Ising-strip cells.  The DMRG target is energy error $\le10^{-2}$; for $n\le12$ the reference is exact diagonalization, and for larger $n$ it is the largest tested DMRG bond dimension.  Here $\chi^\star$ is the smallest tested bond dimension reaching the target, and the MPS-$\K_1$ pilot reports full-$\chi$ validation against dense statevectors.}
\footnotesize
\adjustbox{max width=\columnwidth}{\footnotesize%
\begin{tabular}{lll}
\toprule
\rowcolor{hdrcol} Check & Scope & Result\\
\midrule
DMRG, TFIM strip & $n=6,8,12,20,50$; 3 seeds & $\chi^\star=4$ for every size and seed\\
DMRG, J1-J2 Ising strip & same sizes and seeds & $\chi^\star=4$--$8$ with no growth through $n=50$\\
Exact validation & $n\le12$ & $\chi=16$ agrees with exact energies to numerical precision\\
Canonical MPS-$\K_1$ VQE & $n=6$ pilot & full-$\chi$ energy matches statevector at $10^{-14}$\\
Small-$\chi$ $\K_1$ reproduction & $n=6$ J1-J2 & strict at $\chi=4,8$; TFIM near target but not strict\\
\bottomrule
\end{tabular}}
\end{table}

Here $\chi^\star$ is the smallest tested bond dimension reaching energy error $\le10^{-2}$.  For $n\le12$ the error is measured against exact diagonalization; for $n>12$ it is measured against the largest tested DMRG reference, $\chi=16$.  The table therefore reports bounded-bond accessibility of the target ground-energy references, not a full dequantization of the variational optimization loop.

\begin{appendices}
\renewcommand{\theequation}{S\arabic{equation}}
\renewcommand{\theHequation}{supp.equation.\arabic{equation}}
\renewcommand{\thefigure}{S\arabic{figure}}
\renewcommand{\theHfigure}{supp.figure.\arabic{figure}}
\setcounter{table}{2}
\renewcommand{\thetable}{S\arabic{table}}
\renewcommand{\theHtable}{supp.table.\arabic{table}}

\section{Mechanism Schematics}
\label{app:mechanismschematics}

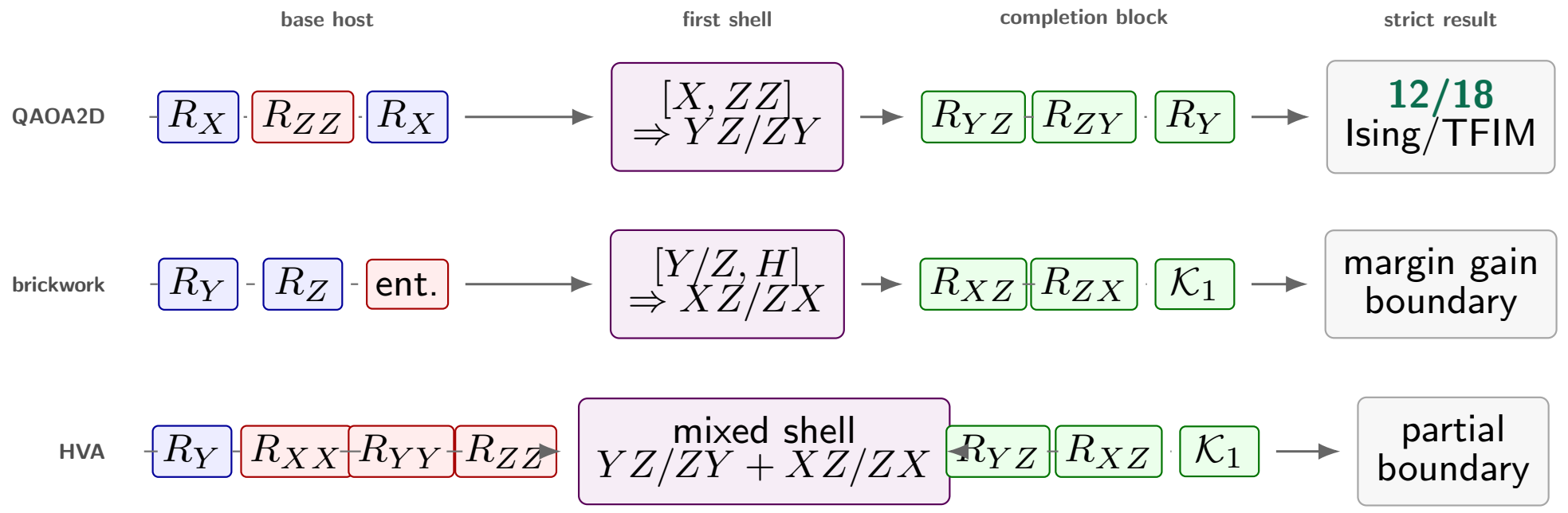
\begin{figure}[!t]
\centering
\resizebox{0.99\columnwidth}{!}{%
\begin{tikzpicture}[
  font=\sffamily\scriptsize,
  gate/.style={draw, rounded corners=1.3pt, minimum width=0.64cm, minimum height=0.40cm, align=center, inner sep=2.0pt, outer sep=2pt},
  basegate/.style={gate, draw=blue!60!black, fill=blue!7},
  hamgate/.style={gate, draw=red!65!black, fill=red!7},
  compgate/.style={gate, draw=green!45!black, fill=green!8},
  shellbox/.style={draw=violet!70!black, fill=violet!7, rounded corners=2pt, align=center, inner sep=4pt, minimum width=1.88cm, minimum height=0.58cm, outer sep=2pt},
  resultbox/.style={draw=black!35, fill=black!3, rounded corners=2pt, align=center, inner sep=4pt, minimum width=1.55cm, minimum height=0.58cm, outer sep=2pt},
  wire/.style={line width=0.42pt, draw=black!55},
  arrow/.style={-{Latex[length=1.8mm]}, line width=0.55pt, draw=black!60, shorten >=2pt, shorten <=2pt},
  title/.style={font=\sffamily\tiny\bfseries, text=black!65, align=center}
]
\node[title] at (1.30,0.80) {base host};
\node[title] at (4.55,0.80) {first shell};
\node[title] at (7.45,0.80) {completion block};
\node[title] at (10.35,0.80) {strict result};

\foreach \y/\name in {0/QAOA2D,-1.36/brickwork,-2.72/HVA} {
  \node[font=\sffamily\tiny\bfseries, anchor=east, text=black!70] at (-0.38,\y) {\name};
}

\node[basegate] (q1) at (0.25,0) {$R_X$};
\node[hamgate] (q2) at (1.10,0) {$R_{ZZ}$};
\node[basegate] (q3) at (1.95,0) {$R_X$};
\node[shellbox] (qs) at (4.55,0) {$[X,ZZ]$\\[-1pt]$\Rightarrow\yzzy$};
\node[compgate] (qc1) at (6.55,0) {$R_{YZ}$};
\node[compgate] (qc2) at (7.45,0) {$R_{ZY}$};
\node[compgate] (qc3) at (8.35,0) {$R_Y$};
\node[resultbox] (qr) at (10.35,0) {\best{12/18}\\Ising/TFIM};
\draw[wire] (-0.08,0) -- (q1.west);
\draw[wire] (q1.east) -- (q2.west);
\draw[wire] (q2.east) -- (q3.west);
\draw[arrow] (q3.east) -- (qs.west);
\draw[arrow] (qs.east) -- (qc1.west);
\draw[wire] (qc1.east) -- (qc2.west);
\draw[wire] (qc2.east) -- (qc3.west);
\draw[arrow] (qc3.east) -- (qr.west);

\node[basegate] (b1) at (0.25,-1.36) {$R_Y$};
\node[basegate] (b2) at (1.10,-1.36) {$R_Z$};
\node[hamgate] (b3) at (1.95,-1.36) {ent.};
\node[shellbox] (bs) at (4.55,-1.36) {$[Y/Z,H]$\\[-1pt]$\Rightarrow XZ/ZX$};
\node[compgate] (bc1) at (6.55,-1.36) {$R_{XZ}$};
\node[compgate] (bc2) at (7.45,-1.36) {$R_{ZX}$};
\node[compgate] (bc3) at (8.35,-1.36) {$\K_1$};
\node[resultbox] (br) at (10.35,-1.36) {margin gain\\boundary};
\draw[wire] (-0.08,-1.36) -- (b1.west);
\draw[wire] (b1.east) -- (b2.west);
\draw[wire] (b2.east) -- (b3.west);
\draw[arrow] (b3.east) -- (bs.west);
\draw[arrow] (bs.east) -- (bc1.west);
\draw[wire] (bc1.east) -- (bc2.west);
\draw[wire] (bc2.east) -- (bc3.west);
\draw[arrow] (bc3.east) -- (br.west);

\node[basegate] (h1) at (0.20,-2.72) {$R_Y$};
\node[hamgate] (h2) at (1.05,-2.72) {$R_{XX}$};
\node[hamgate] (h3) at (1.90,-2.72) {$R_{YY}$};
\node[hamgate] (h4) at (2.75,-2.72) {$R_{ZZ}$};
\node[shellbox] (hs) at (4.85,-2.72) {mixed shell\\[-1pt]$\yzzy+XZ/ZX$};
\node[compgate] (hc1) at (6.75,-2.72) {$R_{YZ}$};
\node[compgate] (hc2) at (7.65,-2.72) {$R_{XZ}$};
\node[compgate] (hc3) at (8.55,-2.72) {$\K_1$};
\node[resultbox] (hr) at (10.45,-2.72) {partial\\boundary};
\draw[wire] (-0.08,-2.72) -- (h1.west);
\draw[wire] (h1.east) -- (h2.west);
\draw[wire] (h2.east) -- (h3.west);
\draw[wire] (h3.east) -- (h4.west);
\draw[arrow] (h4.east) -- (hs.west);
\draw[arrow] (hs.east) -- (hc1.west);
\draw[wire] (hc1.east) -- (hc2.west);
\draw[wire] (hc2.east) -- (hc3.west);
\draw[arrow] (hc3.east) -- (hr.west);

\end{tikzpicture}}
\caption{\label{fig:ansatzhosts}Host-dependent circuit repair.  The algorithm first fixes a shallow host ansatz, computes stripped commutator shells with the target Hamiltonian, and appends a Pauli-rotation block before full reoptimization.  The expanded host evidence shows that QAOA2D exposes the clean $\yzzy$ edge axis, brickwork exposes a different first shell but is rescued by the low-order closure selector, and HVA remains the hardest mixed-shell host.}
\end{figure}

\begin{figure}[!t]
\centering
\resizebox{0.98\columnwidth}{!}{%
\begin{tikzpicture}[
  font=\sffamily\tiny,
  box/.style={draw, rounded corners=2pt, align=center, inner sep=3pt, minimum height=0.45cm, outer sep=2pt},
  motif/.style={box, draw=red!65!black, fill=red!6, text width=1.05cm},
  comm/.style={box, draw=violet!70!black, fill=violet!7, text width=1.35cm},
  axis/.style={box, draw=green!45!black, fill=green!8, text width=1.25cm},
  ctrl/.style={box, draw=black!35, fill=black!3, text width=1.32cm},
  good/.style={box, draw=green!50!black, fill=green!9, text width=1.40cm},
  bad/.style={box, draw=black!35, fill=black!4, text width=1.40cm},
  arrow/.style={-{Latex[length=1.6mm]}, line width=0.50pt, draw=black!60, shorten >=2pt, shorten <=2pt},
  title/.style={font=\sffamily\tiny\bfseries, text=black!65, align=center}
]
\path[use as bounding box] (-0.85,-2.22) rectangle (10.95,1.02);
\node[title] at (2.05,0.80) {(a) Motif-to-axis dictionary};
\node[motif] (zz) at (0,0.25) {$ZZ$\\edge};
\node[comm] (czz) at (1.95,0.25) {$[X,ZZ]$};
\node[axis] (azz) at (3.90,0.25) {$YZ/ZY$\\edge};
\node[motif] (z) at (0,-0.55) {$Z$\\field};
\node[comm] (cz) at (1.95,-0.55) {$[X,Z]$};
\node[axis] (az) at (3.90,-0.55) {local\\$Y$};
\node[motif] (mix) at (0,-1.35) {XXZ/J1-J2\\motifs};
\node[comm] (cmix) at (1.95,-1.35) {$[\G_0,\Hterms]$};
\node[axis] (amix) at (3.90,-1.35) {mixed\\$\K_1$};
\draw[arrow] (zz.east) -- (czz.west);
\draw[arrow] (czz.east) -- (azz.west);
\draw[arrow] (z.east) -- (cz.west);
\draw[arrow] (cz.east) -- (az.west);
\draw[arrow] (mix.east) -- (cmix.west);
\draw[arrow] (cmix.east) -- (amix.west);

\node[title] at (7.85,0.80) {(b) Matched-control interpretation};
\node[bad] (base) at (5.70,0.15) {base\\strict fail};
\node[bad] (depth) at (7.60,0.15) {same-depth\\control};
\node[bad] (wrong) at (9.50,0.15) {wrong-axis\\control};
\node[good, text width=1.55cm] (right) at (9.50,-1.10) {commutator\\axis succeeds};
\draw[arrow] (base.east) -- (depth.west);
\draw[arrow] (depth.east) -- (wrong.west);
\draw[arrow] (wrong.south) -- (right.north);
\end{tikzpicture}}
\caption{\label{fig:dictionary}Motif dictionary and control logic.  The first-shell rule maps Hamiltonian motifs to concrete Pauli-rotation axes before optimization.  A repaired cell is interpreted as evidence for Hamiltonian-aware expressivity repair only when matched-depth, wrong-axis, and random/same-budget controls do not reproduce the strict-window success.}
\end{figure}
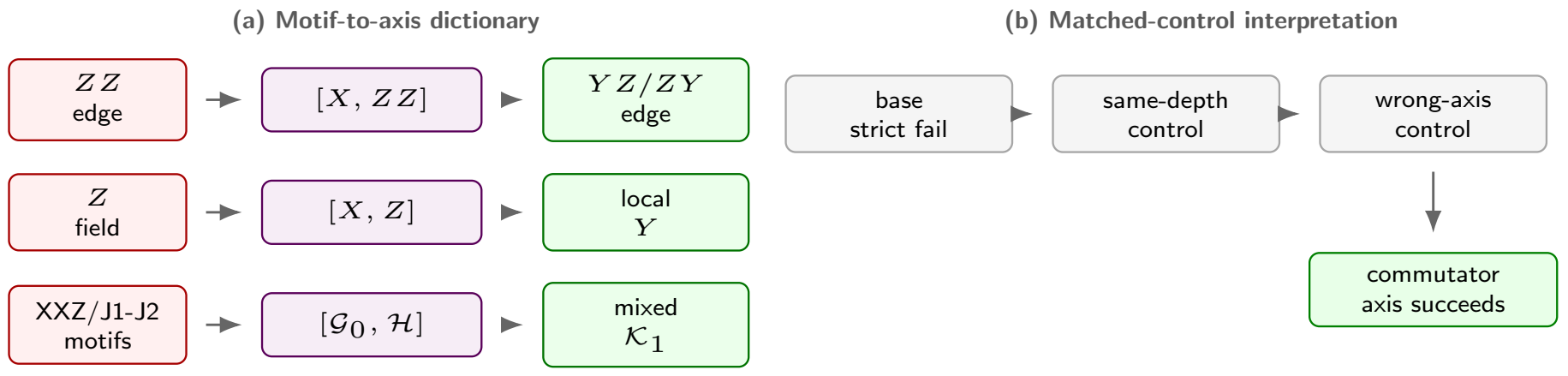

\section{Target Windows and Reported Metrics}

All strict-success counts use the same rule: energy error must be at most $\max(10^{-6},0.1\Delta_{\rm gap})$, where $\Delta_{\rm gap}$ is the spectral gap above the ground-energy manifold.  We report strict margin as target window minus energy error; positive values are strict successes.  Censored cost-to-target equals the function-evaluation cap when a run does not hit the strict target.

\section{Representative Evidence Map}
\label{app:evidencemap}

Table~\ref{tab:mainresults} collects the main positive and boundary regimes discussed in the text.  It is intended as a cross-reference map rather than an additional independent result.

\begin{table}[!t]
\caption{\label{tab:mainresults}Representative evidence map.  Success denotes strict-window success unless otherwise stated.  The table summarizes the main motif-matched positives and the boundary regimes that delimit low-order completion.  Green bold entries mark the best or highlighted reported commutator result within the row.}
\small
\setlength{\tabcolsep}{5pt}
\renewcommand{\arraystretch}{1.16}
\begin{tabular}{@{}lll@{}}
\toprule
\rowcolor{hdrcol}
\begin{minipage}[t]{0.19\textwidth}\textbf{Benchmark}\end{minipage} &
\begin{minipage}[t]{0.23\textwidth}\textbf{Completion rule}\end{minipage} &
\begin{minipage}[t]{0.52\textwidth}\textbf{Representative outcome}\end{minipage}\\
\midrule
\begin{minipage}[t]{0.19\textwidth}\raggedright weighted MIS, $n=8$--$14$ powered scaling\end{minipage}
& \begin{minipage}[t]{0.23\textwidth}\raggedright local $Y$ from $[X,Z]$\end{minipage}
& \begin{minipage}[t]{0.52\textwidth}\raggedright Local-$Y$ rule reaches \best{17/24, 19/24, 11/24, 7/24} strict at $n=8,10,12,14$, while QAOA $p=6$ reaches $0/24,1/24,0/24,0/24$; $P_{\rm opt}$ stays separated from deeper-QAOA and independent-$X$ controls, but strict success decays with size and the powered table is used as an optimal-sampling diagnostic rather than a large-$n$ solver claim (main Fig.~4, Table~\ref{tab:mis_powered_scaling}).\end{minipage}\\
\addlinespace[1pt]
\begin{minipage}[t]{0.19\textwidth}\raggedright QAOA2D adaptive policy\end{minipage}
& \begin{minipage}[t]{0.23\textwidth}\raggedright stop/repeat $\yzzy+Y$ from $[X,ZZ]$\end{minipage}
& \begin{minipage}[t]{0.52\textwidth}\raggedright \best{60/60} strict with 24.7 gates on average, compared with 40 gates for always-repeat.\end{minipage}\\
\addlinespace[1pt]
\begin{minipage}[t]{0.19\textwidth}\raggedright Expanded host matrix\end{minipage}
& \begin{minipage}[t]{0.23\textwidth}\raggedright seven shallow hosts and nine $2\times3$ Hamiltonian settings\end{minipage}
& \begin{minipage}[t]{0.52\textwidth}\raggedright Optimized base reaches $0/189$ strict windows; the default gradient-ranked closure selector reaches \best{133/189}, the trust selector reaches $127/189$, and a non-deployable per-trial control envelope reaches $109/189$ (Table~\ref{tab:closureselectormain}); the strongest single matched control reaches $77/189$ (Table~\ref{tab:control_arm_audit}).\end{minipage}\\
\addlinespace[1pt]
\begin{minipage}[t]{0.19\textwidth}\raggedright 2D Heisenberg J1-J2\end{minipage}
& \begin{minipage}[t]{0.23\textwidth}\raggedright frozen full $\K_1+Y$\end{minipage}
& \begin{minipage}[t]{0.52\textwidth}\raggedright \best{8/9} strict versus $3/9$ for budget-matched random-greedy.\end{minipage}\\
\addlinespace[1pt]
\begin{minipage}[t]{0.19\textwidth}\raggedright Held-out $2\times3$ XXZ\end{minipage}
& \begin{minipage}[t]{0.23\textwidth}\raggedright frozen full $\K_1+Y$\end{minipage}
& \begin{minipage}[t]{0.52\textwidth}\raggedright \best{8/9} strict versus $2/9$ for random-greedy and $0/9$ for noncommutator controls.\end{minipage}\\
\addlinespace[1pt]
\begin{minipage}[t]{0.19\textwidth}\raggedright \emph{Boundary regimes}\end{minipage}
& \begin{minipage}[t]{0.23\textwidth}\raggedright first-shell completion\end{minipage}
& \begin{minipage}[t]{0.52\textwidth}\raggedright MaxCut edge completion reaches only $3/18$ strict and local-$Y$ reaches $0/18$; $2\times4$ XXZ reaches fidelity 0.917 but $0/9$ strict.\end{minipage}\\
\addlinespace[1pt]
\begin{minipage}[t]{0.19\textwidth}\raggedright Tensor-network benchmark\end{minipage}
& \begin{minipage}[t]{0.23\textwidth}\raggedright TeNPy DMRG / canonical MPS\end{minipage}
& \begin{minipage}[t]{0.52\textwidth}\raggedright TFIM strip has $\chi^\star=4$ at all tested sizes; J1-J2 has $\chi^\star=4$--$8$ with no growth through $n=50$ at energy error $\le10^{-2}$; $\K_1$ MPS validation reaches $10^{-14}$.\end{minipage}\\
\bottomrule
\end{tabular}
\end{table}

\section{Expanded Closure-Selector Cell Matrix}
\label{app:closureselectorfull}

\begingroup
\scriptsize
\setlength{\tabcolsep}{3pt}
\renewcommand{\arraystretch}{1.05}
\begin{longtable}{@{}lrrrrrr@{}}
\caption{\label{tab:closureselectorfullmain}Full cell-level expanded host/Hamiltonian matrix.  Each Hamiltonian-setting block contains the seven host ansatz families and three shared seeds, so each entry is a strict-window success count out of three paired trials.  $\K_1+Y$ is the fixed first-shell arm with local-$Y$ companions (the 62/189 arm of Table~\ref{tab:closureselector}); $\K_1+\K_2$ adds implementable second-shell directions.  ``Best ctrl.'' credits, for each seed in the cell, the strongest matched noncommutator control arm; it is the non-deployable per-trial control envelope of Table~\ref{tab:control_arm_audit} restricted to that cell, so its entries can mix arms across seeds.  When this control strictly beats both reported selectors, the red cell reports the winning arm in parentheses: depth = same-family parameter-matched depth, $\xyyx$ = wrong-axis $\xyyx{+}Y$, rand = random edge-axis layer, and ``2$\xyyx$+rand'' denotes a seed-level mixture.  Green marks the reported selector(s) only when they are row-best.  Diagnostic headroom columns are kept out of the main text.}\\
\toprule
\rowcolor{hdrcol}
\textbf{Host ansatz} & \textbf{Base} & $\boldsymbol{\K_1{+}Y}$ & $\boldsymbol{\K_1{+}\K_2}$ & \textbf{Best ctrl.} & \textbf{Ours: grad} & \textbf{Trust}\\
\midrule
\endfirsthead
\caption[]{Full cell-level expanded host/Hamiltonian matrix, continued.}\\
\toprule
\rowcolor{hdrcol}
\textbf{Host ansatz} & \textbf{Base} & $\boldsymbol{\K_1{+}Y}$ & $\boldsymbol{\K_1{+}\K_2}$ & \textbf{Best ctrl.} & \textbf{Ours: grad} & \textbf{Trust}\\
\midrule
\endhead
\midrule
\multicolumn{7}{r}{\emph{continued on next page}}\\
\endfoot
\bottomrule
\endlastfoot
\addlinespace[2pt]
\multicolumn{7}{@{}l}{\textbf{TFIM, $h=0.6$ ($[X,ZZ]\rightarrow YZ/ZY$)}}\\
Product $R_Y$ & 0/3 & 0/3 & 0/3 & 0/3 & 0/3 & 0/3\\
QAOA2D & 0/3 & 0/3 & 0/3 & \ctrlwin{3/3}{depth} & 0/3 & 0/3\\
Brickwork HEA & 0/3 & 0/3 & 0/3 & 0/3 & 0/3 & 0/3\\
HEA $X/Y$+CZ & 0/3 & 0/3 & 0/3 & 0/3 & 0/3 & 0/3\\
HEA $X/Z$+CZ & 0/3 & 0/3 & 0/3 & 0/3 & 0/3 & 0/3\\
HEA $X/Y/Z$+CZ & 0/3 & 0/3 & 0/3 & 0/3 & 0/3 & 0/3\\
HVA-Heisenberg & 0/3 & 0/3 & 0/3 & 0/3 & 0/3 & 0/3\\
\addlinespace[2pt]
\multicolumn{7}{@{}l}{\textbf{TFIM, $h=1.2$ ($[X,ZZ]\rightarrow YZ/ZY$)}}\\
Product $R_Y$ & 0/3 & 0/3 & \ctrlbetter{3/3} & \ctrlwin{3/3}{$\xyyx$} & 0/3 & 0/3\\
QAOA2D & 0/3 & 0/3 & 0/3 & \ctrlwin{3/3}{depth} & 0/3 & 1/3\\
Brickwork HEA & 0/3 & 0/3 & 3/3 & 3/3 & \ours{3/3} & 1/3\\
HEA $X/Y$+CZ & 0/3 & 0/3 & 0/3 & 3/3 & \ours{3/3} & \ours{3/3}\\
HEA $X/Z$+CZ & 0/3 & 0/3 & 0/3 & 3/3 & \ours{3/3} & 0/3\\
HEA $X/Y/Z$+CZ & 0/3 & 0/3 & 0/3 & \ctrlwin{3/3}{2$\xyyx$+rand} & 2/3 & 0/3\\
HVA-Heisenberg & 0/3 & 3/3 & 0/3 & 3/3 & \ours{3/3} & \ours{3/3}\\
\addlinespace[2pt]
\multicolumn{7}{@{}l}{\textbf{J1-J2 Ising, $J_2=0.25$ (edge/diagonal $ZZ$)}}\\
Product $R_Y$ & 0/3 & 0/3 & 0/3 & \ctrlwin{3/3}{$\xyyx$} & 2/3 & 2/3\\
QAOA2D & 0/3 & 0/3 & 3/3 & 3/3 & \ours{3/3} & \ours{3/3}\\
Brickwork HEA & 0/3 & 0/3 & 0/3 & 1/3 & 2/3 & \ours{3/3}\\
HEA $X/Y$+CZ & 0/3 & 0/3 & 0/3 & 2/3 & \ours{3/3} & \ours{3/3}\\
HEA $X/Z$+CZ & 0/3 & 3/3 & 3/3 & 1/3 & \ours{3/3} & \ours{3/3}\\
HEA $X/Y/Z$+CZ & 0/3 & 0/3 & 0/3 & 0/3 & 2/3 & \ours{3/3}\\
HVA-Heisenberg & 0/3 & 2/3 & 0/3 & 1/3 & \ours{2/3} & \ours{2/3}\\
\addlinespace[2pt]
\multicolumn{7}{@{}l}{\textbf{J1-J2 Ising, $J_2=0.45$ (edge/diagonal $ZZ$)}}\\
Product $R_Y$ & 0/3 & 1/3 & 0/3 & 3/3 & \ours{3/3} & \ours{3/3}\\
QAOA2D & 0/3 & 0/3 & 3/3 & 3/3 & \ours{3/3} & \ours{3/3}\\
Brickwork HEA & 0/3 & 0/3 & 3/3 & 3/3 & \ours{3/3} & \ours{3/3}\\
HEA $X/Y$+CZ & 0/3 & 3/3 & 0/3 & 3/3 & \ours{3/3} & \ours{3/3}\\
HEA $X/Z$+CZ & 0/3 & 3/3 & 3/3 & 3/3 & \ours{3/3} & \ours{3/3}\\
HEA $X/Y/Z$+CZ & 0/3 & 2/3 & 0/3 & 3/3 & \ours{3/3} & \ours{3/3}\\
HVA-Heisenberg & 0/3 & 3/3 & 0/3 & 3/3 & 2/3 & \ours{3/3}\\
\addlinespace[2pt]
\multicolumn{7}{@{}l}{\textbf{J1-J2 Ising, $J_2=0.65$ (edge/diagonal $ZZ$)}}\\
Product $R_Y$ & 0/3 & 1/3 & 0/3 & 3/3 & \ours{3/3} & \ours{3/3}\\
QAOA2D & 0/3 & 0/3 & 3/3 & 3/3 & \ours{3/3} & \ours{3/3}\\
Brickwork HEA & 0/3 & 0/3 & 2/3 & 3/3 & \ours{3/3} & \ours{3/3}\\
HEA $X/Y$+CZ & 0/3 & 3/3 & 1/3 & 3/3 & \ours{3/3} & \ours{3/3}\\
HEA $X/Z$+CZ & 0/3 & 3/3 & 3/3 & 3/3 & \ours{3/3} & \ours{3/3}\\
HEA $X/Y/Z$+CZ & 0/3 & 3/3 & 3/3 & 3/3 & \ours{3/3} & \ours{3/3}\\
HVA-Heisenberg & 0/3 & 3/3 & 3/3 & 3/3 & \ours{3/3} & \ours{3/3}\\
\addlinespace[2pt]
\multicolumn{7}{@{}l}{\textbf{Heisenberg J1-J2, $J_2=0.20$ (mixed shell)}}\\
Product $R_Y$ & 0/3 & 0/3 & 0/3 & 1/3 & \ours{3/3} & 1/3\\
QAOA2D & 0/3 & 0/3 & 0/3 & 0/3 & 0/3 & 0/3\\
Brickwork HEA & 0/3 & 0/3 & 3/3 & 2/3 & \ours{3/3} & \ours{3/3}\\
HEA $X/Y$+CZ & 0/3 & 2/3 & 3/3 & 2/3 & \ours{3/3} & \ours{3/3}\\
HEA $X/Z$+CZ & 0/3 & 3/3 & 3/3 & 1/3 & \ours{3/3} & 2/3\\
HEA $X/Y/Z$+CZ & 0/3 & 3/3 & 3/3 & 1/3 & \ours{3/3} & \ours{3/3}\\
HVA-Heisenberg & 0/3 & 0/3 & 0/3 & 1/3 & \ours{1/3} & \ours{1/3}\\
\addlinespace[2pt]
\multicolumn{7}{@{}l}{\textbf{Heisenberg J1-J2, $J_2=0.35$ (mixed shell)}}\\
Product $R_Y$ & 0/3 & 0/3 & 0/3 & 0/3 & 2/3 & \ours{3/3}\\
QAOA2D & 0/3 & 0/3 & 0/3 & 2/3 & \ours{3/3} & 0/3\\
Brickwork HEA & 0/3 & 0/3 & 3/3 & 1/3 & \ours{3/3} & 1/3\\
HEA $X/Y$+CZ & 0/3 & 2/3 & 3/3 & 2/3 & \ours{3/3} & \ours{3/3}\\
HEA $X/Z$+CZ & 0/3 & 3/3 & 3/3 & 2/3 & 2/3 & \ours{3/3}\\
HEA $X/Y/Z$+CZ & 0/3 & 3/3 & 3/3 & 3/3 & \ours{3/3} & \ours{3/3}\\
HVA-Heisenberg & 0/3 & 0/3 & 0/3 & 0/3 & 0/3 & \ours{1/3}\\
\addlinespace[2pt]
\multicolumn{7}{@{}l}{\textbf{XXZ J1-J2, $J_2=0.30,\Delta=0.50$ (anisotropic shell)}}\\
Product $R_Y$ & 0/3 & 0/3 & 0/3 & 0/3 & 0/3 & \ours{1/3}\\
QAOA2D & 0/3 & 0/3 & 0/3 & \ctrlwin{1/3}{rand} & 0/3 & 0/3\\
Brickwork HEA & 0/3 & 0/3 & 3/3 & 3/3 & \ours{3/3} & \ours{3/3}\\
HEA $X/Y$+CZ & 0/3 & 3/3 & 3/3 & 2/3 & \ours{3/3} & \ours{3/3}\\
HEA $X/Z$+CZ & 0/3 & 3/3 & 3/3 & 3/3 & \ours{3/3} & \ours{3/3}\\
HEA $X/Y/Z$+CZ & 0/3 & 2/3 & 2/3 & 3/3 & \ours{3/3} & \ours{3/3}\\
HVA-Heisenberg & 0/3 & 0/3 & 0/3 & 0/3 & \ours{2/3} & \ours{2/3}\\
\addlinespace[2pt]
\multicolumn{7}{@{}l}{\textbf{XXZ J1-J2, $J_2=0.50,\Delta=1.50$ (anisotropic shell)}}\\
Product $R_Y$ & 0/3 & 0/3 & 0/3 & 0/3 & 1/3 & \ours{2/3}\\
QAOA2D & 0/3 & 0/3 & 0/3 & 0/3 & \ours{3/3} & 2/3\\
Brickwork HEA & 0/3 & 0/3 & 3/3 & 0/3 & 2/3 & \ours{3/3}\\
HEA $X/Y$+CZ & 0/3 & 1/3 & 3/3 & 1/3 & \ours{3/3} & \ours{3/3}\\
HEA $X/Z$+CZ & 0/3 & 3/3 & 3/3 & 0/3 & \ours{3/3} & \ours{3/3}\\
HEA $X/Y/Z$+CZ & 0/3 & 1/3 & 1/3 & 0/3 & \ours{3/3} & \ours{3/3}\\
HVA-Heisenberg & 0/3 & 0/3 & 0/3 & \ctrlwin{1/3}{rand} & 0/3 & 0/3\\
\end{longtable}
\endgroup

\section{Early 2D Transfer and Budget Gates}

The first lattice-transfer experiments were broad.  At a small budget $B=8$ on six $2\times3$ cells and three seeds, every method had zero strict successes, but geometry-aware completion still beat same-family depth.  The later budget-to-first-success sweep, summarized in Table~\ref{tab:budget2d}, showed why the paper moved away from one-step or QITE-style selectors \cite{motta2020qite} and toward family/axis structure: AGP/CD first produced strict successes on QAOA-like cells, while fixed lattice-edge completion became the strongest generalist by $B=20$.

\begin{table}[!t]
\caption{\label{tab:budget2d}2D budget-to-first-success sweep on six $2\times3$ cells and three seeds.  The table reports strict successes out of 18 trials.}
\adjustbox{max width=\textwidth}{\footnotesize%
\begin{tabular}{lrrrrrrr}
\toprule
\rowcolor{hdrcol} Budget & best method & best & AGP/CD & fixed 2D & one-step & trust-QITE & gradient\\
\midrule
$B=12$ & AGP/CD & 2 & 2 & 0 & 0 & 0 & 0\\
$B=16$ & AGP/CD & 3 & 3 & 0 & 0 & 0 & 0\\
$B=20$ & fixed 2D & 6 & 3 & 6 & 1 & 1 & 0\\
\bottomrule
\end{tabular}}
\end{table}

On the QAOA2D subset, AGP/CD initially had a co-scaled budget advantage over the old fixed $\xyyx+Y$ template: at $B=10,12,14,16,18,20,24$, AGP/CD reached $7,8,10,10,10,10,10$ strict successes out of 20, while fixed $\xyyx+Y$ reached zero.  This result motivated the counterdiabatic/AGP branch, but it did not survive axis-matched controls below.

\section{Optimizer and Base-Reachability Controls}

The optimizer/init kill gate used 18 heterogeneous cells.  Base default failed on 16 cells; random-scan warm starts, Powell, same-budget L-BFGS, and a 3x-NFEV base overbudget control rescued none of those 16 failures.  Structured repair rescued six failures under the same NFEV cap.  A follow-up aggressive base search on the six repair-only cells used random scan plus top-k L-BFGS, random scan plus Powell, and differential evolution plus refinement.  The best base ansatz reached zero of six strict windows, while the repaired ansatz reached all six reference margins.

\begin{table}[!t]
\caption{\label{tab:basecontrols}Optimizer, initialization, and base-reachability controls.}
\adjustbox{max width=\textwidth}{\footnotesize%
\begin{tabular}{lrrrr}
\toprule
\rowcolor{hdrcol} Gate & cells/trials & strict success & mean margin & conclusion\\
\midrule
base default & 18 cells & 2/18 & -1.826 & shallow base floor\\
best base optimizer/init control & 16 failed cells & 0 rescued & -1.811 & no optimizer/init rescue\\
structured repair, same NFEV & 18 cells & 8/18 & -0.066 & six repair-only rescues\\
aggressive base search & 6 repair-only cells, 54 trials & 0/54 & -1.182 & best-reachable shortfall\\
repair reference on same cells & 6 cells & 6/6 & +0.115 & reachable after augmentation\\
\bottomrule
\end{tabular}}
\end{table}

\section{Resource-Matched Architecture Controls}

On the six repair-only cells from the optimizer/init gate, same-family depth+1 and same-family parameter-matched depth rescued zero cells.  Brickwork parameter-matched controls also failed on the applicable non-brickwork case.  Heterogeneous random/generic Pauli controls rescued four of six cells, which weakens a uniqueness claim for the old $\xyyx+Y$ template but strengthens the broader conclusion that shallow failures require heterogeneous operator augmentation rather than same-family scaling.

\begin{table}[!t]
\caption{\label{tab:resourcecontrols}Resource-matched architecture controls on the six repair-only cells.  The old $\xyyx+Y$ template is not unique, but same-family depth and parameter scaling do not explain the repairs.}
\adjustbox{max width=\textwidth}{\footnotesize%
\begin{tabular}{lrr}
\toprule
\rowcolor{hdrcol} Arm & strict success & mean margin\\
\midrule
$\xyyx+Y$ repair & 12/12 & +0.115\\
$\xxyy+Y$ same-count control & 5/12 & -0.093\\
random same-count Pauli layer, best & 4/12 & -0.065\\
generic local+NN same-count & 4/12 & -0.095\\
Hamiltonian-term same-count & 4/12 & -0.339\\
same-family parameter-matched depth & 0/12 & -0.619\\
same-family depth+1 & 0/12 & -0.758\\
\bottomrule
\end{tabular}}
\end{table}

\section{Algorithmic Baselines That Did Not Become the Main Method}

Several candidate algorithms were tested and retired before the commutator-axis framing.  Table~\ref{tab:negativealgos} records these gates to make the negative evidence explicit.  The important pattern is that local ranking and QITE-style objectives \cite{motta2020qite} can improve early descent or loose targets, but they do not beat simple structured completions after full reoptimization to strict low-energy windows.

\begin{table}[!t]
\caption{\label{tab:negativealgos}Negative algorithm gates retained as baselines.}
\adjustbox{max width=\textwidth}{\footnotesize%
\begin{tabular}{lrrrrl}
\toprule
\rowcolor{hdrcol} Gate & setting & method success & best/control success & key margin/cost result & decision\\
\midrule
BHAC beam, $B=4$ & 6 repair-only cells, 2 seeds & 5/12 & 6/12 ADAPT & BHAC margin -0.037 & no pivot\\
BHAC beam, $B=6$ & 6 repair-only cells, 1 seed & 4/6 & 4/6 ADAPT & below best budget control & no pivot\\
ADAPT-lite, $B=8$ & 6 cells, 3 seeds & 18/18 & 17/18 one-step & CI vs one-step crosses 0 & baseline only\\
trust-QITE full reopt & 6 cells, 3 seeds & 6/18 & 12/18 fixed & loses to fixed by -0.074 margin & no-go\\
trust-QITE strict cost & 6 cells, 3 seeds & 1/3 rate & fixed/best wins & best-control cost CI negative & no-go\\
raw QITE coverage & 324 trials & 0.056 frozen rate & -- & coverage-gain Spearman -0.078 & no-go\\
\bottomrule
\end{tabular}}
\end{table}

\section{Matched and Generous ADAPT Stress Details}

The matched ADAPT stress test was extended with a generous companion arm that receives up to five times the fixed-rule NFEV cap and six additional insertions.  Table~\ref{tab:adaptgenerous} shows the cost-to-target sweep.  The generous ADAPT arm steadily improves as the cap increases and becomes competitive, but it still requires a much larger censored target cost.  We did not complete the 18000-cap point because the full-pool sequential reoptimization sweep became the dominant runtime; the monotone 6000, 12000, and 30000 points are sufficient to show that the finite-budget gap is not a narrow cap artifact.

\begin{table}[!t]
\caption{\label{tab:adaptgenerous}Generous full-pool ADAPT cap sweep.  Fixed $\K_1$ always uses a 6000 post-base cap; ADAPT receives the listed cap and six extra insertions.  The same full-pool sequential candidate set and stopping rule as in Table~\ref{tab:adaptmatch} are used.}
\adjustbox{max width=\textwidth}{\footnotesize%
\begin{tabular}{llrrrrr}
\toprule
\rowcolor{hdrcol} Group & ADAPT cap & fixed strict & ADAPT strict & fixed cost & ADAPT cost & ADAPT/fixed\\
\midrule
application QUBO & 6000 & 0.583 & 0.083 & 2727 & 5982 & 2.19\\
application QUBO & 12000 & 0.583 & 0.375 & 2727 & 10103 & 3.70\\
application QUBO & 30000 & 0.583 & 0.542 & 2727 & 19736 & 7.24\\
QAOA2D & 6000 & 0.667 & 0.167 & 2171 & 5038 & 2.32\\
QAOA2D & 12000 & 0.667 & 0.333 & 2171 & 9915 & 4.57\\
QAOA2D & 30000 & 0.667 & 0.500 & 2171 & 19712 & 9.08\\
\bottomrule
\end{tabular}}
\end{table}

\section{Mechanism Diagnostics}

For six QAOA2D out-of-sample cells, the algebraic closure $[{\rm local}\ X,{\rm lattice}\ ZZ]$ generated lattice-edge $\yzzy$ coverage in 6/6 cells.  For J1-J2 cells, target-H closure additionally generated diagonal $\yzzy$ terms.  Tangent coverage diagnostics were informative but not sufficient: edge+diagonal $\yzzy+Y$ had high ground-residual coverage, while local $Y$ alone had almost zero first-order coverage.  The successful repeated layer is therefore a finite-depth reoptimization mechanism, not a naive one-step tangent rule.

\begin{table}[!t]
\caption{\label{tab:mechanism}Mechanism diagnostics.  Algebraic closure identifies the edge $\yzzy$ axis, while first-order tangent coverage alone does not rank the best finite-depth repair.}
\adjustbox{max width=\textwidth}{\footnotesize%
\begin{tabular}{lrrr}
\toprule
\rowcolor{hdrcol} Family or diagnostic & cells covered & ground-residual coverage & energy-residual coverage\\
\midrule
QAOA closure $[X,ZZ]$ generates lattice $\yzzy$ & 6/6 & -- & --\\
target-H closure generates lattice $\yzzy$ & 6/6 & -- & --\\
target-H closure generates diagonal $\yzzy$ on J1-J2 & 2/2 families & -- & --\\
QAOA base generators & -- & 0.857 & 0.919\\
Hamiltonian terms & -- & 0.873 & 1.000\\
edge $\yzzy$ & -- & 0.467 & 0.422\\
edge+diagonal $\yzzy+Y$ & -- & 0.921 & 0.954\\
edge $\xxyy+Y$ & -- & 0.712 & 0.800\\
full edge-axis pool & -- & 0.951 & 0.998\\
\bottomrule
\end{tabular}}
\end{table}

\section{Base-Ansatz Swap Gate}
\label{app:ansatzswap}

The completion rule depends on both the Hamiltonian and the base generator set.  To test whether the paper is merely transferring one hand-picked QAOA axis, we expanded the ansatz-swap gate to six $2\times3$ Hamiltonian settings---TFIM at two fields, frustrated J1-J2 Ising at two couplings, Heisenberg J1-J2, and XXZ J1-J2---and three shallow hosts: QAOA2D, a brickwork hardware-efficient-like circuit, and a Heisenberg HVA host.  For every Hamiltonian-host pair we ran three shared seeds and compared the optimized base, the automatic stripped first shell plus local $Y$, fixed $\yzzy+Y$, fixed $XZ/ZX+Y$, same-family depth, Hamiltonian-term layers, local-$Y$, and wrong-axis controls under the same $14$-gate and $3000$-NFEV cap.

\begin{figure}[!t]
\centering
\resizebox{0.74\columnwidth}{!}{%
\begin{tikzpicture}[
  font=\sffamily\tiny,
  box/.style={draw, rounded corners=2pt, align=center, inner sep=4pt, minimum height=0.50cm},
  hbox/.style={box, draw=black!45, fill=black!3, text width=2.05cm},
  host/.style={box, draw=blue!60!black, fill=blue!6, text width=1.82cm},
  shell/.style={box, draw=violet!70!black, fill=violet!7, text width=1.90cm},
  successbox/.style={box, draw=green!45!black, fill=green!8, text width=1.90cm},
  bnd/.style={box, draw=orange!80!black, fill=orange!9, text width=1.90cm},
  arrow/.style={-{Latex[length=1.8mm]}, line width=0.55pt, draw=black!60, shorten >=2pt, shorten <=2pt},
  note/.style={font=\sffamily\tiny, align=center, text=black!62}
]
\node[hbox] (ham) at (0,0) {same $H$ family\\six $2{\times}3$ cells};
\node[host] (q) at (-2.55,-1.05) {QAOA2D\\$X,ZZ$ host};
\node[host] (b) at (0,-1.05) {brickwork\\$Y/Z+$ ent.};
\node[host] (hva) at (2.55,-1.05) {HVA\\$XX/YY/ZZ$};
\node[shell] (qk) at (-2.55,-2.10) {$\K_1$ motif\\$\yzzy$};
\node[shell] (bk) at (0,-2.10) {$\K_1$ motif\\$XZ/ZX$};
\node[shell] (hk) at (2.55,-2.10) {$\K_1$ motif\\mixed axes};
\node[successbox] (qr) at (-2.55,-3.17) {strict repair\\Ising/TFIM\\\best{12/12}};
\node[bnd] (br) at (0,-3.17) {margin gain\\strict boundary};
\node[bnd] (hr) at (2.55,-3.17) {partial gain\\strict boundary};
\node[note, text width=5.15cm] at (0,-4.03)
{Scope: host changes $\K_1$; one block closes only QAOA2D Ising/TFIM.};
\draw[arrow] (ham) -- (q);
\draw[arrow] (ham) -- (b);
\draw[arrow] (ham) -- (hva);
\draw[arrow] (q) -- (qk);
\draw[arrow] (b) -- (bk);
\draw[arrow] (hva) -- (hk);
\draw[arrow] (qk) -- (qr);
\draw[arrow] (bk) -- (br);
\draw[arrow] (hk) -- (hr);
\end{tikzpicture}}
\caption{\label{fig:ansatzscope}How to read the ansatz-swap matrix.  Holding the Hamiltonian family fixed while changing the shallow host changes the stripped first-shell motif and therefore the appropriate completion layer.  This is a scope/host-dependence test: QAOA2D exposes the $\yzzy$ motif that gives strict Ising/TFIM repair, whereas brickwork and HVA expose different or mixed motifs and remain one-block boundary cases.}
\end{figure}
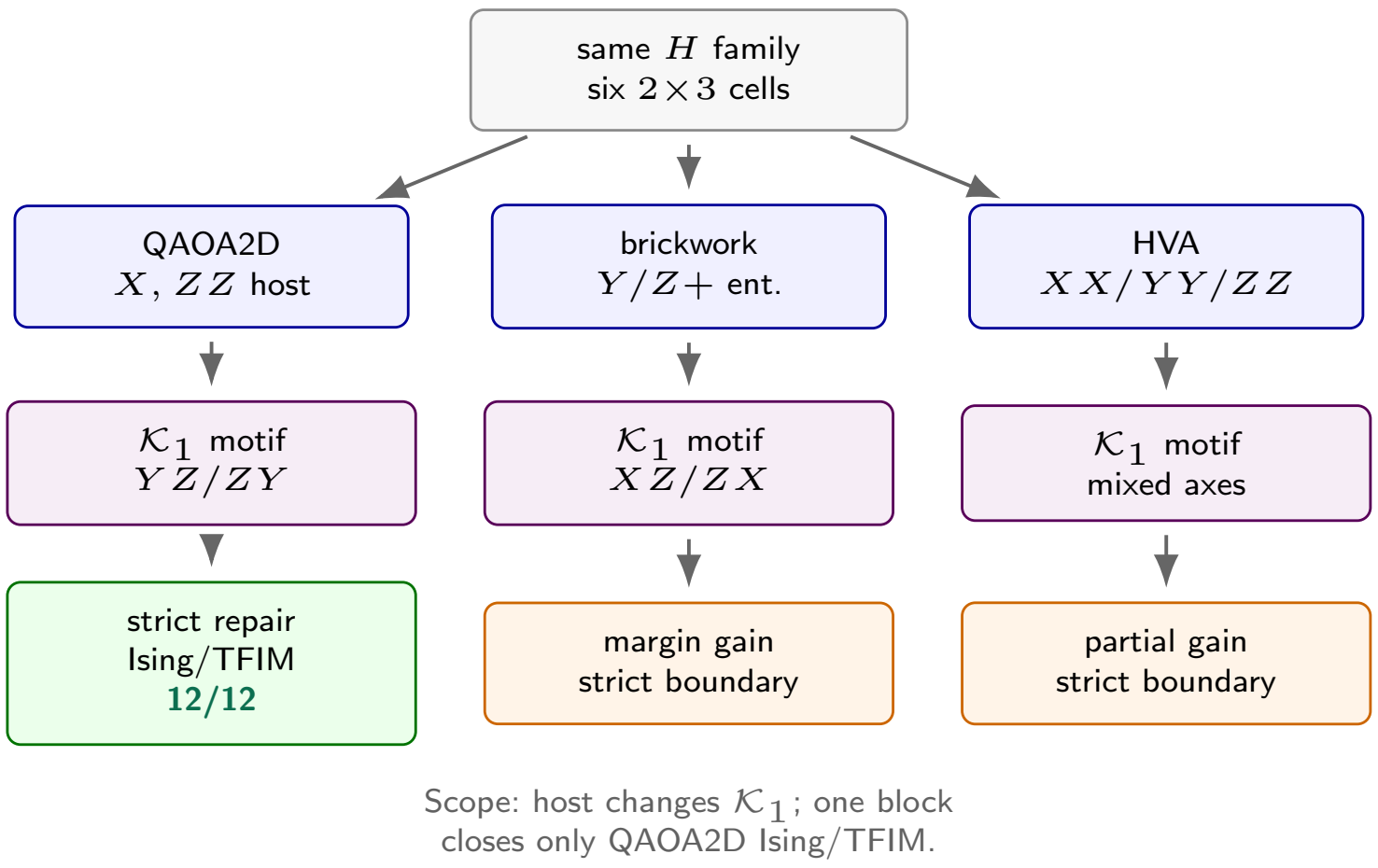

Figure~\ref{fig:ansatzscope} gives the intended reading guide for this experiment.  The ansatz-swap gate is a \emph{scope test}: it asks how the first-shell prediction changes when the shallow host is changed while the Hamiltonian family is held fixed.  It should not be read as a claim that the same raw $\K_1$ block solves every host.  In the QAOA2D Ising/TFIM rows, the host exposes the clean $\yzzy$ motif used by the main QAOA repair.  In the brickwork and HVA rows, the first shell changes and the one-block repair mostly improves margins without closing the strict window.  Those rows are therefore boundary evidence for the current single-block implementation, not negative evidence for the commutator principle itself.

\begin{table}[!t]
\caption{\label{tab:ansatzswap}Same-Hamiltonian base-ansatz dependence matrix over six $2\times3$ Hamiltonian settings, three shallow hosts, and three seeds.  This is a scope test, not a universality table.  ``Shell'' lists the number of stripped first-shell edge terms of type $(\yzzy, XZ/ZX)$, and success entries are strict-window counts out of three shared seeds.  QAOA2D exposes the clean $\yzzy$ motif and is repaired on the Ising/TFIM cells; brickwork and HVA expose different or mixed shells and remain one-block boundary cases, even when the automatic first-shell arm improves the mean margin.}
\adjustbox{max width=\textwidth}{\scriptsize%
\begin{tabular}{llrrrrrrr}
\toprule
\rowcolor{hdrcol} Hamiltonian & host & shell & base & auto $\K_1{+}Y$ & fixed $\yzzy{+}Y$ & fixed $XZ/ZX{+}Y$ & depth & $\Delta m_{\K_1-\mathrm{base}}$\\
\midrule
Heisenberg $J_2{=}0.35$ & brickwork & $(0,14)$ & 0/3 & 0/3 & 0/3 & 0/3 & 0/3 & $+3.221$\\
Heisenberg $J_2{=}0.35$ & HVA & $(0,14)$ & 0/3 & 0/3 & 0/3 & 0/3 & 0/3 & $+1.218$\\
Heisenberg $J_2{=}0.35$ & QAOA2D & $(14,0)$ & 0/3 & 0/3 & 0/3 & 0/3 & 0/3 & $+4.545$\\
\addlinespace[1pt]
J1-J2 Ising $J_2{=}0.25$ & brickwork & $(0,14)$ & 0/3 & 0/3 & 0/3 & 0/3 & 0/3 & $+0.025$\\
J1-J2 Ising $J_2{=}0.25$ & HVA & $(14,14)$ & 0/3 & 0/3 & 1/3 & 0/3 & 0/3 & $+0.762$\\
J1-J2 Ising $J_2{=}0.25$ & QAOA2D & $(14,0)$ & 0/3 & 0/3 & \best{3/3} & 0/3 & \best{3/3} & $+0.266$\\
\addlinespace[1pt]
J1-J2 Ising $J_2{=}0.65$ & brickwork & $(0,14)$ & 0/3 & 0/3 & 0/3 & 0/3 & 0/3 & $+0.331$\\
J1-J2 Ising $J_2{=}0.65$ & HVA & $(14,14)$ & 0/3 & 0/3 & 1/3 & 0/3 & 0/3 & $+0.627$\\
J1-J2 Ising $J_2{=}0.65$ & QAOA2D & $(14,0)$ & 0/3 & \best{3/3} & \best{3/3} & 0/3 & \best{3/3} & $+0.067$\\
\addlinespace[1pt]
TFIM $h{=}0.8$ & brickwork & $(0,14)$ & 0/3 & 0/3 & 0/3 & 0/3 & 0/3 & $+0.025$\\
TFIM $h{=}0.8$ & HVA & $(14,14)$ & 0/3 & 0/3 & 0/3 & 0/3 & 0/3 & $+0.000$\\
TFIM $h{=}0.8$ & QAOA2D & $(14,0)$ & 0/3 & \best{3/3} & \best{3/3} & 0/3 & 0/3 & $+3.093$\\
\addlinespace[1pt]
TFIM $h{=}1.2$ & brickwork & $(0,14)$ & 0/3 & 0/3 & 0/3 & 0/3 & 0/3 & $+0.052$\\
TFIM $h{=}1.2$ & HVA & $(14,14)$ & 0/3 & 0/3 & 2/3 & 0/3 & 0/3 & $+0.078$\\
TFIM $h{=}1.2$ & QAOA2D & $(14,0)$ & 0/3 & \best{3/3} & \best{3/3} & 0/3 & 0/3 & $+4.550$\\
\addlinespace[1pt]
XXZ $J_2{=}0.50,\Delta{=}1.5$ & brickwork & $(0,14)$ & 0/3 & 0/3 & 0/3 & 0/3 & 0/3 & $+2.880$\\
XXZ $J_2{=}0.50,\Delta{=}1.5$ & HVA & $(0,14)$ & 0/3 & 0/3 & 0/3 & 0/3 & 0/3 & $+1.532$\\
XXZ $J_2{=}0.50,\Delta{=}1.5$ & QAOA2D & $(14,0)$ & 0/3 & 0/3 & 0/3 & 0/3 & 0/3 & $+5.577$\\
\bottomrule
\end{tabular}}
\end{table}

We then ran a dedicated host-specific higher-order diagnostic to separate three questions that Table~\ref{tab:ansatzswap} mixes: whether the correct host-dependent axis is selected, whether a single block is deep enough, and whether the first commutator shell is already the right algebraic order.  The rule was fixed before optimization.  The first-shell block uses $\K_1=[\G_0,H]_{\rm stripped}$ relative to the host generator set $\G_0$; the higher-order block adds the second closure $\K_2=[\K_1,\G_0\cup H]_{\rm stripped}$, but inserts only the local and two-body rotations supported by the same circuit model.  The diagnostic uses the same six $2\times3$ Hamiltonian settings, the same three hosts, and three shared seeds as Table~\ref{tab:ansatzswap}, increases the insertion budget to 56 gates, and keeps matched controls in every cell.  Table~\ref{tab:hostspecific} shows the full Hamiltonian-host matrix.  The ansatz-dependence experiment should therefore be read as follows: first-shell repair is the cheapest algebraic layer, while $\K_1+\K_2$ tests whether apparent failures are missing a second low-order commutator shell.

\begin{table}[!t]
\caption{\label{tab:hostspecific}Full host-specific higher-order matrix on the same six Hamiltonian settings and three hosts as Table~\ref{tab:ansatzswap}.  The first-shell host rule is fixed by $\K_1=[\G_0,H]_{\rm stripped}$; the higher-order rule inserts $\K_1+\K_2$ with $\K_2=[\K_1,\G_0\cup H]_{\rm stripped}$, restricted to implementable local/two-body rotations.  Entries are strict-window counts out of three shared seeds under a 56-gate insertion budget and a 6000-evaluation cap.  The shell column reports $(|\K_1|,|\K_2^{\le2}|)$, where $|\K_2^{\le2}|$ is the implementable local/two-body subset.  ``Best comm.'' is the best predeclared commutator arm among first-shell and higher-order variants; ``Best ctrl.'' is the strongest matched noncommutator control; ``Off-rule'' is the strongest off-motif commutator axis.}
\adjustbox{max width=\textwidth}{\scriptsize%
\begin{tabular}{llcrrrrrrr}
\toprule
\rowcolor{hdrcol} Hamiltonian & host & shell & base & $\K_1$ host & $\K_1{+}\K_2$ & best comm. & best ctrl. & off-rule & $\Delta m_{\rm comm-ctrl}$\\
\midrule
Heisenberg $J_2{=}0.35$ & brickwork & $(44,22)$ & 0/3 & 0/3 & \best{3/3} & \best{3/3} & 2/3 & \best{3/3} & $+0.142$\\
Heisenberg $J_2{=}0.35$ & HVA & $(118,12)$ & 0/3 & 0/3 & 0/3 & 0/3 & 1/3 & \best{3/3} & $-1.480$\\
Heisenberg $J_2{=}0.35$ & QAOA2D & $(86,12)$ & 0/3 & 2/3 & \best{3/3} & \best{3/3} & 1/3 & \best{3/3} & $+0.350$\\
\addlinespace[1pt]
J1-J2 Ising $J_2{=}0.25$ & brickwork & $(22,55)$ & 0/3 & 0/3 & 0/3 & 0/3 & \best{3/3} & \best{3/3} & $-0.097$\\
J1-J2 Ising $J_2{=}0.25$ & HVA & $(130,34)$ & 0/3 & 2/3 & 0/3 & 2/3 & 2/3 & \best{3/3} & $+0.017$\\
J1-J2 Ising $J_2{=}0.25$ & QAOA2D & $(22,11)$ & 0/3 & \best{3/3} & \best{3/3} & \best{3/3} & \best{3/3} & \best{3/3} & $+0.002$\\
\addlinespace[1pt]
J1-J2 Ising $J_2{=}0.65$ & brickwork & $(22,55)$ & 0/3 & 0/3 & 2/3 & 2/3 & \best{3/3} & \best{3/3} & $-0.059$\\
J1-J2 Ising $J_2{=}0.65$ & HVA & $(130,34)$ & 0/3 & \best{3/3} & \best{3/3} & \best{3/3} & \best{3/3} & \best{3/3} & $+0.006$\\
J1-J2 Ising $J_2{=}0.65$ & QAOA2D & $(22,11)$ & 0/3 & \best{3/3} & \best{3/3} & \best{3/3} & \best{3/3} & \best{3/3} & $+0.009$\\
\addlinespace[1pt]
TFIM $h{=}0.8$ & brickwork & $(14,35)$ & 0/3 & 0/3 & 0/3 & 0/3 & 0/3 & 0/3 & $+0.001$\\
TFIM $h{=}0.8$ & HVA & $(106,34)$ & 0/3 & 0/3 & 0/3 & 0/3 & 0/3 & 0/3 & $+0.000$\\
TFIM $h{=}0.8$ & QAOA2D & $(14,7)$ & 0/3 & \best{3/3} & \best{3/3} & \best{3/3} & \best{3/3} & \best{3/3} & $-0.000$\\
\addlinespace[1pt]
TFIM $h{=}1.2$ & brickwork & $(14,35)$ & 0/3 & 0/3 & \best{3/3} & \best{3/3} & \best{3/3} & \best{3/3} & $+0.017$\\
TFIM $h{=}1.2$ & HVA & $(106,34)$ & 0/3 & \best{3/3} & 0/3 & \best{3/3} & \best{3/3} & \best{3/3} & $+0.005$\\
TFIM $h{=}1.2$ & QAOA2D & $(14,7)$ & 0/3 & \best{3/3} & \best{3/3} & \best{3/3} & \best{3/3} & \best{3/3} & $-0.000$\\
\addlinespace[1pt]
XXZ $J_2{=}0.50,\Delta{=}1.5$ & brickwork & $(44,22)$ & 0/3 & 0/3 & \best{3/3} & \best{3/3} & 0/3 & \best{3/3} & $+0.363$\\
XXZ $J_2{=}0.50,\Delta{=}1.5$ & HVA & $(118,12)$ & 0/3 & 0/3 & 0/3 & 0/3 & 0/3 & \best{2/3} & $-1.149$\\
XXZ $J_2{=}0.50,\Delta{=}1.5$ & QAOA2D & $(86,12)$ & 0/3 & 0/3 & 0/3 & 0/3 & \best{1/3} & 0/3 & $-0.520$\\
\bottomrule
\end{tabular}}
\end{table}

The matrix has three implications.  First, the construction is genuinely ansatz-aware: with the same Hamiltonian family, QAOA2D exposes $\yzzy$, brickwork exposes $XZ/ZX$, and HVA exposes a larger mixed shell.  Second, apparent first-shell failures need not be failures of the commutator principle.  Brickwork is the clearest example: the first-shell host rule remains at $0/18$, but the $\K_1+\K_2$ closure reaches $11/18$ strict successes and beats the strongest matched noncommutator control on $11/18$ paired trials.  QAOA2D also improves slightly, from $14/18$ under the host first shell to $15/18$ with the higher-order closure.  Third, the old ``off-rule'' column is not merely a nuisance control: it shows that useful axes often lie nearby in the same low-order commutator-closure neighborhood, especially on Heisenberg/XXZ rows.  This motivates the final closure-selector experiment in Table~\ref{tab:closureselector}.

\begin{table}[!t]
\caption{\label{tab:closureselector}Detailed appendix version of the expanded commutator-closure selector matrix summarized in Table~\ref{tab:closureselectormain}.  The closure pool is fixed before optimization from implementable $\K_1$, implementable $\K_2$, and low-order companion axes; the default selector ranks this pool by parameter-shift gradient magnitude at the optimized base.  ``Envelope'' credits, for each host--Hamiltonian cell and seed, the better strict outcome of the two predeclared selectors (per-trial aggregation, which is why it can exceed the cell-level maximum of the two selector columns in Table~\ref{tab:closureselectorfullmain}); it is reported as diagnostic headroom, not as the default algorithm.  The control-envelope column credits, for each host--Hamiltonian cell and seed, the strongest matched noncommutator arm among same-family depth, Hamiltonian layers, local-$Y$-only layers, wrong-axis layers, and random edge-axis layers before aggregation.  Entries are strict-window successes out of 27 paired trials per host, from nine $2\times3$ Hamiltonian settings and three shared seeds, with a 56-gate insertion budget and a 6000-evaluation cap.}
\adjustbox{max width=\textwidth}{\scriptsize%
\begin{tabular}{lrrrrrrr}
\toprule
\rowcolor{hdrcol} Host & base & $\K_1{+}Y$ & $\K_1{+}\K_2$ & grad selector & trust selector & selector envelope & control envelope\\
\midrule
product & 0/27 & 2/27 & 3/27 & 14/27 & 15/27 & \best{18/27} & 13/27\\
QAOA2D & 0/27 & 0/27 & 9/27 & 15/27 & 12/27 & 16/27 & \best{18/27}\\
brickwork & 0/27 & 0/27 & 20/27 & 22/27 & 20/27 & \best{24/27} & 16/27\\
HEA RX/RY+CZ & 0/27 & 14/27 & 13/27 & \best{24/27} & \best{24/27} & \best{24/27} & 18/27\\
HEA RX/RZ+CZ & 0/27 & 21/27 & 21/27 & 23/27 & 20/27 & \best{24/27} & 16/27\\
HEA XYZ+CZ & 0/27 & 14/27 & 12/27 & 22/27 & 21/27 & \best{23/27} & 16/27\\
HVA Heisenberg & 0/27 & 11/27 & 3/27 & 13/27 & 15/27 & \best{16/27} & 12/27\\
\midrule
\textbf{Total} & \textbf{0/189} & \textbf{62/189} & \textbf{81/189} & \best{133/189} & \textbf{127/189} & \textbf{145/189} & \textbf{109/189}\\
\bottomrule
\end{tabular}}
\end{table}

\noindent\textbf{Aggregate audit of the noncommutator controls.}\par\noindent
The main text emphasizes the per-trial control envelope because it is the most conservative comparator: for each host--Hamiltonian cell and seed it chooses the strongest noncommutator control before aggregating.  Table~\ref{tab:control_arm_audit} makes the single-control baseline in the main figure auditable from the same 189-trial run.  The strongest deployable single noncommutator control is the random edge-axis arm at $77/189$; this arm is a seeded same-count layer that samples local $Y_i$ axes and two-qubit edge Pauli axes from the available local/edge supports under the same 56-rotation and 6000-evaluation caps.  The envelope reaches $109/189$ by allowing the control family to vary from trial to trial.  The envelope is not a single deployable ansatz family, so it is not assigned a unique mean fidelity or cost.

\noindent\textbf{$\K_1\cup\K_2$-only selector ablation.}\par\noindent
To separate the contribution of the strict commutator shells from the matched low-order companion axes in the reported closure pool, we reran the full 189-trial matrix with two additional selectors.  These selectors use the same hosts, Hamiltonians, seeds, 56-rotation insertion cap, 6000-evaluation cap, initialization pool, and full reoptimization protocol as the reported selectors, but the ranking pool is restricted to implementable $\K_1\cup\K_2$ axes only.  They therefore remove the extra low-order companion axes $\mathcal Q$ used by the reported $\mathcal C_{\le2}=\mathrm{impl}(\K_1\cup\K_2\cup\mathcal Q)$ selector.
The ablation, reported in Table~\ref{tab:k1k2_selector_ablation}, has two uses.  First, it shows that low-order commutator closure alone is already a strong repair signal: the $\K_1\cup\K_2$-only gradient selector reaches $104/189$, just below the conservative per-trial noncommutator envelope at $109/189$ and well above the fixed unranked $\K_1+\K_2$ block.  Second, it shows why the reported method uses the slightly broader implementable closure pool: adding the matched companion axes and reranking increases the gradient selector from $104/189$ to $133/189$.  The companion axes are therefore not an unrelated generic layer; they are fixed same-support Pauli directions that make the front-loaded closure selector more robust across hosts.

\noindent\textbf{Locality and shallow-block audit.}\par\noindent
We also audit the selected repair blocks themselves, because a useful shallow-circuit repair should not silently turn into an arbitrary all-to-all or high-weight operator layer.  Table~\ref{tab:locality_audit} is computed directly from the selected Pauli-rotation sequences in the 189-trial layer files.  The geometric range is the largest Manhattan distance between qubits in a selected Pauli string on the $2\times3$ row-major lattice; one-qubit terms have range zero.  The coloring depth is a greedy support-overlap coloring of the repair block, giving a conservative parallel-depth upper bound for the selected rotations alone.  It is not a hardware-native compilation depth, and the full circuit depth also includes the host ansatz.  All selected repair rotations in this audit have Pauli weight at most two; the fraction of three-local or higher-weight rotations is $0\%$ for every row.  Thus the reported repair is not an all-to-all multi-qubit layer.  Most selected terms are nearest-neighbor or next-nearest-neighbor on the $2\times3$ lattice, with only about one percent of rotations reaching lattice range three.  For asymptotic claims we do not rely on this finite $2\times3$ audit alone: Appendix~\ref{app:simulability} separately proves the bounded-light-cone statement for fixed-depth, bounded-range local scheduling.

Table~\ref{tab:closureselector} changes the interpretation of the host-swap study.  The robust object is not a single hand-coded axis such as $\yzzy+Y$, nor the raw first shell alone.  It is the budgeted low-order closure space.  The default gradient-ranked selector improves strict success from $0/189$ for the optimized base to $133/189$, while the conservative per-trial control envelope reaches $109/189$.  The gain is largest for product, brickwork, and hardware-efficient hosts, where the first shell alone misses useful companion directions but the closure selector recovers them before full reoptimization.  The main boundary is also explicit: QAOA2D is mixed in this larger matrix, because the TFIM rows are easy enough for matched controls, even though J1-J2 Ising remains solved by the closure.  This is the right scope of the final method: a front-loaded commutator-closure selector can repair many gradient-healthy shallow-host arrival failures, but it is not guaranteed to beat the envelope of all resource-matched controls on every host/Hamiltonian pair.

\section{The boundary is a depth-threshold effect, not a spanning wall}
\label{app:depththreshold}

We tested directly whether the single-block boundary---a single first-shell layer occasionally missing the strict window, e.g.\ on $2\times4$ cells---reflects a structural limit of the first shell or a finite-depth optimization threshold.  Three diagnostics on the QAOA2D out-of-sample cells (exact statevector, $n\le8$, single seed) point to the latter.

First, static tangent-span coverage of the low-energy residual does not separate the cells that succeed with a single block from those that do not: the first shell $\K_1$ gives mean ground-residual coverage $0.41$ for both groups, and adding the second commutator shell $\K_2$ (or the full edge-axis pool) saturates coverage to $0.97$--$1.0$ for \emph{both} groups, with group separation $\le0.03$ (Table~\ref{tab:coversat}).  Second, a per-block trainability signal (mean-squared gradient of the completed ansatz at the warm start) overlaps between the two groups and is non-separating.  Third, and decisively, when repeated first-shell blocks are appended and \emph{all} parameters are reoptimized, the depth needed to enter the strict window is $1$ for five of the six cells and $2$ for the sixth (Table~\ref{tab:depththreshold}); a tight evaluation budget gives the same depth-$1$ errors as a generous one, so the effect is not budget starvation.

The depth-$1$ error sits in a narrow band ($0.0035$--$0.0202$) that grows only mildly with system size, while the strict window $0.1\Delta_{\rm gap}$ is set by the spectral gap.  Success at depth $1$ is thus governed by whether the near-uniform depth-$1$ residual falls below the gap-defined window; the only depth-$1$ miss ($2\times4$ TFIM, error $0.0202$ vs window $0.0136$) is marginal and is closed by a single additional block.  We conclude that the empirical boundary is a depth-fixable threshold effect, not a structural reachability or trainability wall, and that a static ``when does $\K_1$ span the low-energy directions'' criterion is not the right characterization; the open quantitative question is the scaling of the depth-$1$ residual.

\begin{table}[!t]
\caption{\label{tab:coversat}Standard expressivity/reachability diagnostics do \emph{not} separate the cells that reach the strict window (success: four $2\times3$ cells) from those that miss it with a single block (boundary: two $2\times4$ cells); exact statevector.  Coverage is the span-coverage of the named residual by the tangent directions of the listed generator set.}
\footnotesize
\adjustbox{max width=\columnwidth}{\footnotesize%
\begin{tabular}{lrrl}
\toprule
\rowcolor{hdrcol} diagnostic & success & boundary & separates?\\
\midrule
$\K_1$ tangent coverage (ground residual) & 0.41 & 0.41 & no ($+0.005$)\\
$\K_1\cup\K_2$ coverage (ground residual) & 1.00 & 0.97 & no ($+0.03$)\\
full edge-axis pool (ground residual) & 0.99 & 0.96 & no ($+0.03$)\\
$\K_1\cup\K_2$ coverage (energy residual) & 1.00 & 1.00 & no ($0.00$)\\
depth-1 trainability $\langle|\nabla|^2\rangle$ & 0.006--0.027 & 0.009--0.013 & no (overlap)\\
\bottomrule
\end{tabular}}
\end{table}

\begin{table}[!t]
\caption{\label{tab:depththreshold}Depth-to-reach under full reoptimization (repeated first-shell blocks, generous budget; exact statevector, single seed).  ${\rm err}_d$ is the best energy error at depth $d$; $d^\star$ is the first depth within the strict window $\max(10^{-6},0.1\Delta_{\rm gap})$.  Reachability is $d^\star\le2$ for every cell.}
\footnotesize
\adjustbox{max width=\columnwidth}{\footnotesize%
\begin{tabular}{lrrrrrr}
\toprule
\rowcolor{hdrcol} cell & $\Delta_{\rm gap}$ & window & ${\rm err}_0$ & ${\rm err}_1$ & ${\rm err}_2$ & $d^\star$\\
\midrule
$2\times3$ TFIM $h{=}0.8$ & 0.041 & 0.0041 & 1.50 & 0.0035 & 0.0000 & 1\\
$2\times3$ TFIM $h{=}1.6$ & 0.856 & 0.0855 & 0.33 & 0.0084 & 0.0000 & 1\\
$2\times3$ J1J2 $j_2{=}0.25$ & 0.269 & 0.0269 & 0.48 & 0.0105 & 0.0003 & 1\\
$2\times3$ J1J2 $j_2{=}0.65$ & 0.895 & 0.0895 & 0.11 & 0.0084 & 0.0079 & 1\\
$2\times4$ TFIM $h{=}1.2$ & 0.136 & 0.0136 & 1.23 & 0.0202 & 0.0004 & \best{2}\\
$2\times4$ J1J2 $j_2{=}0.45$ & 0.375 & 0.0375 & 0.30 & 0.0139 & 0.0006 & 1\\
\bottomrule
\end{tabular}}
\end{table}

\section{Pauli-path loss-estimation scaling}
\label{app:paulipath}

As an independent classical check on Theorem~\ref{thm:closurelightcone}, we evaluated the completed-circuit loss on the TFIM strips by low-weight Pauli (Heisenberg) back-propagation, truncating the back-propagated Hamiltonian to Pauli weight $\le w$ in the spirit of LOWESA-type and noiseless Pauli-propagation expectation estimation \cite{fontana2023noisy,rudolph2023lowesa,angrisani2025classically}.  At full weight the estimate reproduces the exact statevector energy to $\sim10^{-12}$.  Table~\ref{tab:paulipath} reports the truncation behavior: the weight $w^\star$ needed for fixed additive accuracy ($\le10^{-2}$) does not grow with $n$ (it is $6$ across $n=8$--$12$), and at fixed weight the number of retained Pauli strings---the classical cost---grows polynomially in $n$, consistent with the linear-time light-cone bound of Theorem~\ref{thm:closurelightcone}.

\begin{table}[!t]
\caption{\label{tab:paulipath}Pauli-path loss estimation on the geometrically local TFIM strips.  $w^\star$ is the smallest Pauli weight reaching energy error $\le10^{-2}$ (exact-validated for $n\le12$); ``terms'' is the number of retained Pauli strings (the classical cost).  At fixed weight the count grows polynomially in $n$.  Dashes indicate that $w^\star$ is not reported for $n>12$ because exact statevector validation is available only for $n\le12$.}
\adjustbox{max width=\columnwidth}{\footnotesize%
\begin{tabular}{rrrrr}
\toprule
\rowcolor{hdrcol} $n$ & $w^\star$ & terms@$w^\star$ & terms@$w{=}4$ & terms@$w{=}5$\\
\midrule
6 & 6 & 3{,}171 & 1{,}871 & 3{,}356\\
8 & 6 & 25{,}517 & 5{,}919 & 19{,}838\\
10 & 6 & 83{,}382 & 12{,}942 & 63{,}494\\
12 & 6 & 126{,}718 & 23{,}639 & 146{,}991\\
16 & --- & --- & 58{,}837 & 483{,}525\\
20 & --- & --- & 114{,}969 & 1{,}105{,}496\\
\bottomrule
\end{tabular}}
\end{table}

\section{QAOA2D Axis Controls and Out-of-Sample Generalization}

The original AGP/CD story was replaced by the simpler axis-completion story after the axis-matched gate in Table~\ref{tab:axiscontrols}.  Fixed edge $\yzzy+Y$ solved all 20 paired pilot trials at $B=14$, whereas AGP/CD solved 9/20.  On six out-of-sample QAOA2D cells at the same budget, fixed edge $\yzzy+Y$ remained the best method, solving all four new $2\times3$ parameter-shifted cells and reaching relaxed success on the $2\times4$ cells.

\begin{table}[!t]
\caption{\label{tab:axiscontrols}Axis-matched and out-of-sample QAOA2D controls at $B=14$.  The ``2x'' and ``4x'' columns report success under energy windows relaxed by factors of two and four relative to the strict window.}
\adjustbox{max width=\textwidth}{\footnotesize%
\begin{tabular}{lrrrrr}
\toprule
\rowcolor{hdrcol} Setting/method & strict & 2x & 4x & mean margin & fidelity\\
\midrule
pilot fixed edge $\yzzy+Y$ & 1.000 & 1.000 & 1.000 & +0.0328 & 0.9990\\
pilot AGP/CD zero-init & 0.450 & 0.950 & 1.000 & -0.0056 & 0.9897\\
pilot H-stripped gradient & 0.000 & 0.500 & 0.500 & -0.1758 & 0.9364\\
OOS fixed edge $\yzzy+Y$ & 0.667 & 0.667 & 1.000 & +0.0120 & 0.9937\\
OOS AGP/CD & 0.333 & 0.444 & 0.556 & -0.1372 & 0.9472\\
OOS fixed edge $\xxyy+Y$ & 0.167 & 0.167 & 0.333 & -0.3937 & 0.8639\\
OOS fixed edge $\xyyx+Y$ & 0.000 & 0.167 & 0.278 & -0.5822 & 0.7336\\
\bottomrule
\end{tabular}}
\end{table}

\begin{table}[!t]
\caption{\label{tab:k1agp}Matched $\K_1$ versus explicit AGP/CD on the QAOA2D out-of-sample set.  Both completion arms use 14 inserted gates, 16 total parameters, and the same optimization protocol.}
\adjustbox{max width=\textwidth}{\footnotesize%
\begin{tabular}{lrrrrr}
\toprule
\rowcolor{hdrcol} Method & strict & mean margin & mean error & fidelity & target cost\\
\midrule
base QAOA2D & 0/18 & -0.615 & 0.658 & 0.781 & 283\\
fixed edge $\yzzy+Y$ & \best{12/18} & \best{+0.012} & \best{0.031} & \best{0.994} & \best{2171}\\
explicit AGP/CD layer & 6/18 & -0.137 & 0.180 & 0.946 & 4269\\
\bottomrule
\end{tabular}}
\end{table}

\section{2x4 Budget, Augmentation, and Repeated-Layer Rescue}

The $2\times4$ tests are the main size stress for QAOA2D.  A single $\yzzy+Y$ layer is strongest among single-layer axes but does not solve TFIM $2\times4$ at $B\leq 40$.  Flat augmentations with local $X/Z$ or $\xxyy$ tails do not close the remaining strict gap.  Repeating the structured axis block does close it, and the effect survives resource-matched repeated-axis and matched-parameter QAOA controls.

\begin{table}[!t]
\caption{\label{tab:qaoa2d2x4}QAOA2D $2\times4$ stress tests.  Strict is averaged over paired seeds/cells for each row.  The ``2x'' and ``4x'' columns report success under energy windows relaxed by factors of two and four relative to the strict window.}
\adjustbox{max width=\textwidth}{\footnotesize%
\begin{tabular}{lrrrrr}
\toprule
\rowcolor{hdrcol} Gate/method & budget & strict & 2x & 4x & mean margin\\
\midrule
single $\yzzy+Y$, two $2\times4$ cells & 14 & 0.000 & 0.000 & 1.000 & -0.0517\\
single $\yzzy+Y$, two $2\times4$ cells & 22 & 0.500 & 1.000 & 1.000 & +0.0085\\
single $\yzzy+Y$, two $2\times4$ cells & 28 & 0.500 & 1.000 & 1.000 & +0.0085\\
TFIM $\yzzy+Y$ plus flat tails & 40 & 0.000 & 1.000 & 1.000 & -0.0066\\
TFIM repeated $\yzzy+Y$ & 40 & 1.000 & -- & -- & +0.0134\\
TFIM commutator closure & 56 & 1.000 & -- & -- & +0.0022\\
single $\yzzy+Y$ with larger cap & 56 & 0.000 & -- & -- & -0.0066\\
\bottomrule
\end{tabular}}
\end{table}

\begin{table}[!t]
\caption{\label{tab:repeatresource}Resource-matched repeated-axis controls on QAOA2D $2\times4$ cells.}
\adjustbox{max width=\textwidth}{\footnotesize%
\begin{tabular}{lrrrr}
\toprule
\rowcolor{hdrcol} Budget/method & strict & mean margin & mean fidelity & censored cost\\
\midrule
$B=40$ repeated $\yzzy+Y$ & 1.000 & +0.0251 & -- & 882.5\\
$B=40$ parameter-matched QAOA & 0.667 & -0.0026 & -- & 13805.3\\
$B=40$ repeated $\xyyx+Y$ & 0.000 & -0.1034 & 0.5779 & --\\
$B=40$ repeated $\xxyy+Y$ & 0.000 & -0.2415 & 0.7561 & --\\
$B=56$ repeated $\yzzy+Y$ & 1.000 & +0.0134 & -- & 1476\\
$B=56$ parameter-matched QAOA & 0.333 & -0.3277 & -- & 25629\\
$B=56$ repeated $\xyyx+Y$ & 0.000 & -0.0562 & 0.5280 & --\\
$B=56$ repeated $\xxyy+Y$ & 0.000 & -0.0730 & 0.5058 & --\\
\bottomrule
\end{tabular}}
\end{table}

\section{Adaptive Stop/Repeat Details}

The 10-seed cap sweep tested screening caps $C\in\{500,650,800,1000,1500\}$ on 60 QAOA2D trials.  Caps 800, 1000, and 1500 matched always-repeat strict success.  Cap 800 repeated on only 1/6 trials and reduced average gates from 40 to 24.7, but did not reliably reduce mean function evaluations.  We therefore use adaptive stop/repeat as a sparse-circuit policy, not as an optimizer-efficiency claim.

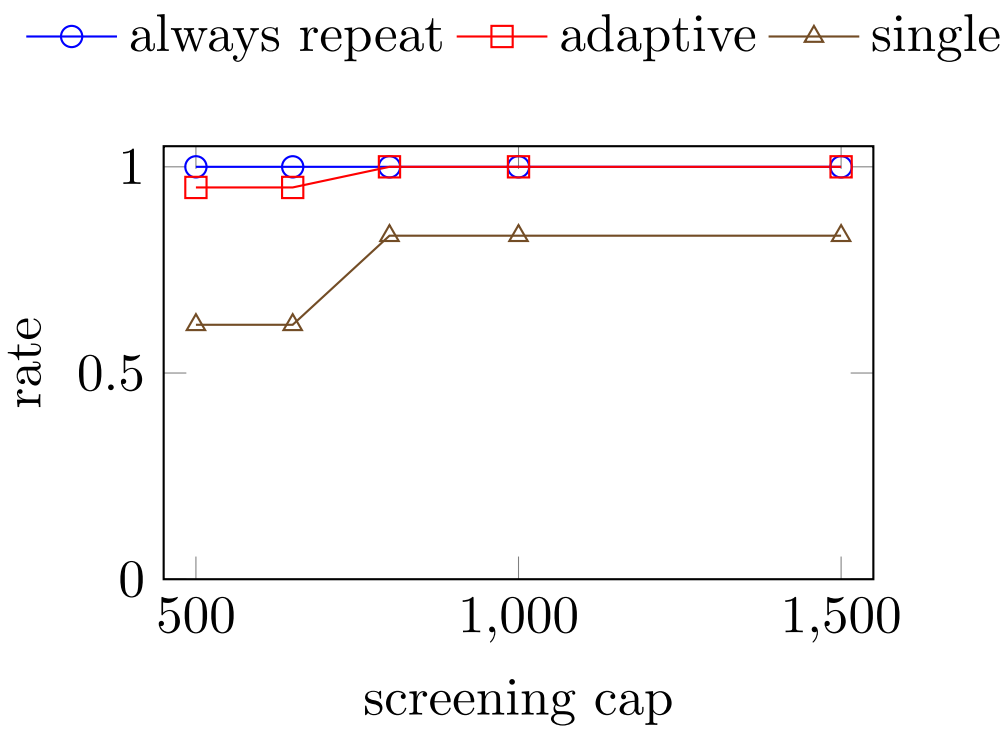
\begin{figure}[!t]
\centering
\begin{tikzpicture}
\begin{axis}[
  width=0.48\textwidth,
  height=0.34\textwidth,
  xlabel={screening cap},
  ylabel={rate},
  xmin=450,xmax=1550,
  ymin=0,ymax=1.05,
  legend style={at={(0.5,1.15)},anchor=south,legend columns=3,draw=none},
]
\addplot+[mark=o] coordinates {(500,1.0) (650,1.0) (800,1.0) (1000,1.0) (1500,1.0)};
\addplot+[mark=square] coordinates {(500,0.95) (650,0.95) (800,1.0) (1000,1.0) (1500,1.0)};
\addplot+[mark=triangle] coordinates {(500,0.617) (650,0.617) (800,0.833) (1000,0.833) (1500,0.833)};
\legend{always repeat,adaptive,single}
\end{axis}
\end{tikzpicture}
\caption{\label{fig:adaptive}Fixed-axis adaptive stop/repeat sweep over 60 QAOA2D trials.  Cap-800 adaptive matches always-repeat strict success while repeating on only $1/6$ of trials, reducing average gates from 40 to 24.7.  This is a sparsification policy for the fixed $\yzzy+Y$ motif, not the broad closure-selector algorithm summarized in the main text.}
\end{figure}

\begin{table}[!t]
\caption{\label{tab:adaptivecaps}Adaptive stop/repeat screening-cap sweep over 60 QAOA2D trials.  Cost is censored first-hit NFEV; negative savings means adaptive is more expensive than always-repeat.}
\adjustbox{max width=\textwidth}{\footnotesize%
\begin{tabular}{rrrrrrr}
\toprule
\rowcolor{hdrcol} cap & adaptive strict & single strict & repeat strict & adaptive cost & repeat cost & repeat rate\\
\midrule
500 & 0.950 & 0.617 & 1.000 & 705.3 & 661.8 & --\\
650 & 0.950 & 0.617 & 1.000 & 762.8 & 661.8 & --\\
800 & 1.000 & 0.833 & 1.000 & 668.0 & 661.8 & 0.167\\
1000 & 1.000 & 0.833 & 1.000 & 701.4 & 661.8 & --\\
1500 & 1.000 & 0.833 & 1.000 & 784.7 & 661.8 & --\\
\bottomrule
\end{tabular}}
\end{table}

\section{Held-Out Transfer Details}

The 2D Heisenberg transfer uses $J_2\in\{0.20,0.35,0.65\}$, three shared seeds, and a 50-gate budget.  Full $\K_1+Y$ reaches 8/9 strict successes; random-greedy reaches 3/9.  The held-out $2\times3$ XXZ transfer uses $(J_2,\Delta)\in\{(0.30,0.50),(0.50,1.50),(0.70,0.75)\}$, three shared seeds, and the same 50-gate budget.  Full $\K_1+Y$ reaches 8/9 strict successes; random-greedy reaches 2/9; all structured noncommutator controls reach 0/9.  The $2\times4$ XXZ scaling stress uses the same held-out cells on $2\times4$ grids and reaches high fidelity but zero strict successes.

\begin{table}[!t]
\caption{\label{tab:transferdetail}Held-out transfer and scaling details.}
\adjustbox{max width=\textwidth}{\footnotesize%
\begin{tabular}{llrrrr}
\toprule
\rowcolor{hdrcol} Family & method & strict & mean margin & mean error & fidelity\\
\midrule
2x3 Heisenberg J1-J2 & base & 0/9 & -8.174 & 8.473 & $\sim0$\\
2x3 Heisenberg J1-J2 & $\K_1+Y$ & 8/9 & +0.128 & 0.170 & 0.979\\
2x3 Heisenberg J1-J2 & budget random-greedy & 3/9 & -0.093 & 0.392 & 0.922\\
2x3 XXZ J1-J2 & base & 0/9 & -7.195 & 7.427 & $\sim0$\\
2x3 XXZ J1-J2 & $\K_1+Y$ & 8/9 & +0.061 & 0.171 & 0.974\\
2x3 XXZ J1-J2 & budget random-greedy & 2/9 & -0.059 & 0.291 & 0.952\\
2x3 XXZ J1-J2 & Hamiltonian layer & 0/9 & -2.672 & 2.904 & $\sim0$\\
2x3 XXZ J1-J2 & old $\yzzy+Y$ & 0/9 & -0.464 & 0.696 & 0.860\\
2x4 XXZ J1-J2 & base & 0/9 & -13.541 & 13.738 & 0.002\\
2x4 XXZ J1-J2 & full $\K_1+Y$, high NFEV & 0/9 & -0.197 & 0.394 & 0.917\\
2x4 XXZ J1-J2 & old $\yzzy+Y$ & 0/9 & -0.757 & 0.954 & 0.796\\
2x4 XXZ J1-J2 & random-greedy sample & 0/9 & -1.818 & 2.015 & 0.488\\
\bottomrule
\end{tabular}}
\end{table}

\section{Application-QUBO Screening Details}
\label{app:qubo}

The application-QUBO harness uses exact diagonal energies for fielded QAOA cost layers
\[
H_C=\sum_{(i,j)}J_{ij}Z_iZ_j+\sum_i h_iZ_i,
\]
with an $X$ mixer.  Weighted MIS uses the QUBO
$-\sum_i w_ix_i+A\sum_{(i,j)\in E}x_ix_j$, MVC uses
$\sum_i x_i+A\sum_{(i,j)\in E}(1-x_i)(1-x_j)$, and weighted MaxCut is the $ZZ$-only baseline.  The QUBO runs report optimal-subspace overlap rather than a single ground-state fidelity because diagonal optima can be degenerate.

\begin{figure}[!t]
\centering
\includegraphics[width=0.82\textwidth]{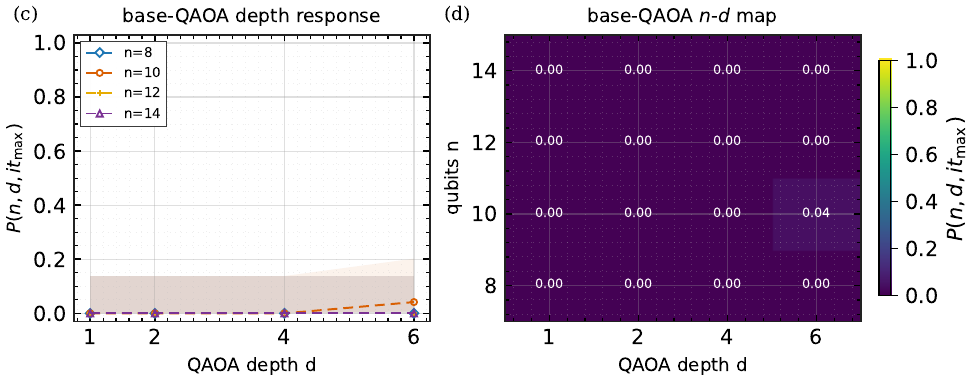}
\caption{\label{fig:mis_reachability_phase_baseqaoa}Appendix panels for the structured weighted-MIS reachability atlas in Fig.~\ref{fig:mis_reachability_phase}.  Panel (c) shows the base-QAOA depth response, and panel (d) shows the base-QAOA $n$-$d$ map at the replayed NFEV budgets.  These diagnostics explain why the main text emphasizes the local-field completion arm: base QAOA remains outside the strict low-energy window up to $p=6$ for almost all tested cells, while the commutator-predicted local-$Y$ completion shifts the arrival boundary upward.}
\end{figure}

\begin{figure}[!t]
\centering
\includegraphics[width=0.90\textwidth]{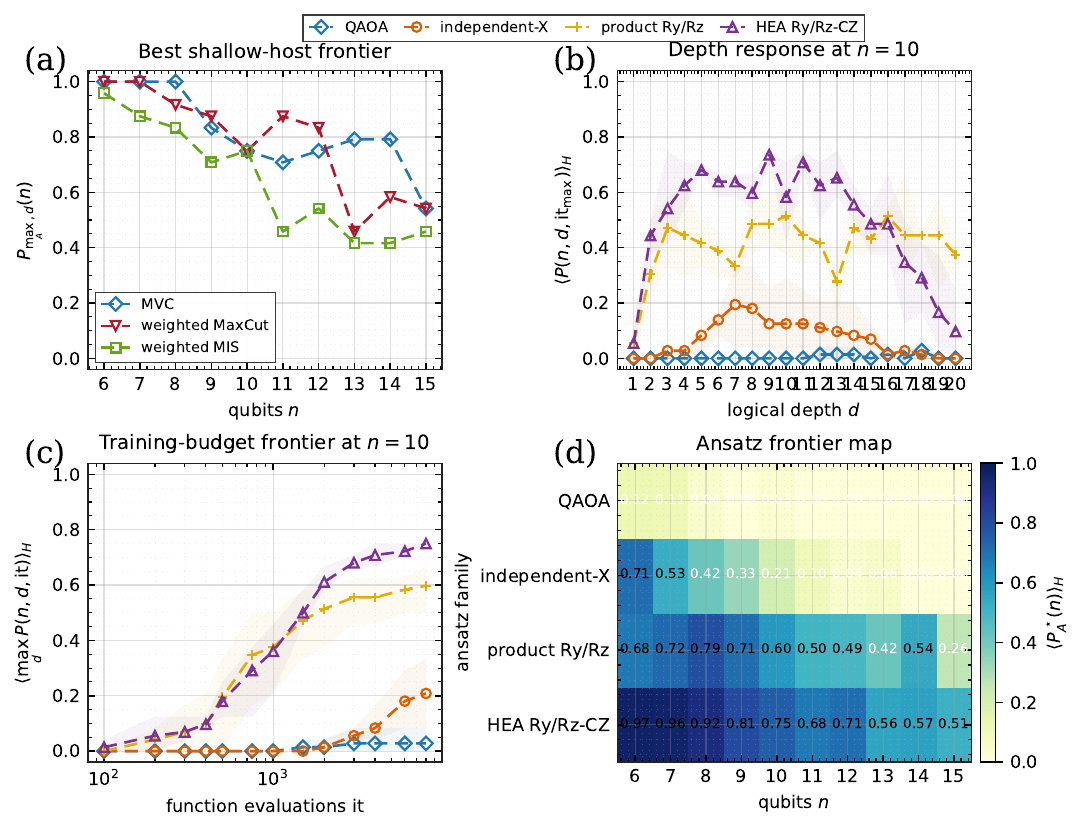}
\caption{\label{fig:ansatz_zoo_compact}Compact two-by-two summary of the integrated powered ansatz-zoo reachability sweep in main Fig.~3.  Panel (a) collapses over ordinary ansatz families to show the best shallow-host size frontier for each Hamiltonian family over $n=6,\ldots,15$.  Panels (b) and (c) use the dense $n=10$ depth--budget slice to show the average depth response and NFEV-budget frontier with a min--max envelope over Hamiltonian families, and panel (d) summarizes the average ansatz-family frontier map.  This compact view is useful for reading the global trend, while main Fig.~3 keeps the family-by-family decomposition in the main text.}
\end{figure}

\begin{figure}[!t]
\centering
\includegraphics[width=0.98\textwidth]{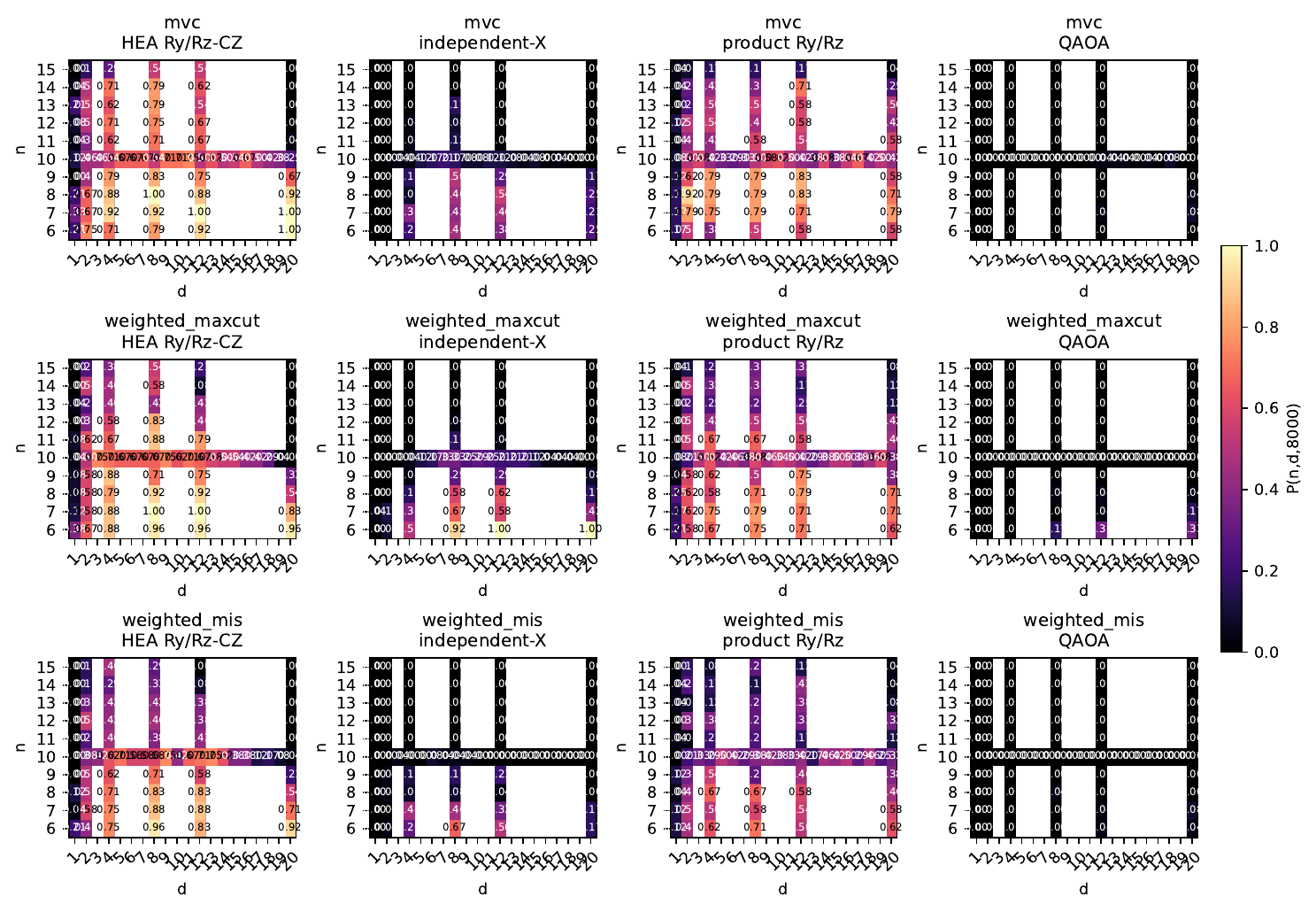}
\caption{\label{fig:ansatz_zoo_decomposed}Decomposed integrated multi-family ansatz-zoo reachability sweep supporting main Fig.~3.  Each row is a Hamiltonian family and each column is an ordinary shallow ansatz family.  Entries show the strict-window arrival probability $P_A(n,d,\mathrm{it}_{\max})$ over every integer size $n=6,\ldots,15$ and logical depth at the largest replayed budget, with the $n=10$ row densified in depth.  The decomposition separates effects collapsed in the main frontier plot: QAOA remains low across most of the tested grid, independent-$X$ gains local mixer freedom but loses rapidly with size, product circuits can solve many diagonal cells without entanglement, and HEA hosts are strongest but still show nontrivial depth and size thresholds.}
\end{figure}

\begin{figure}[!t]
\centering
\includegraphics[width=0.87\textwidth]{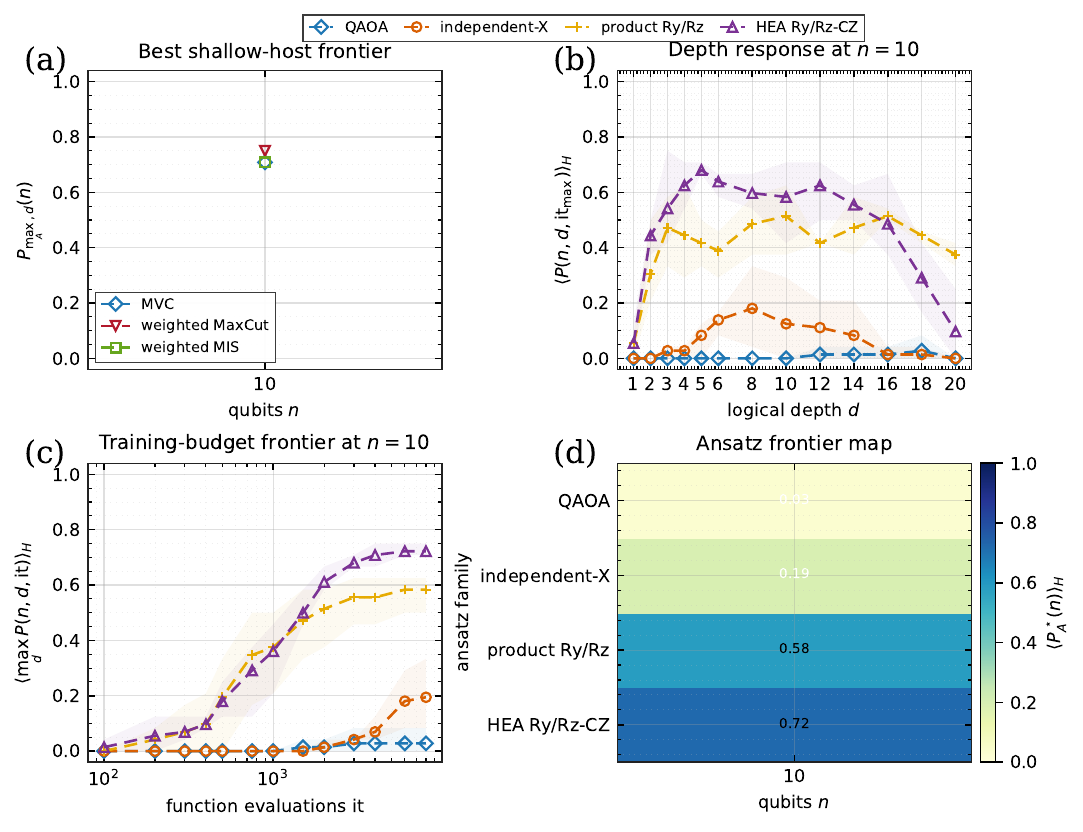}
\caption{\label{fig:dense_n10_compact}Dense $n=10$ depth--budget reachability slice supporting Table~\ref{tab:dense_n10_frontier}.  Compared with the balanced continuous sweep, this slice adds intermediate depths and budgets while keeping the same families, ansatz hosts, six instances, and four shared seeds.  The strongest frontier remains HEA/product-dominated, while QAOA and independent-$X$ stay lower even when the depth grid is made much denser.}
\end{figure}

\begin{figure}[!t]
\centering
\includegraphics[width=0.315\textwidth]{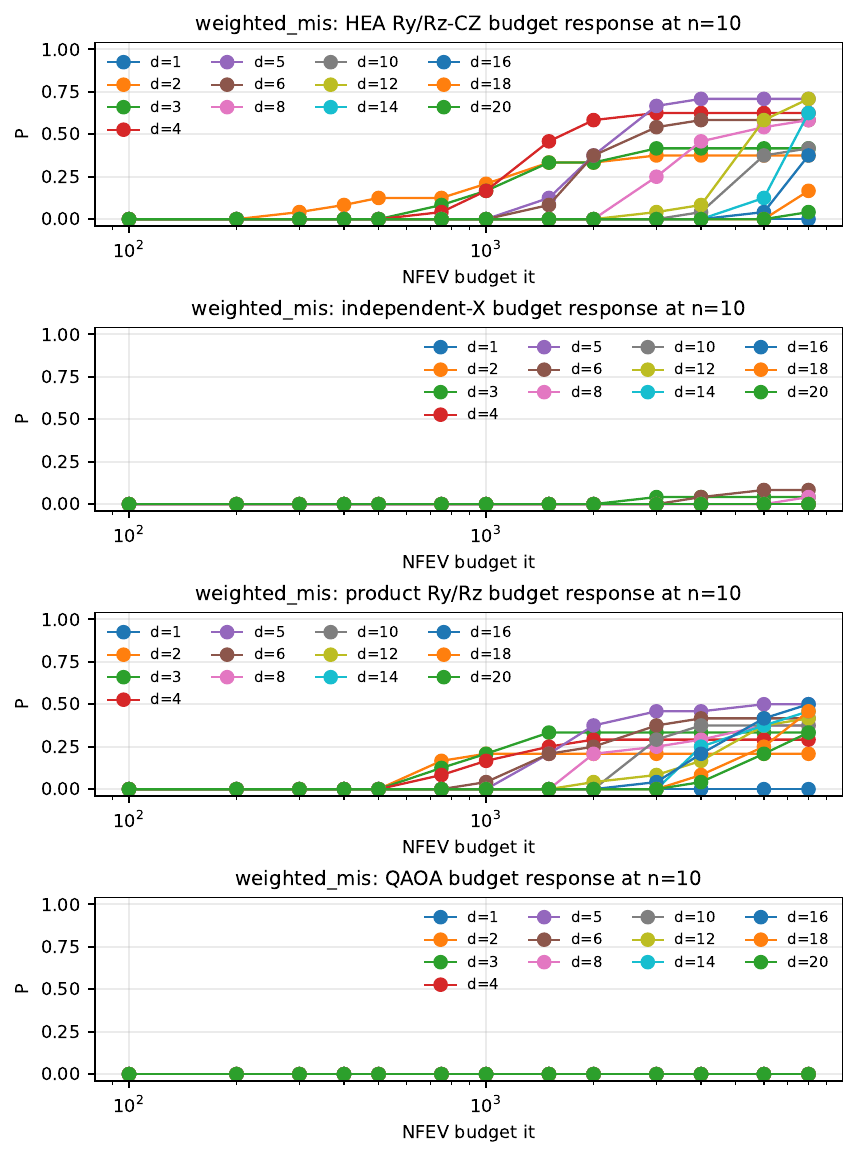}\hfill
\includegraphics[width=0.315\textwidth]{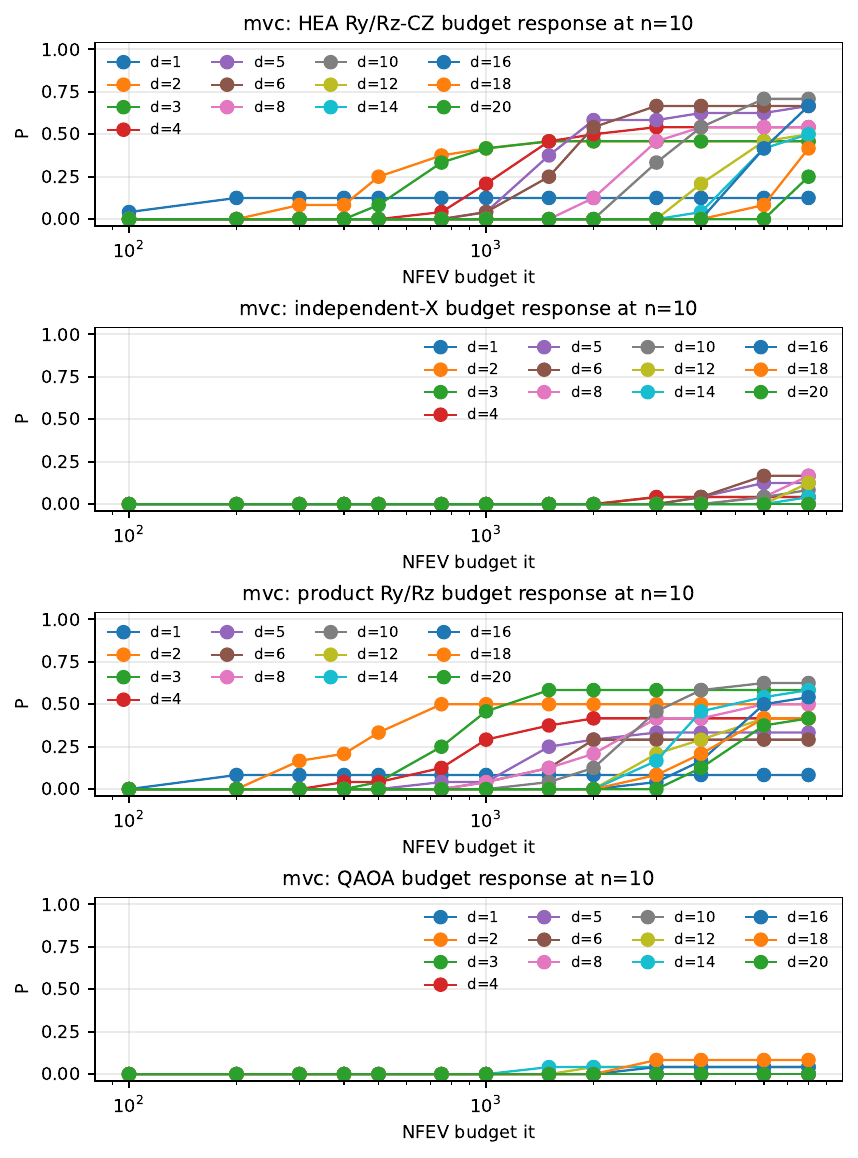}\hfill
\includegraphics[width=0.315\textwidth]{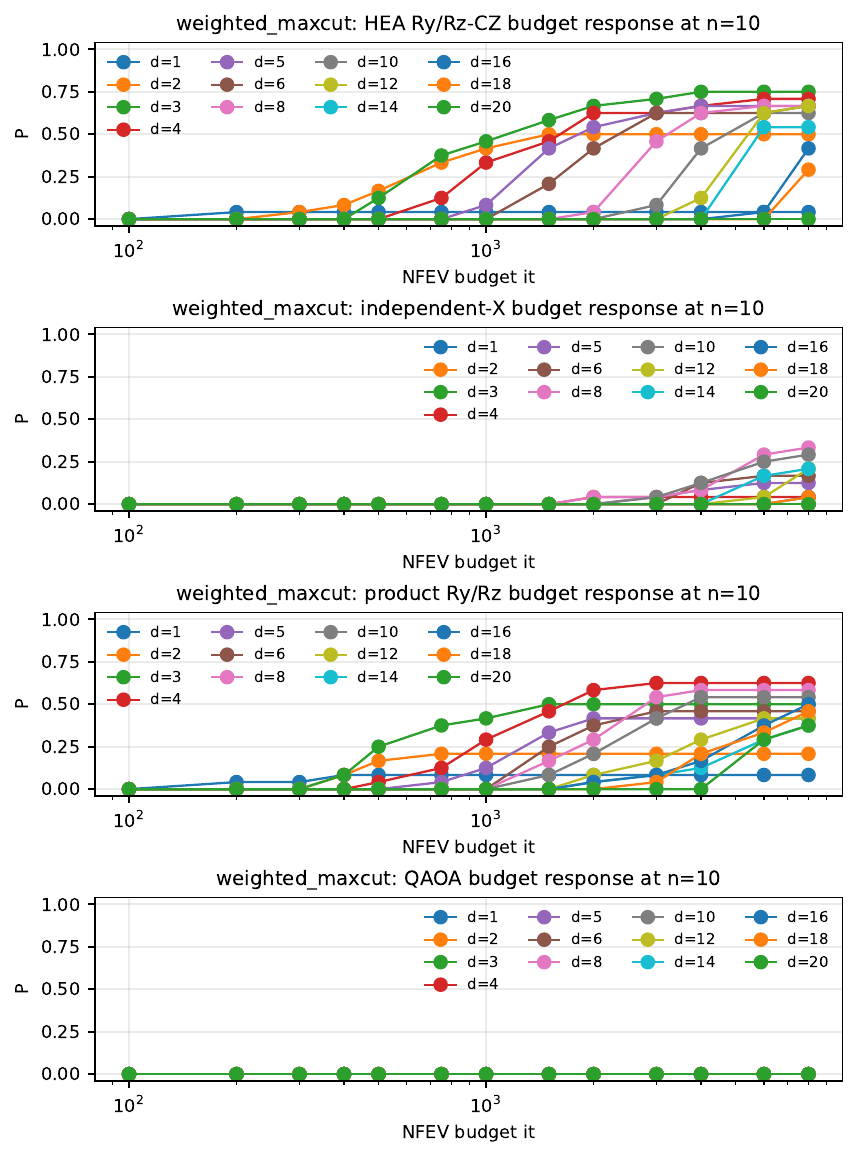}
\caption{\label{fig:dense_n10_budget}Dense $n=10$ budget curves for weighted MIS, MVC, and weighted MaxCut.  For each ansatz family and budget, the curve takes the best tested depth on the dense grid.  These panels show that extra optimizer budget improves some hosts but does not turn the weak QAOA frontier into the HEA/product frontier, supporting the reachability interpretation rather than a pure optimizer-budget explanation.}
\end{figure}

The dense $n=10$ slice in Table~\ref{tab:dense_n10_frontier} is visualized in Figs.~\ref{fig:dense_n10_compact} and~\ref{fig:dense_n10_budget}.  It uses the same three Hamiltonian families and four ansatz hosts as the main-text ansatz-zoo frontier, but densifies the logical-depth grid to every integer $d=1,\ldots,20$ and the replayed budget grid to $\{100,200,300,400,500,750,1000,1500,2000,3000,4000,6000,8000\}$.  This is a local robustness check on the depth frontier rather than a new algorithmic arm.

\begin{figure}[!t]
\centering
\includegraphics[width=0.98\textwidth]{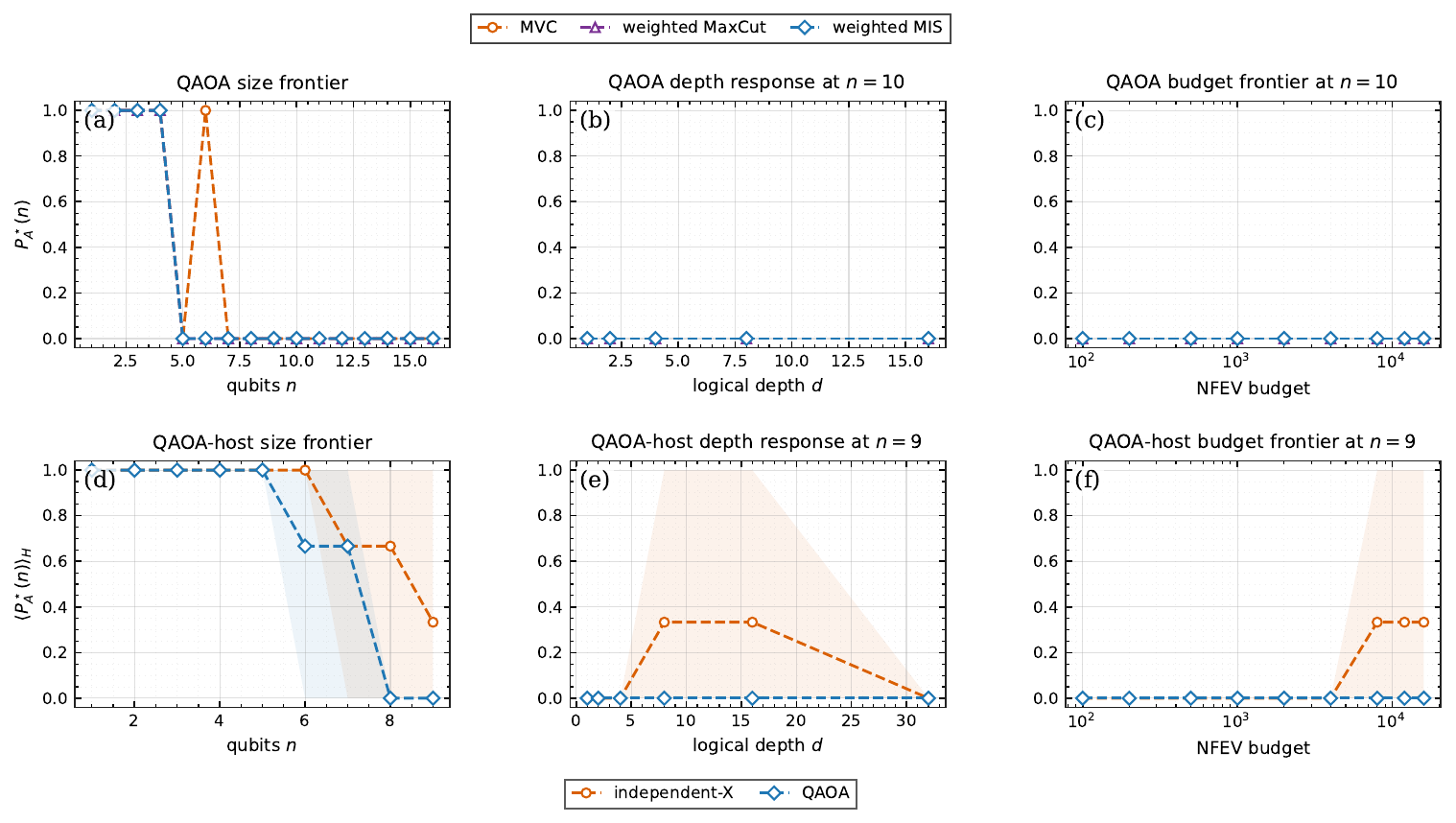}
\caption{\label{fig:extended_reachability_combined}Combined extended reachability pilots supporting main Fig.~3.  Top row: a QAOA-only high-size stress pass over weighted MIS, MVC, and weighted MaxCut for every $n=1,\ldots,16$, depths $d\in\{1,2,4,8,16\}$, and replayed budgets up to $1.6\times10^4$ function evaluations.  Bottom row: a long-depth QAOA-host slice over $n=1,\ldots,9$, QAOA and independent-$X$ hosts, depths $d\in\{1,2,4,8,16,32\}$, and the same replayed-budget range; curves average over Hamiltonian families and translucent bands show the family min--max envelope.  These pilots are not powered statistical comparisons: the high-size row uses one instance and one seed per family-size cell, and the long-depth row is a complete but narrower host slice.  Their purpose is to show that strict-window arrival is easy at very small $n$ but remains sharply size-, depth-, budget-, family-, and host-dependent even when the QAOA-style template is pushed to higher depths or given independent local mixer angles.}
\end{figure}

To extend the phenomenology beyond the balanced $n\le 14$ grid in the main text, we also ran two narrower scaling pilots and combine them in Fig.~\ref{fig:extended_reachability_combined}.  The top row uses a single QAOA host over every size $n=1,\ldots,16$ with depths up to $d=16$ and replayed budgets up to $1.6\times 10^4$ function evaluations.  This is not used as a powered statistical claim because it uses one instance and one seed per family-size cell; its role is to show that the small-$n$ easy regime gives way quickly to a near-zero strict-window arrival regime under the same shallow QAOA parameterization.  The bottom row keeps two QAOA-style hosts, QAOA and independent-$X$, and pushes the depth grid to $d=32$ on the complete $n=1,\ldots,9$ slice.  Together these pilots support the same interpretation as main Fig.~3: shallow low-energy arrival is a structured finite-size phenomenon, not a monotone guarantee that increasing depth or local mixer freedom will steadily close the strict low-energy window.

\begin{table}[!t]
\caption{\label{tab:qubo_gate}Application-QUBO gate summary with the front-loaded rule.  The rule is local $Y$ for fielded QUBOs and edge $\yzzy$ for pure/near-pure $ZZ$ costs; controls exclude commutator-axis arms.}
\adjustbox{max width=\textwidth}{\footnotesize%
\begin{tabular}{llrrrrrr}
\toprule
\rowcolor{hdrcol} Family & size & repair-only & healthy grad & full $\K_1$ & fixed rule & best control & rule-control margin\\
\midrule
weighted MIS & $n=8$ & 12 & 12/12 & 9/12 & 11/12 & 6/12 & +0.274\\
weighted MIS & $n=10$ & 6 & 6/6 & 2/6 & 5/6 & 0/6 & +1.242\\
MVC & $n=8$ & 12 & 12/12 & 6/12 & 6/12 & 0/12 & +0.237\\
MVC & $n=10$ & 6 & 6/6 & 0/6 & 2/6 & 0/6 & +0.075\\
weighted MaxCut & $n=8$ & 12 & 12/12 & 0/12 & 0/12 & 0/12 & -0.004\\
weighted MaxCut & $n=10$ & 6 & 6/6 & 0/6 & 0/6 & 0/6 & -0.101\\
\bottomrule
\end{tabular}}
\end{table}

\begin{table}[!t]
\caption{\label{tab:qubo_methods}Application-QUBO method highlights.  Cost is censored first-hit NFEV.}
\adjustbox{max width=\textwidth}{\footnotesize%
\begin{tabular}{llrrrr}
\toprule
\rowcolor{hdrcol} Family/size & method & strict & mean margin & opt. overlap & cost\\
\midrule
weighted MIS $n=8$ & QAOA $p=1$ & 0/12 & -5.741 & 0.057 & 6000\\
weighted MIS $n=8$ & QAOA $p=2$ & 0/12 & -4.023 & 0.143 & 6000\\
weighted MIS $n=8$ & QAOA $p=3$ & 0/12 & -3.475 & 0.186 & 6000\\
weighted MIS $n=8$ & QAOA $p=4$ & 0/12 & -2.513 & 0.309 & 6000\\
weighted MIS $n=8$ & independent-$X$ $p=1$ & 0/12 & -4.075 & 0.208 & 6000\\
weighted MIS $n=8$ & independent-$X$ $p=2$ & 2/12 & -1.565 & 0.703 & 5250\\
weighted MIS $n=8$ & full $\K_1$ & 9/12 & -0.283 & 0.795 & 1859\\
weighted MIS $n=8$ & fixed-rule local $Y$ & 11/12 & +0.149 & 0.995 & 1286\\
weighted MIS $n=8$ & local $X$ / local $Z$ & 0/12 & -3.792/-5.741 & 0.278/0.057 & 6000\\
weighted MIS $n=8$ & random local $X/Z$ & 0/12 & -4.932 & 0.181 & 6000\\
MVC $n=8$ & QAOA $p=1$ & 0/12 & -1.986 & 0.063 & 6000\\
MVC $n=8$ & QAOA $p=2$ & 0/12 & -1.487 & 0.144 & 6000\\
MVC $n=8$ & full $\K_1$ & 6/12 & -0.148 & 0.754 & 3115\\
MVC $n=8$ & fixed-rule local $Y$ & 6/12 & -0.272 & 0.690 & 3111\\
MVC $n=8$ & local $X$ / local $Z$ & 0/12 & -1.924/-1.986 & 0.109/0.063 & 6000\\
MVC $n=8$ & random local $X/Z$ & 0/12 & -1.971 & 0.074 & 6000\\
MaxCut $n=8$ & QAOA $p=1$ & 0/12 & -5.846 & 0.081 & 6000\\
MaxCut $n=8$ & QAOA $p=2$ & 0/12 & -5.370 & 0.109 & 6000\\
MaxCut $n=8$ & fixed-rule edge $\yzzy$ & 0/12 & -0.846 & 0.566 & 6000\\
weighted MIS $n=10$ & QAOA $p=3$ & 0/6 & -4.538 & 0.013 & 7000\\
weighted MIS $n=10$ & QAOA $p=4$ & 0/6 & -3.966 & 0.024 & 7000\\
weighted MIS $n=10$ & independent-$X$ $p=1$ & 0/6 & -4.773 & 0.188 & 7000\\
weighted MIS $n=10$ & independent-$X$ $p=2$ & 1/6 & -2.234 & 0.486 & 6504\\
weighted MIS $n=10$ & fixed-rule local $Y$ & 5/6 & +0.010 & 0.959 & 1536\\
MVC $n=10$ & fixed-rule local $Y$ & 2/6 & -0.649 & 0.480 & 4814\\
MaxCut $n=10$ & fixed-rule edge $\yzzy$ & 0/6 & -1.824 & 0.258 & 7000\\
\bottomrule
\end{tabular}}
\end{table}

\begin{table}[!t]
\caption{\label{tab:qubo_scaling}Powered weighted-MIS scaling detail corresponding to main Fig.~4.  Each row averages over 24 trials except where stated otherwise; cost is censored NFEV-to-target.}
\adjustbox{max width=\textwidth}{\footnotesize%
\begin{tabular}{llrrrr}
\toprule
\rowcolor{hdrcol} $n$ & method & strict & $P_{\rm opt}$ & margin & cost\\
\midrule
8 & QAOA $p=4$ & 0/24 & 0.122 & -3.154 & 6000\\
8 & QAOA $p=6$ & 0/24 & 0.218 & -2.415 & 6000\\
8 & independent-$X$ $p=2$ & 2/24 & 0.518 & -1.837 & 5597\\
8 & local-$Y$ rule & \best{17/24} & \best{0.811} & \best{-0.094} & \best{2324}\\
8 & full $\K_1$ & 9/24 & 0.539 & -0.976 & 4024\\
\addlinespace[1pt]
10 & QAOA $p=4$ & 0/24 & 0.121 & -3.946 & 6000\\
10 & QAOA $p=6$ & 1/24 & 0.153 & -3.414 & 5962\\
10 & independent-$X$ $p=2$ & 0/24 & 0.301 & -3.550 & 6000\\
10 & local-$Y$ rule & \best{19/24} & \best{0.889} & \best{-0.213} & \best{1717}\\
10 & full $\K_1$ & 11/24 & 0.564 & -1.620 & 3647\\
\addlinespace[1pt]
12 & QAOA $p=4$ & 0/24 & 0.038 & -6.949 & 6000\\
12 & QAOA $p=6$ & 0/24 & 0.086 & -6.363 & 6000\\
12 & independent-$X$ $p=2$ & 1/24 & 0.099 & -9.501 & 5841\\
12 & local-$Y$ rule & \best{11/24} & \best{0.561} & \best{-0.662} & \best{3727}\\
12 & full $\K_1$ & 5/24 & 0.278 & -2.244 & 5220\\
\addlinespace[1pt]
14 & QAOA $p=4$ & 0/24 & 0.012 & -9.913 & 6000\\
14 & QAOA $p=6$ & 0/24 & 0.015 & -9.359 & 6000\\
14 & independent-$X$ $p=2$ & 0/24 & 0.052 & -9.044 & 6000\\
14 & local-$Y$ rule & \best{7/24} & 0.364 & \best{-1.126} & \best{4625}\\
14 & full $\K_1$ & 6/24 & \best{0.468} & -1.698 & 5150\\
\bottomrule
\end{tabular}}
\end{table}

The powered scaling stress preserves and strengthens the qualitative ordering of the application runs: the front-loaded local-$Y$ completion improves the energy margin, strict success, and optimal-subspace overlap over QAOA $p=4,p=6$ and independent-$X$ controls at every tested size.  The result is still size-sensitive.  Strict success decays by $n=14$, and full $\K_1$ overtakes the fixed local-$Y$ completion in $P_{\rm opt}$ at $n=14$ while remaining slightly worse in strict success and cost.  We therefore use this as a powered finite-size scaling diagnostic, not as a theorem that one commutator block solves large weighted MIS.

\section{Expanded Host/Hamiltonian Aggregate Matrix}
\label{app:aggregatematrix}

Table~\ref{tab:closureselectormain} gives the aggregate strict-window counts summarized in the main text, resolving the same paired benchmark along both design axes: the Hamiltonian-family dimension (Panel A) and the host-ansatz dimension (Panel B). The full cell-level matrix is Table~\ref{tab:closureselectorfullmain}, and the per-host selector breakdown is Table~\ref{tab:closureselector} in Appendix~\ref{app:ansatzswap}.

\begin{table}[!t]
\caption{\label{tab:closureselectormain}Main matrix-style expanded host/Hamiltonian result.  Panel A aggregates over the seven host ansatz families for each Hamiltonian family; Panel B aggregates over the nine Hamiltonian settings for each host ansatz.  Entries are strict-window successes under shared seeds and a matched 56-gate insertion budget.  The ``Ctrl. env.'' column is the conservative non-deployable per-trial control envelope: for each host--Hamiltonian cell and seed it reports the strongest matched noncommutator arm among the tested controls before aggregation.  The final two columns are the paper's two predeclared commutator-closure selectors: a gradient-ranked selector and a trust-region energy-descent selector, both followed by full reoptimization.  Green marks a reported selector only when it is row-best; light red marks a matched noncommutator control that strictly exceeds both reported selectors in that aggregate.  Diagnostic headroom analyses are deferred to Appendix~\ref{app:ansatzswap}.}
\footnotesize
\setlength{\tabcolsep}{3.5pt}
\renewcommand{\arraystretch}{1.10}
\adjustbox{max width=\textwidth}{%
\begin{tabular}{@{}llrrrr@{}}
\multicolumn{6}{@{}l}{\textbf{Panel A: Hamiltonian-family dimension.}}\\[2pt]
\toprule
\rowcolor{hdrcol}
\textbf{Hamiltonian family} & \textbf{Concrete setting and dominant commutator motif} & \textbf{Base} & \textbf{Ctrl. env.} & \textbf{Ours: grad closure} & \textbf{Trust closure}\\
\midrule
\begin{minipage}[t]{0.15\textwidth}\raggedright TFIM\end{minipage} &
  \begin{minipage}[t]{0.46\textwidth}\raggedright $2\times3$ Ising grid with transverse field, $h\in\{0.6,1.2\}$; edge motif $[X,ZZ]\!\rightarrow\!\yzzy$.  This is the main mixed/boundary family because matched controls outperform both reported selectors on this family.\end{minipage} &
0/42 & \ctrlbetter{24/42} & 14/42 & 8/42\\
\addlinespace[1pt]
\begin{minipage}[t]{0.15\textwidth}\raggedright J1-J2 Ising\end{minipage} &
\begin{minipage}[t]{0.46\textwidth}\raggedright nearest-edge and diagonal $ZZ$ couplings with $J_2\in\{0.25,0.45,0.65\}$; Ising edge/diagonal commutator axes dominate.\end{minipage} &
0/63 & 53/63 & 58/63 & \ours{61/63}\\
\addlinespace[1pt]
\begin{minipage}[t]{0.15\textwidth}\raggedright Heisenberg J1-J2\end{minipage} &
\begin{minipage}[t]{0.46\textwidth}\raggedright $XX+YY+ZZ$ edge and diagonal couplings with $J_2\in\{0.20,0.35\}$; mixed commutator shell beyond a single Ising axis.\end{minipage} &
0/42 & 18/42 & \ours{32/42} & 27/42\\
\addlinespace[1pt]
\begin{minipage}[t]{0.15\textwidth}\raggedright XXZ J1-J2\end{minipage} &
\begin{minipage}[t]{0.46\textwidth}\raggedright anisotropic $XX+YY+\Delta ZZ$ couplings, $(J_2,\Delta)\in\{(0.30,0.50),(0.50,1.50)\}$; mixed shell with anisotropy-dependent axes.\end{minipage} &
0/42 & 14/42 & 29/42 & \ours{31/42}\\
\midrule
\textbf{Total} & seven hosts $\times$ nine Hamiltonian settings $\times$ three seeds &
\textbf{0/189} & \textbf{109/189} & \ours{133/189} & \textbf{127/189}\\
\bottomrule
\end{tabular}}
\vspace{4pt}

\adjustbox{max width=\textwidth}{%
\begin{tabular}{@{}llrrrr@{}}
\multicolumn{6}{@{}l}{\textbf{Panel B: host-ansatz dimension.}}\\[2pt]
\toprule
\rowcolor{hdrcol}
\textbf{Host ansatz} & \textbf{Circuit content} & \textbf{Base} & \textbf{Ctrl. env.} & \textbf{Ours: grad closure} & \textbf{Trust closure}\\
\midrule
Product rotations & independent local $R_Y$ only & 0/27 & 13/27 & 14/27 & \ours{15/27}\\
QAOA2D & $H$, shared $R_{ZZ}$ cost and $R_X$ mixer & 0/27 & \ctrlbetter{18/27} & 15/27 & 12/27\\
Brickwork HEA & local $R_Y/R_Z$ plus checkerboard CZ & 0/27 & 16/27 & \ours{22/27} & 20/27\\
HEA $X/Y$+CZ & local $R_X/R_Y$ plus checkerboard CZ & 0/27 & 18/27 & \ours{24/27} & \ours{24/27}\\
HEA $X/Z$+CZ & local $R_X/R_Z$ plus checkerboard CZ & 0/27 & 16/27 & \ours{23/27} & 20/27\\
HEA $X/Y/Z$+CZ & local $R_X/R_Y/R_Z$ plus checkerboard CZ & 0/27 & 16/27 & \ours{22/27} & 21/27\\
HVA-Heisenberg & local $R_Y$ plus shared $XX/YY/ZZ$ edge and diagonal rotations & 0/27 & 12/27 & 13/27 & \ours{15/27}\\
\midrule
\textbf{Total} & seven hosts $\times$ nine Hamiltonian settings $\times$ three seeds &
\textbf{0/189} & \textbf{109/189} & \ours{133/189} & \textbf{127/189}\\
\bottomrule
\end{tabular}}
\end{table}

\section{Frozen-Rule Transfer}
\label{app:transferfig}

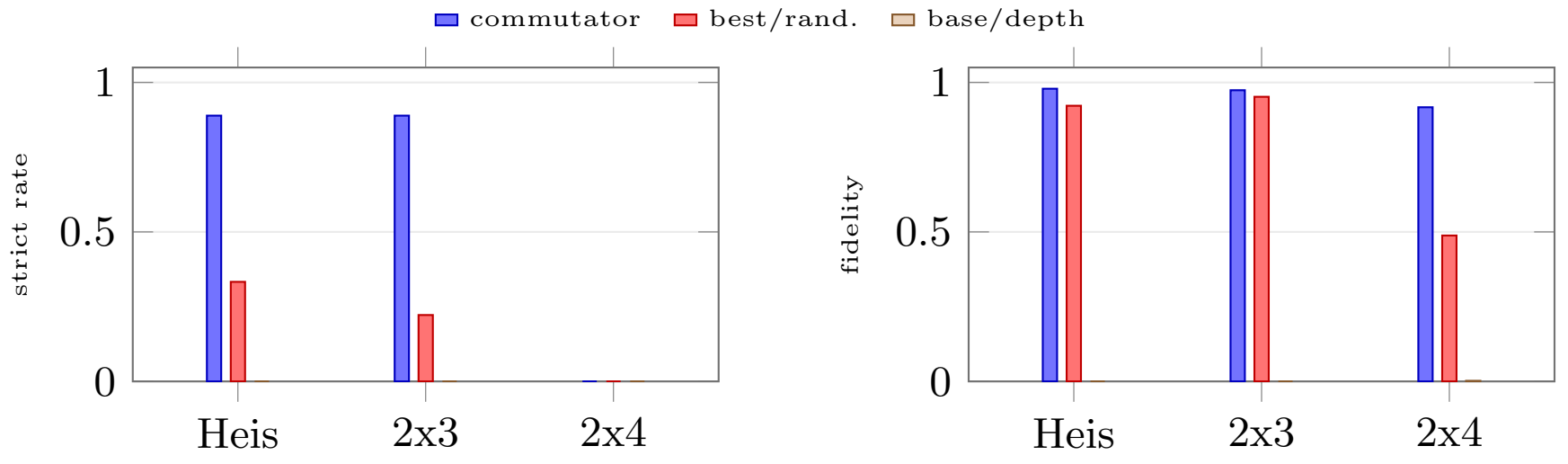
\begin{figure}[!t]
\centering
\begin{minipage}{0.92\columnwidth}
\centering
{\tiny
\tikz{\draw[fill=blue!55, draw=blue!75!black] (0,0) rectangle (0.16,0.08);} commutator\quad
\tikz{\draw[fill=red!55, draw=red!75!black] (0,0) rectangle (0.16,0.08);} best/rand.\quad
\tikz{\draw[fill=brown!35, draw=brown!70!black] (0,0) rectangle (0.16,0.08);} base/depth
}
\vspace{2pt}

\begin{minipage}{0.49\linewidth}
\centering
\begin{tikzpicture}
\begin{axis}[
  width=\linewidth,
  height=0.66\linewidth,
  ybar,
  ymin=0,ymax=1.05,
  ylabel={strict rate},
  ylabel style={font=\tiny},
  symbolic x coords={Heis,2x3,2x4},
  xtick=data,
  bar width=3pt,
  enlarge x limits=0.28,
  ymajorgrids=true,
  grid style={black!8},
  axis line style={black!55},
  tick style={black!55},
  tick label style={font=\scriptsize},
]
\addplot+[fill=blue!55, draw=blue!75!black] coordinates {(Heis,0.889) (2x3,0.889) (2x4,0.0)};
\addplot+[fill=red!55, draw=red!75!black] coordinates {(Heis,0.333) (2x3,0.222) (2x4,0.0)};
\addplot+[fill=brown!35, draw=brown!70!black] coordinates {(Heis,0.0) (2x3,0.0) (2x4,0.0)};
\end{axis}
\end{tikzpicture}
\end{minipage}
\hfill
\begin{minipage}{0.49\linewidth}
\centering
\begin{tikzpicture}
\begin{axis}[
  width=\linewidth,
  height=0.66\linewidth,
  ybar,
  ymin=0,ymax=1.05,
  ylabel={fidelity},
  ylabel style={font=\tiny},
  symbolic x coords={Heis,2x3,2x4},
  xtick=data,
  bar width=3pt,
  enlarge x limits=0.28,
  ymajorgrids=true,
  grid style={black!8},
  axis line style={black!55},
  tick style={black!55},
  tick label style={font=\scriptsize},
]
\addplot+[fill=blue!55, draw=blue!75!black] coordinates {(Heis,0.979) (2x3,0.974) (2x4,0.917)};
\addplot+[fill=red!55, draw=red!75!black] coordinates {(Heis,0.922) (2x3,0.952) (2x4,0.488)};
\addplot+[fill=brown!35, draw=brown!70!black] coordinates {(Heis,0.0) (2x3,0.0) (2x4,0.002)};
\end{axis}
\end{tikzpicture}
\end{minipage}
\end{minipage}
\caption{\label{fig:transfer}Frozen first-shell transfer and scaling.  The fixed first-shell completion transfers to 2D Heisenberg and held-out 2x3 XXZ with 8/9 strict successes.  On 2x4 XXZ, the same frozen first-shell completion is boundary-positive: fidelity remains high but strict success is not reached.  This transfer figure supports the commutator mechanism; it is complementary to, and narrower than, the broad closure-selector aggregate summarized in the main text.}
\end{figure}

Figure~\ref{fig:transfer} shows the frozen first-shell transfer summarized in the main text: the fixed first-shell completion transfers to 2D Heisenberg and held-out $2\times3$ XXZ with $8/9$ strict successes each, and is boundary-positive on $2\times4$ XXZ, where fidelity remains high but strict success is not reached. The corresponding numerical detail is in Table~\ref{tab:transferdetail}.

\section{Application-QUBO Reachability, Screening, and Scaling Detail}
\label{app:appqubodetail}

\begin{figure}[!t]
\centering
\includegraphics[width=\columnwidth]{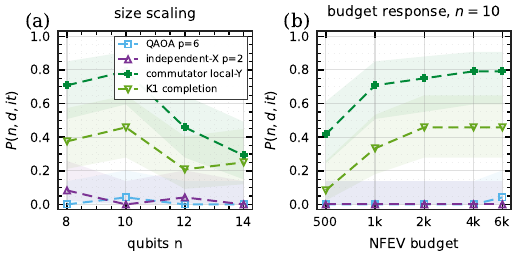}
\caption{\label{fig:mis_reachability_phase}Structured weighted-MIS reachability after adding the fixed local-field completion arm, redrawn with the same scaling-plot style as main Fig.~3.  The sweep replays the powered weighted-MIS trials at smaller function-evaluation budgets using recorded first-hit NFEV, yielding $P(n,d,\mathrm{it})$.  Panel (a) shows size scaling at the maximum budget, and panel (b) shows budget response at $n=10$.  Transparent bands denote Wilson binomial confidence intervals over the 24 paired trials contributing to each curve point.  Base-QAOA depth and heat-map diagnostics are given in Appendix~\ref{app:qubo}, Fig.~\ref{fig:mis_reachability_phase_baseqaoa}.}
\end{figure}

This appendix collects the application-QUBO evidence summarized in the main text. Main Fig.~3 is the powered pre-completion phenomenology, and Table~\ref{tab:dense_n10_frontier} gives its dense $n=10$ depth--budget slice. Tables~\ref{tab:qubo_main} and~\ref{tab:qubo_broad} give the front-loaded-rule screen and its balanced redraw; Fig.~\ref{fig:mis_reachability_phase} is the structured weighted-MIS reachability atlas after adding the local-field completion arm; and Tables~\ref{tab:mis_powered_scaling} and~\ref{tab:mis_depth} give the powered weighted-MIS scaling summary and the earlier focused baselines. Further decomposed atlases, dense-slice plots, and budget curves are in Appendix~\ref{app:qubo}.

The QUBO instance generator is seeded and paired across all method arms.  For weighted MIS and minimum vertex cover, the six-instance powered sweeps cycle through Erdős--Rényi sparse graphs with edge probability $0.28$, Erdős--Rényi denser graphs with edge probability $0.48$, random geometric graphs with radius $0.48$ for $n\le8$ and $0.40$ otherwise, and an irregular graph family formed from an Erdős--Rényi graph with edge probability $0.36$ plus a star-tail backbone.  Weighted MaxCut uses the same sparse, denser, and geometric graph families, replacing the irregular family by a regularized ring-plus-chords family.  Integer node or edge weights are drawn uniformly from $\{1,2,3,4\}$.  Weighted MIS uses QUBO penalty $A=\max_i w_i+1$ for edge violations, minimum vertex cover uses unit selection cost and penalty $A=2$, and weighted MaxCut is the corresponding $ZZ$-only Ising objective.  The generated edges, weights, Ising fields $h_i$, and couplings $J_{ij}$ are archived in the instance/trial CSV files used to make the figures.

Table~\ref{tab:qubo_main} gives the application-QUBO screen, and Table~\ref{tab:qubo_broad} gives a balanced $n=8$ redraw.  The screen is not the main application claim: the balanced redraw lowers weighted MIS to $10/18$ and gives a rule-minus-best-control margin interval that crosses zero.  The robust application evidence therefore comes from the powered $P_{\rm opt}$ scaling separation below, not from the initial $11/12$ screen.  Before adding any commutator completion, we ask the phenomenological question posed by the main-text arrival-probability definition: how does low-energy arrival depend on Hamiltonian family, system size, ansatz family, logical depth, and training budget?  Main Fig.~3 answers this question with no completion arm.  It sweeps three diagonal Ising/QUBO families---weighted MIS, minimum vertex cover, and weighted MaxCut---and four shallow ansatz families: global-angle QAOA, an independent-$X$ QAOA variant, a separable product Ry/Rz circuit, and a line-brickwork HEA Ry/Rz-CZ circuit.  The powered trend pass uses every integer size $n=6,\ldots,15$, six graph instances and four shared optimizer seeds per family-size cell, one multistart per optimization cell, and replayed budgets up to $8000$ function evaluations.  For the size frontier we use $d\in\{1,2,4,8,12,20\}$; for the $n=10$ depth-response and budget-frontier columns, we use every integer depth $d=1,\ldots,20$ and 13 replayed budgets from $100$ to $8000$ function evaluations.  Each plotted family--size--ansatz--depth point aggregates 24 paired trials, and main Fig.~3 contains 21312 optimized cells after adding the $n=15$ slice and replacing the sparse $n=10$ slice with the dense one.  This sweep tests whether low-energy-arrival failure is merely a weak-baseline artifact.  It is not.  QAOA is weak across the three families, while product and HEA hosts substantially raise the arrival frontier.  Averaged over the 30 family-size frontiers, the best strict-window arrival probability is $0.743$ for HEA Ry/Rz-CZ, $0.571$ for product Ry/Rz, $0.239$ for independent-$X$, and $0.031$ for global-angle QAOA.  At $n=15$, the strongest shallow-host frontier still reaches $11/24$ trials on weighted MIS, $13/24$ on MVC, and $13/24$ on weighted MaxCut, but all three families remain in a partial-success regime.  The frontier remains depth-, Hamiltonian-, and budget-sensitive: independent-$X$ decays rapidly beyond $n=8$--$10$, product circuits plateau at partial success, and even the strongest HEA frontier falls to partial success at $n=15$.  Stronger shallow ansatz families therefore improve, but do not erase, the empirical arrival boundary.  Appendix~\ref{app:qubo} gives a compact summary view plus the decomposed ansatz-by-ansatz heatmaps and budget curves.
The dense $n=10$ slice checks that the main-text depth response is not a sparse-grid artifact.  It sweeps every integer depth $d=1,\ldots,20$ and replay budgets up to $8000$ function evaluations, again using six instances and four shared seeds per family--ansatz--depth point.  This slice contributes 5760 optimized cells and replaces the sparse $n=10$ rows in main Fig.~3.  Table~\ref{tab:dense_n10_frontier} reports the best strict-window frontier at the largest budget.  The dense slice preserves the qualitative ordering from the continuous sweep: HEA and product hosts are substantially stronger than QAOA, independent-$X$ gives partial local-mixer lift, and the best depth is family- and ansatz-dependent rather than monotone.  For example, at $n=10$ HEA reaches $18/24$ on weighted MIS, $18/24$ on MVC, and $18/24$ on weighted MaxCut, whereas global-angle QAOA reaches $0/24$, $2/24$, and $0/24$.  Appendix~\ref{app:qubo} shows the same dense frontier as standalone plots.
\begin{table}[!t]
\caption{\label{tab:dense_n10_frontier}Dense $n=10$ depth--budget reachability slice before commutator completion.  For each Hamiltonian family and ansatz family, the table reports the best tested depth at budget $8000$, strict-window arrivals out of 24 paired trials, the Wilson interval for $P_A$, and the mean optimal-subspace overlap and energy error at the cap.  This slice densifies the depth grid of main Fig.~3 without changing the algorithmic arms.}
\footnotesize
\setlength{\tabcolsep}{4pt}
\adjustbox{max width=\textwidth}{\footnotesize%
\begin{tabular}{llrrrrrr}
\toprule
\rowcolor{hdrcol} Family & ansatz & best $d$ & strict & $P_A$ & Wilson CI & opt. overlap & energy error\\
\midrule
weighted MIS & HEA Ry/Rz-CZ & 9 & \best{18/24} & \best{0.750} & $[0.551,0.880]$ & 0.789 & 0.463\\
weighted MIS & product Ry/Rz & 5 & 12/24 & 0.500 & $[0.314,0.686]$ & 0.500 & 1.625\\
weighted MIS & independent-$X$ & 6 & 2/24 & 0.083 & $[0.023,0.259]$ & 0.390 & 3.252\\
weighted MIS & QAOA & 14 & 0/24 & 0.000 & $[0.000,0.138]$ & 0.121 & 9.952\\
\addlinespace[2pt]
MVC & HEA Ry/Rz-CZ & 11 & \best{18/24} & \best{0.750} & $[0.551,0.880]$ & 0.749 & 0.253\\
MVC & product Ry/Rz & 17 & 16/24 & 0.667 & $[0.467,0.820]$ & 0.667 & 0.333\\
MVC & independent-$X$ & 7 & 5/24 & 0.208 & $[0.092,0.405]$ & 0.574 & 1.556\\
MVC & QAOA & 18 & 2/24 & 0.083 & $[0.023,0.259]$ & 0.276 & 3.981\\
\addlinespace[2pt]
weighted MaxCut & HEA Ry/Rz-CZ & 3 & \best{18/24} & \best{0.750} & $[0.551,0.880]$ & 0.750 & 0.333\\
weighted MaxCut & product Ry/Rz & 4 & 15/24 & 0.625 & $[0.427,0.788]$ & 0.625 & 0.875\\
weighted MaxCut & independent-$X$ & 8 & 8/24 & 0.333 & $[0.180,0.533]$ & 0.552 & 2.263\\
weighted MaxCut & QAOA & 19 & 0/24 & 0.000 & $[0.000,0.138]$ & 0.186 & 7.270\\
\bottomrule
\end{tabular}}
\end{table}
\begin{table}[!t]
\caption{\label{tab:qubo_main}Application-QUBO transfer under a front-loaded sublayer rule fixed before the reported matched-init runs.  All entries are repair-only trials: QAOA $p=1$ and $p=2$ both fail the strict window.  The gradient signal $g$ is the mean squared finite-difference energy gradient at random QAOA $p=1$ parameters; the retained-cell threshold is $10^{-4}$.  These rows form a small screen ($4$ instances with up to $3$ seeds per cell); the balanced $n=8$ redraw in Table~\ref{tab:qubo_broad} gives lower strict rates and is treated as the robust small-$n$ estimate.}
\footnotesize
\adjustbox{max width=\columnwidth}{\footnotesize%
\begin{tabular}{llrrrlllr}
\toprule
\rowcolor{hdrcol} Family & $n$ & trials & $g_{\min}$ & $\bar g$ & base $p{=}1,2$ & rule & ctrl. & $\Delta m$\\
\midrule
weighted MIS & 8 & 12 & 21.9 & 142.7 & 0/12,0/12 & 11/12 & 6/12 & +0.274\\
weighted MIS & 10 & 6 & 105.8 & 1056.6 & 0/6,0/6 & 5/6 & 0/6 & +1.242\\
\addlinespace[2pt]
MVC & 8 & 12 & 4.04 & 56.7 & 0/12,0/12 & 6/12 & 0/12 & +0.237\\
MVC & 10 & 6 & 12.2 & 95.9 & 0/6,0/6 & 2/6 & 0/6 & +0.075\\
\addlinespace[2pt]
weighted MaxCut & 8 & 12 & 28.0 & 54.0 & 0/12,0/12 & 0/12 & 0/12 & -0.004\\
weighted MaxCut & 10 & 6 & 37.0 & 94.7 & 0/6,0/6 & 0/6 & 0/6 & -0.101\\
\bottomrule
\end{tabular}}
\end{table}
\begin{table}[!t]
\caption{\label{tab:qubo_broad}Balanced redraw at $n=8$ (six instances $\times$ three seeds $=18$ trials per family), refining the screen of Table~\ref{tab:qubo_main}.  ``rule'' is the predeclared completion (local $Y$ for MIS/MVC; edge $\yzzy$ for MaxCut).  $\Delta m$ is the per-instance margin of the rule minus the \emph{strongest} control, with a bootstrap $95\%$ CI; parameter-matched QAOA reaches $0/18$ in every family.}
\footnotesize
\adjustbox{max width=\columnwidth}{\footnotesize%
\begin{tabular}{lrrrrr}
\toprule
\rowcolor{hdrcol} Family & rule & edge $\xyyx$ & edge $\xxyy$ & indep-$X$ $p2$ & $\Delta m$ (rule$-$best ctrl)\\
\midrule
weighted MIS & \best{10/18} & 2/18 & 3/18 & 2/18 & $-0.06\,[-0.60,0.43]$\\
MVC & 3/18 & 3/18 & 3/18 & 0/18 & $-0.05\,[-0.34,0.20]$\\
weighted MaxCut & 3/18 & 0/18 & 0/18 & 0/18 & $-0.25\,[-0.54,0.00]$\\
\bottomrule
\end{tabular}}
\end{table}
The powered result asks whether a structured QAOA-family repair can shift this boundary.  Main Fig.~4, Fig.~\ref{fig:mis_reachability_phase}, and Table~\ref{tab:mis_powered_scaling} report the weighted-MIS scaling gate with six instances and four shared seeds at each integer $n=8,\ldots,14$, a common $6000$-evaluation cap, and QAOA controls up to $p=6$; Fig.~\ref{fig:mis_reachability_phase} and Tables~\ref{tab:qubo_scaling} and~\ref{tab:mis_powered_scaling} display the even sizes $n=8,10,12,14$.  The sublayer is fixed before training: fielded QUBOs select local $Y$ from $[X,Z]$.  Local-$Y$ remains the strict-success winner among the QAOA-style structured-family controls at every tested size and maintains a large optimal-subspace sampling advantage over QAOA up to $p=6$ and independent-$X$.  Together with main Fig.~3, this result distinguishes two questions.  Generic product and HEA hosts can already be strong on diagonal weighted MIS; the commutator local-$Y$ arm asks whether a Hamiltonian-aware, fixed, shallow completion can move a problem-informed QAOA host toward that reachability regime without abandoning the host.  Replaying the recorded first-hit NFEV at smaller budgets gives an empirical arrival-probability sweep $P(n,d,\mathrm{it})$ for the structured family.  Figure~\ref{fig:mis_reachability_phase} shows the two primary arrival views: $P$ versus system size $n$ at the maximum budget and $P$ versus NFEV budget at $n=10$.  The corresponding base-QAOA depth response and $n$-$d$ map are deferred to Appendix~\ref{app:qubo}, Fig.~\ref{fig:mis_reachability_phase_baseqaoa}.  The local-field commutator completion reaches $P_0=0.5$ by $\mathrm{it}=1000$ at $n=8$ and $n=10$, whereas base QAOA never reaches $P_0=0.5$ for any tested depth or budget.  The scaling caveat is equally clear: strict success decays from $17/24$ and $19/24$ at $n=8,10$ to $11/24$ and $7/24$ at $n=12,14$.  The fielded-QUBO result is therefore not a large-$n$ solution claim.  It is an optimal-sampling scaling diagnostic, in the style of Ref.~\cite{zhou2020qaoa}, showing that the commutator-predicted local axis substantially increases the probability of sampling an optimal bitstring while strict-window reachability remains size-sensitive.
\begin{table}[!t]
\caption{\label{tab:mis_powered_scaling}Powered weighted-MIS scaling summary.  Entries show strict successes and mean optimal-subspace overlap $P_{\rm opt}$ over 24 trials per size.  Cost is censored NFEV-to-target for the local-$Y$ rule.  The ``full $\K_1$'' columns denote the full first-shell arm, and ``rule gates'' is the number of inserted local-$Y$ repair rotations in the fixed rule.}
\footnotesize
\setlength{\tabcolsep}{4pt}
\adjustbox{max width=\textwidth}{\footnotesize%
\begin{tabular}{rrrrrrrrrrr}
\toprule
\rowcolor{hdrcol}
$n$ & QAOA $p=6$ & $P_{\rm opt}$ & indep-$X$ & $P_{\rm opt}$ & local-$Y$ rule & $P_{\rm opt}$ & full $\K_1$ & $P_{\rm opt}$ & rule cost & rule gates\\
\midrule
8  & 0/24 & 0.218 & 2/24 & 0.518 & \best{17/24} & \best{0.811} & 9/24  & 0.539 & \best{2324} & 8\\
10 & 1/24 & 0.153 & 0/24 & 0.301 & \best{19/24} & \best{0.889} & 11/24 & 0.564 & \best{1717} & 10\\
12 & 0/24 & 0.086 & 1/24 & 0.099 & \best{11/24} & \best{0.561} & 5/24  & 0.278 & \best{3727} & 12\\
14 & 0/24 & 0.015 & 0/24 & 0.052 & \best{7/24}  & 0.364 & 6/24  & \best{0.468} & \best{4625} & 14\\
\bottomrule
\end{tabular}}
\end{table}
\begin{table}[!t]
\caption{\label{tab:mis_depth}Earlier weighted-MIS focused baselines.  Deeper same-family QAOA and independent-$X$ mixers test whether the local-$Y$ gain is only more depth or local parameter freedom.  Cost is censored first-hit NFEV, with non-strict runs censored at the evaluation budget.  These smaller screens motivated the powered scaling gate in main Fig.~4 and Table~\ref{tab:mis_powered_scaling}.}
\footnotesize
\adjustbox{max width=\columnwidth}{\footnotesize%
\begin{tabular}{llrrrrr}
\toprule
\rowcolor{hdrcol} $n$ & method & params & strict & margin & opt. overlap & cost\\
\midrule
8 & fixed-rule local $Y$ & 10 & \best{11/12} & \best{+0.149} & \best{0.995} & \best{1286}\\
8 & QAOA $p=3$ & 6 & 0/12 & -3.475 & 0.186 & 6000\\
8 & QAOA $p=4$ & 8 & 0/12 & -2.513 & 0.309 & 6000\\
8 & independent-$X$ $p=1$ & 9 & 0/12 & -4.075 & 0.208 & 6000\\
8 & independent-$X$ $p=2$ & 18 & 2/12 & -1.565 & 0.703 & 5250\\
\addlinespace[2pt]
10 & fixed-rule local $Y$ & 12 & \best{5/6} & \best{+0.010} & \best{0.959} & \best{1536}\\
10 & QAOA $p=3$ & 6 & 0/6 & -4.538 & 0.013 & 7000\\
10 & QAOA $p=4$ & 8 & 0/6 & -3.966 & 0.024 & 7000\\
10 & independent-$X$ $p=1$ & 11 & 0/6 & -4.773 & 0.188 & 7000\\
10 & independent-$X$ $p=2$ & 22 & 1/6 & -2.234 & 0.486 & 6504\\
\bottomrule
\end{tabular}}
\end{table}
Table~\ref{tab:mis_depth} records the focused baseline screen that motivated the powered run.  These depth and mixer baselines test two simple alternatives to the local-$Y$ rule: more QAOA layers and local $X$-axis parameter freedom.  Increasing QAOA depth to $p=3$ or $p=4$ does not reach any strict weighted-MIS target.  Giving the $X$ mixer independent per-qubit angles helps more than global-angle QAOA, but even independent-$X$ $p=2$ reaches only $2/12$ and $1/6$ strict successes while using more parameters than fixed-rule local $Y$ (18 versus 10 at $n=8$, and 22 versus 12 at $n=10$).  The powered gate strengthens this conclusion with QAOA $p=6$ and 24 trials per size: same-family depth and local $X$-axis freedom improve the base but do not reproduce the commutator-predicted $Y$-axis lift.

\begin{figure}[!t]
\centering
\includegraphics[width=\textwidth]{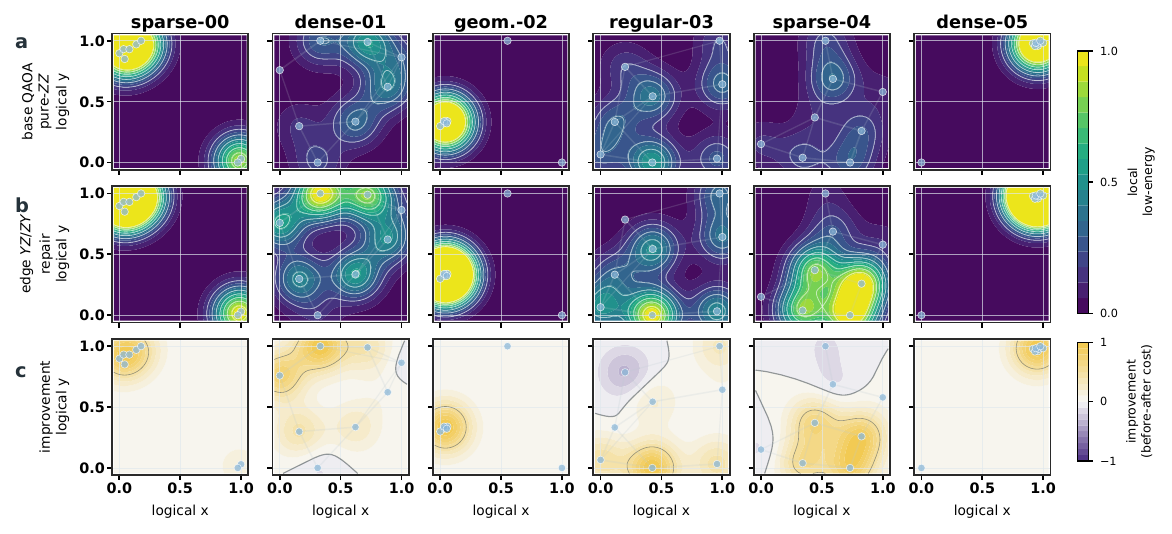}
\caption{\label{fig:supp_spatial_maxcut}Spatial diagnostic for pure-$ZZ$ weighted-MaxCut instances.  Points denote QUBO variables/qubits; logical $x,y$ are graph-layout coordinates and contours are interpolated site-allocated values.  \textbf{a}, Base QAOA local-energy maps after optimization.  \textbf{b}, The same instances after adding and reoptimizing the edge $YZ/ZY$ commutator layer predicted by $[X_i,Z_iZ_j]$.  \textbf{c}, Site-allocated local improvement map, computed as the local contribution before repair minus that after repair.  Brighter color in \textbf{a} and \textbf{b} denotes lower allocated local Hamiltonian contribution within a fixed instance; yellow in \textbf{c} marks locally improved regions and purple marks regions that worsened.  These MaxCut cells are boundary cases: the edge commutator layer improves the energy but does not generally enter the strict window.}
\end{figure}

\begin{figure}[!t]
\centering
\includegraphics[width=\textwidth]{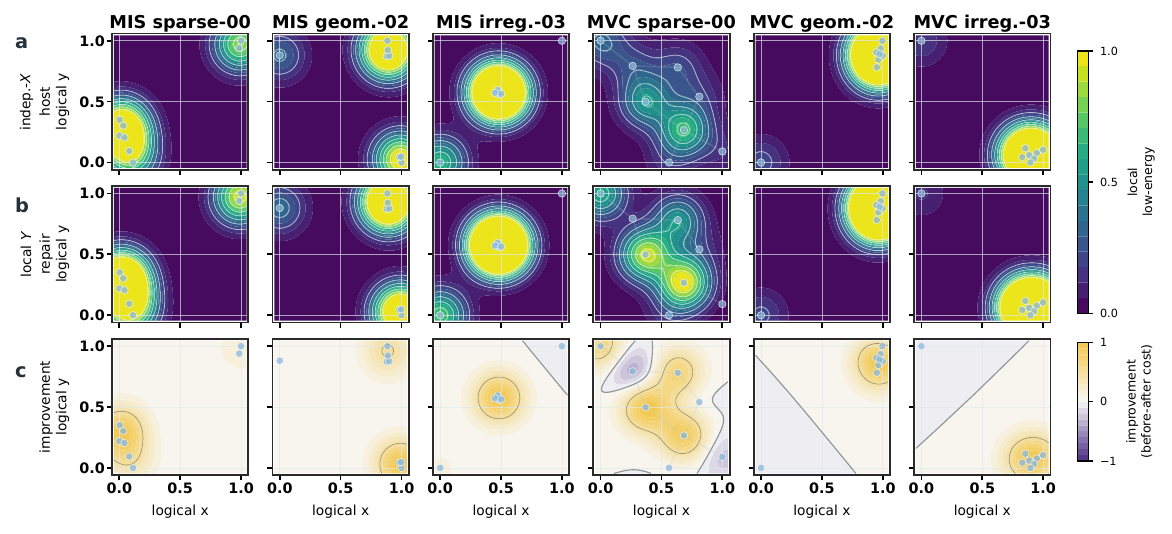}
\caption{\label{fig:supp_spatial_indepx}Spatial diagnostic for a different shallow host on the fielded-QUBO instances.  Points denote QUBO variables/qubits; logical $x,y$ are graph-layout coordinates and contours are interpolated site-allocated values.  \textbf{a}, Optimized independent-$X$ QAOA-style host, in which each qubit has its own mixer angle.  \textbf{b}, The same host after adding and reoptimizing the local-$Y_i$ commutator layer predicted by the field term $[X_i,Z_i]$.  \textbf{c}, Site-allocated local improvement map.  The figure shows that the local-field commutator motif is not tied to a single global-angle QAOA host: when the host still contains local $X_i$ mixer directions, the same $Z_i$ field terms predict local $Y_i$ repair axes.}
\end{figure}

Figures~\ref{fig:supp_spatial_maxcut} and~\ref{fig:supp_spatial_indepx} extend the main-text spatial diagnostic to a different Hamiltonian family and a different QAOA-style host.  They are interpretability plots, not additional powered scaling benchmarks: each panel allocates the diagonal Hamiltonian expectation to sites, compares the optimized base and repaired circuits on the same graph layout, and displays where the repair lowers or raises local Hamiltonian contributions.  Points denote QUBO variables/qubits; the logical $x,y$ axes are fixed graph-layout coordinates rather than circuit parameters, and the contours interpolate the site-allocated values across that layout.

\FloatBarrier
\section{Negative and ADAPT-Style Controls}
\label{app:negcontrols}

Table~\ref{tab:negative} summarizes the controls that separate expressivity repair from optimizer and parameter-count explanations, and Table~\ref{tab:adaptmatch} reports the matched full-pool controls discussed in the main text; the full generous-cap ADAPT sweep is Table~\ref{tab:adaptgenerous}.  The full-pool candidate set consists of all implementable one- and two-qubit Pauli rotations on the supports used by the corresponding stress-test circuit model, not only the commutator closure; it strictly contains the fixed commutator axes.  The one-shot full local Pauli-pool control ranks this larger pool once at the optimized base, inserts a gate-matched block, and then fully reoptimizes under the same cap as fixed $\K_1$.  The sequential ADAPT control instead inserts one operator at a time, with reoptimization after each insertion, until the matched final gate count is reached or the evaluation cap is exhausted.  The generous ADAPT arm receives up to six additional insertions and the 30000-evaluation cap.  No gradient-norm early stopping is used.  A sequential ADAPT search restricted to exactly the same $\mathcal C_{\le2}$ pool as the reported selector was not run and remains an open control.

\begin{table}[!t]
\caption{\label{tab:negative}Controls separating expressivity repair from optimizer and parameter-count explanations.}
\footnotesize
\begin{tabular}{lll}
\toprule
\rowcolor{hdrcol} Gate & Main result & Interpretation\\
\midrule
Optimizer/init & 0/16 rescued & not weak optimization\\
Aggressive base & 0/6 reached & best-reachable shortfall\\
Same-family depth & 0/6 rescued & not more parameters\\
Generic hetero & 4/6 rescued & old $\xyyx$ not unique\\
Tangent diagnostic & local $Y$ low & not one-step selector\\
\bottomrule
\end{tabular}
\end{table}
\begin{table}[!t]
\caption{\label{tab:adaptmatch}Direct full-pool stress tests with explicit cost accounting.  The one-shot full local Pauli-pool arm uses the same optimized bases, block-insertion protocol, inserted-gate budget, and final reoptimization cap as fixed $\K_1$; unlike fixed $\K_1$, it additionally performs one-shot gradient screening over the full implementable local/edge Pauli pool.  The matched ADAPT arm uses the same total post-base NFEV cap of 6000 and the same final inserted-gate budget as fixed $\K_1$; insertions are sequential and each insertion is followed by reoptimization.  The generous ADAPT arm is a stress test with five times the NFEV cap and up to six extra insertions.  Base rows report only the evaluations used to obtain the fixed optimized starting point and do not have a post-base first-hit target cost.  Repair/control rows report censored first-hit NFEV to the strict window after that optimized base is fixed.  Fixed $\K_1$ is a predeclared algebraic motif and has zero selector evaluations; one-shot full pool pays the listed front-loaded candidate-scoring cost.  ``Actual evals'' is the mean number of objective evaluations actually spent in the post-base optimization, plus selector evaluations when present.  For sequential ADAPT rows, selection and reoptimization are interleaved, so the selector cost is marked ``seq.'' and included in the actual-evaluation column.}
\footnotesize
\adjustbox{max width=\textwidth}{\scriptsize\begin{tabular}{llrrrrrrr}
\toprule
\rowcolor{hdrcol} Group & Method & Strict & Margin & Overlap/fid. & Target NFEV & Base evals & Sel. NFEV & Actual evals\\
\midrule
Application QUBO & base QAOA $p=1$ & 0/24 & -4.865 & 0.140 & -- & 372 & -- & 372\\
Application QUBO & fixed $\K_1$ & \best{14/24} & \best{-0.343} & \best{0.689} & \best{2727} & -- & 0 & 3178\\
Application QUBO & one-shot full local Pauli pool & 0/24 & -1.916 & 0.455 & 6000 & -- & 286 & 4193\\
Application QUBO & full-pool ADAPT, matched & 1/24 & -1.591 & 0.473 & 5995 & -- & seq. & 5716\\
Application QUBO & full-pool ADAPT, generous & 13/24 & -0.399 & 0.737 & 19736 & -- & seq. & 20361\\
\addlinespace
QAOA2D & base QAOA2D & 0/18 & -0.615 & 0.781 & -- & 283 & -- & 283\\
QAOA2D & fixed $\K_1$ & \best{12/18} & \best{+0.012} & \best{0.994} & \best{2171} & -- & 0 & 5841\\
QAOA2D & one-shot full local Pauli pool & 3/18 & -0.357 & 0.867 & 5006 & -- & 286 & 4313\\
QAOA2D & full-pool ADAPT, matched & 3/18 & -0.143 & 0.941 & 5037 & -- & seq. & 5086\\
QAOA2D & full-pool ADAPT, generous & 9/18 & -0.029 & 0.981 & 19712 & -- & seq. & 20329\\
\bottomrule
\end{tabular}}
\end{table}
Table~\ref{tab:negative} summarizes the negative-control logic, and Table~\ref{tab:adaptmatch} addresses the closest full-pool baselines.  The one-shot full local Pauli-pool control is the most direct generic-pool alternative to the reported front-loaded block: it uses the same base states, inserted-gate count, block-insertion protocol, and final reoptimization cap as fixed $\K_1$; unlike the fixed algebraic motif, it performs one-shot gradient scoring over the full local/edge Pauli pool.  Across 42 paired stress-test trials, fixed $\K_1$ beats this one-shot full-pool control on strict margin in 39 cases, improves the mean margin by $1.058$ (median $0.599$), and reduces censored target cost by about $3.1\times10^3$ evaluations.  The matched ADAPT-style control is also stringent: it sees a larger full Pauli pool and reoptimizes after every insertion, but is matched to the fixed rule by final gate count and by the same total post-base function-evaluation cap.  Fixed $\K_1$ beats matched ADAPT on strict margin in 36 cases, improves the mean margin by $0.780$, and reduces censored target cost by about $3.1\times10^3$ evaluations.  A generous ADAPT stress test with a $30000$-evaluation cap and up to six additional insertions becomes competitive, but still remains below fixed $\K_1$ in strict success (13/24 versus 14/24 on application QUBOs and 9/18 versus 12/18 on QAOA2D) while requiring $7.2\times$ and $9.1\times$ larger censored target cost, respectively.  The comparison does not argue against adaptive ansatz construction.  It shows that low-order commutator closure can supply a coherent first adaptive layer at much lower selection cost.  These experiments are finite-budget falsification gates: the repairs are not explained by generic one-shot local-pool ranking or by one-gate-at-a-time gradient selection under matched or substantially relaxed resources.  They also leave open a full 189-trial one-shot local-pool sweep and the narrower control of a sequential search restricted to exactly the same $\mathcal C_{\le2}$ pool.

\FloatBarrier
\section{Selector Reproducibility and Locality Audits}
\label{app:selector_audits}

This section collects the four audit tables used by the main text to make the 189-trial selector comparison reproducible and to check that the repair block remains bounded-local.  They are collected here so that the main text can cite them as a single block (Tables S34--S37).

\begin{table}[!t]
\caption{\label{tab:selector_repro_checklist}Reproducibility checklist for the 189-trial selector matrix.}
\footnotesize
\setlength{\tabcolsep}{5pt}
\renewcommand{\arraystretch}{1.12}
\adjustbox{max width=\textwidth}{%
\begin{tabular}{p{0.22\textwidth}p{0.68\textwidth}}
\toprule
\rowcolor{hdrcol}
\textbf{Item} & \textbf{Specification}\\
\midrule
Geometry and size & $2\times3$ lattice, i.e., six-qubit exact statevector optimization and dense exact diagonalization.\\
Hamiltonian settings & TFIM $h\in\{0.6,1.2\}$; J1-J2 Ising $J_2\in\{0.25,0.45,0.65\}$; Heisenberg J1-J2 $J_2\in\{0.20,0.35\}$; XXZ $(J_2,\Delta)\in\{(0.30,0.50),(0.50,1.50)\}$.\\
Hosts & Product rotations, QAOA2D, brickwork $R_Y/R_Z$-CZ, HEA $R_X/R_Y$-CZ, HEA $R_X/R_Z$-CZ, HEA $R_X/R_Y/R_Z$-CZ, and HVA-Heisenberg.\\
Seeds & Three shared seeds, 20261611, 20261612, and 20261613, applied across method arms within each host--Hamiltonian cell.\\
Optimizer & Bounded L-BFGS-B with parameter bounds $[-\pi,\pi]$, tolerances \texttt{ftol=1e-10} and \texttt{gtol=1e-8}.\\
Initialization and budget & Seeded multistart initialization with up to five initial points per completion arm when available: the inherited optimized-base parameters with new repair angles initialized by the selector rule, plus seeded random perturbations within the same parameter bounds.  The listed seed fixes both the perturbations and any randomized control-arm choices.  The cap of 6000 objective evaluations is shared across starts within an arm rather than reset for each start, with at most 260 L-BFGS-B iterations per start before the remaining cap is exhausted.\\
Selector accounting & Candidate-scoring evaluations construct the front-loaded block and are not counted in the censored target-cost column.  The target-cost column reports objective evaluations during the final full reoptimization after the selected block is inserted.  We use $N_{\rm total}=N_{\rm selection}+N_{\rm reoptimization}$ for end-to-end post-base accounting, with $N_{\rm selection}$ estimated from the archived candidate-pool size and reported separately from the final 6000-evaluation reoptimization cap.  Both the gradient and trust selectors use two shifted energy evaluations per candidate in this accounting.  ADAPT-style controls report their total post-base sequential selection/reoptimization cost separately in Table~\ref{tab:adaptmatch}.\\
Gate budget & The reported selectors and matched noncommutator controls use the same 56-Pauli-rotation insertion cap.  Overlapping local rotations are interpreted as separate Pauli rotations; circuit scheduling depth depends on the support-overlap coloring and is therefore audited separately in Table~\ref{tab:locality_audit}.\\
Candidate pool & Implementable one- and two-qubit rotations from $\K_1\cup\K_2$, plus the fixed same-support companion set $\mathcal Q$: local $Y_i$ axes on local-field supports, edge $Y_iZ_j/Z_iY_j$ axes on $ZZ$ supports exposed by $X$ mixers, and edge $X_iZ_j/Z_iX_j$ axes on the corresponding brickwork/HEA companion supports.  Duplicates, signs, scalar factors, host axes, and Hamiltonian axes are stripped before ranking.\\
\bottomrule
\end{tabular}}
\end{table}

\begin{table}[!t]
\caption{\label{tab:control_arm_audit}Aggregate audit of the individual noncommutator controls in the same 189-trial run.  The random edge-axis layer is a seeded same-count layer that samples from local $Y_i$ axes and edge two-qubit Pauli axes $\{XX,YY,XY,YX,XZ,ZX,YZ,ZY\}$ on available supports under the same 56-rotation and 6000-evaluation caps; because this sampling space includes axes such as $YZ/ZY$, occasional coincidences with commutator axes only strengthen the control.  The non-deployable per-trial control envelope credits, for each host--Hamiltonian cell and seed, whichever noncommutator control has the best strict outcome before aggregation; it is therefore not assigned a unique mean fidelity or cost.}
\footnotesize
\setlength{\tabcolsep}{4.5pt}
\renewcommand{\arraystretch}{1.10}
\adjustbox{max width=\textwidth}{%
\begin{tabular}{lrrrr}
\toprule
\rowcolor{hdrcol}
\textbf{Arm} & \textbf{Strict} & \textbf{Mean margin} & \textbf{Mean fidelity} & \textbf{Mean cost}\\
\midrule
Optimized base, no completion & 0/189 & -2.709 & 0.338 & 1699\\
Hamiltonian-term layer & 5/189 & -1.018 & 0.532 & 5863\\
Local-$Y$-only layer & 0/189 & -1.713 & 0.434 & 6000\\
Same-family parameter-matched depth & 27/189 & -1.049 & 0.610 & 5549\\
Wrong-axis $XX/YY{+}Y$ & 3/189 & -0.816 & 0.589 & 5906\\
Wrong-axis $XY/YX{+}Y$ & 62/189 & -0.278 & 0.844 & 4711\\
Random edge-axis layer & \textbf{77/189} & -0.050 & 0.876 & 4670\\
\midrule
Per-trial control envelope & \textbf{109/189} & -0.029 & -- & --\\
\bottomrule
\end{tabular}}
\end{table}

\begin{table}[!t]
\caption{\label{tab:k1k2_selector_ablation}Selector ablation and post-base cost accounting for the 189-trial closure-selector matrix.  Pool size is the mean archived candidate count over the 63 host--Hamiltonian cells before applying the 56-rotation insertion cap.  $N_{\rm sel}$ is the corresponding mean front-loaded selector cost, estimated as two shifted energy evaluations per candidate for both gradient and trust ranking.  $N_{\rm reopt}$ is the archived mean censored first-hit cost during final reoptimization over the 189 paired trials; $N_{\rm total}=N_{\rm sel}+N_{\rm reopt}$.  The $\K_1\cup\K_2$-only selectors use the same hosts, Hamiltonians, seeds, initialization pool, and final reoptimization protocol as the reported selectors but remove the fixed companion-axis set $\mathcal Q$.  The per-trial control envelope is not one deployable arm and therefore has no unique pool size or cost.}
\footnotesize
\setlength{\tabcolsep}{3.1pt}
\renewcommand{\arraystretch}{1.10}
\adjustbox{max width=\textwidth}{%
\begin{tabular}{lrrrrrrr}
\toprule
\rowcolor{hdrcol}
\textbf{Selector or comparator} & \textbf{Strict} & \textbf{Pool} & $\boldsymbol{N_{\rm sel}}$ & $\boldsymbol{N_{\rm reopt}}$ & $\boldsymbol{N_{\rm total}}$ & \textbf{Margin} & \textbf{Fidelity}\\
\midrule
Reported gradient selector, $\K_1\cup\K_2\cup\mathcal Q$ & \textbf{133/189} & 92.3 & 185 & 3620 & 3805 & -0.031 & 0.891\\
Reported trust selector, $\K_1\cup\K_2\cup\mathcal Q$ & 127/189 & 92.3 & 185 & 3671 & 3856 & -0.034 & 0.890\\
Gradient selector, $\K_1\cup\K_2$ only & 104/189 & 73.2 & 146 & 3987 & 4133 & -0.380 & 0.797\\
Trust selector, $\K_1\cup\K_2$ only & 98/189 & 73.2 & 146 & 3972 & 4118 & -0.400 & 0.791\\
Fixed unranked $\K_1+\K_2$ block & 81/189 & -- & 0 & 4450 & 4450 & -0.365 & 0.760\\
Per-trial noncommutator control envelope & 109/189 & -- & -- & -- & -- & -0.029 & --\\
\bottomrule
\end{tabular}}
\end{table}

\begin{table}[!t]
\caption{\label{tab:locality_audit}Shallow/locality audit of the selected repair blocks.  ``Rot.'' reports mean inserted Pauli rotations with the maximum in parentheses.  The geometric range is the largest Manhattan distance between qubits in a selected Pauli string on the $2\times3$ row-major lattice.  The coloring depth is a greedy support-overlap coloring of the repair block alone and is a conservative parallel-depth upper bound, not a hardware-native compiled depth.}
\footnotesize
\setlength{\tabcolsep}{3.2pt}
\renewcommand{\arraystretch}{1.10}
\adjustbox{max width=\textwidth}{%
\begin{tabular}{lrrrrrrr}
\toprule
\rowcolor{hdrcol}
\textbf{Repair block} & \textbf{Rot.} & \textbf{1-local} & \textbf{2-local} & \textbf{range $\le1$} & \textbf{range $\le2$} & \textbf{max range} & \textbf{color depth}\\
\midrule
Reported gradient selector & 55.7 (56) & 20.4\% & 79.6\% & 80.0\% & 99.2\% & 3 & 22.6 (31)\\
Reported trust selector & 55.7 (56) & 21.5\% & 78.5\% & 80.4\% & 99.2\% & 3 & 22.4 (31)\\
Fixed $\K_1+\K_2+Y$ block & 48.5 (56) & 8.0\% & 92.0\% & 72.1\% & 99.2\% & 3 & 22.0 (30)\\
$\K_1\cup\K_2$-only gradient selector & 46.5 (56) & 2.8\% & 97.2\% & 70.9\% & 99.0\% & 3 & 21.0 (31)\\
$\K_1\cup\K_2$-only trust selector & 46.5 (56) & 3.1\% & 96.9\% & 71.2\% & 99.0\% & 3 & 21.0 (32)\\
\bottomrule
\end{tabular}}
\end{table}

\FloatBarrier
\section{Classical-Accessibility and Loss-Evaluation Bounds}
\label{app:simulability}

This appendix states and proves the locality and light-cone results used in the main-text limitation section. Lemmas~\ref{lem:shelllocal} and~\ref{lem:iteratedshelllocal} bound the geometric range of bounded-order commutator shells; Theorem~\ref{thm:closurelightcone} converts this into a linear-time classical evaluation of the loss for fixed-depth local hosts; Corollary~\ref{cor:noadv} rules out an asymptotic loss-evaluation advantage in that regime; and Remark~\ref{rem:dense} delimits dense, high-degree, and non-geometric regimes where this particular geometric light-cone argument does not apply.  These statements concern loss and gradient evaluation under fixed-depth locality assumptions; they are not claims of full output-distribution simulation or a complete dequantization of every possible optimization trajectory.

\begin{figure}[!t]
\centering
\begin{minipage}[t]{0.60\textwidth}
\centering
\textbf{a}\\[-1pt]
\includegraphics[width=\linewidth]{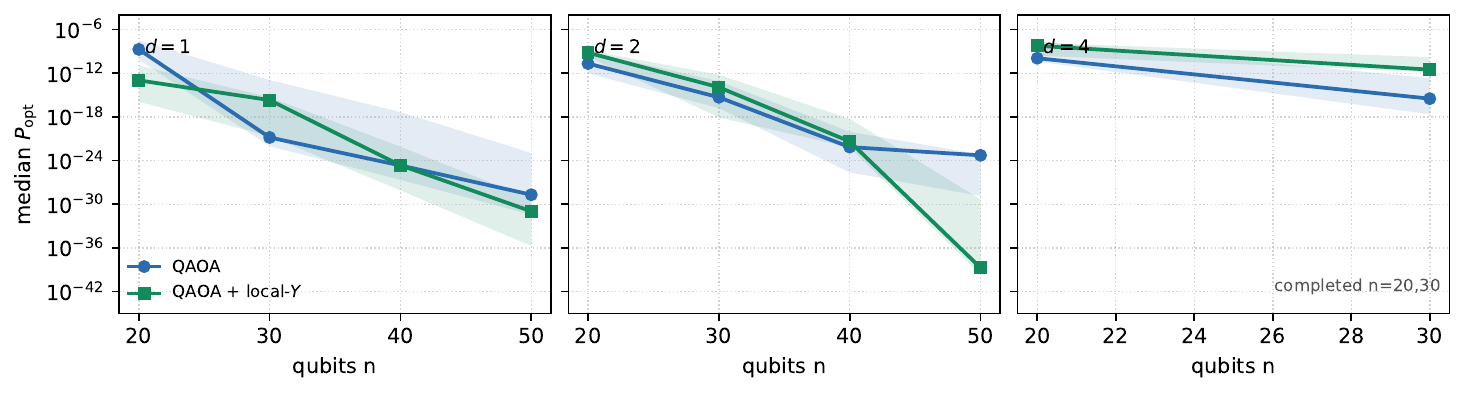}
\end{minipage}\hfill
\begin{minipage}[t]{0.37\textwidth}
\centering
\textbf{b}\\[-2pt]
\begin{tikzpicture}
\begin{axis}[
  width=\linewidth,
  height=0.58\linewidth,
  ybar=1.5pt,
  ymin=0,ymax=9.8,
  ylabel={$\chi^\star$},
  xlabel={qubits $n$},
  symbolic x coords={6,8,12,20,50},
  xtick=data,
  ytick={0,4,8},
  bar width=3.7pt,
  enlarge x limits=0.18,
  tick label style={font=\scriptsize},
  label style={font=\scriptsize},
  legend style={at={(0.03,0.98)},anchor=north west,draw=none,font=\tiny,row sep=1pt},
  ymajorgrids=true,
  grid style={black!10},
  axis line style={black!55},
  tick style={black!55},
  nodes near coords,
  nodes near coords style={font=\tiny, text=black, yshift=0.8pt},
  every node near coord/.append style={/pgf/number format/fixed,/pgf/number format/precision=0},
]
\addplot+[fill=blue!18, draw=blue!70!black, line width=0.45pt] coordinates
  {(6,4) (8,4) (12,4) (20,4) (50,4)};
\addplot+[fill=green!24, draw=green!45!black, line width=0.45pt] coordinates
  {(6,4) (8,4) (12,8) (20,8) (50,4)};
\legend{TFIM,J1-J2}
\end{axis}
\end{tikzpicture}
\end{minipage}
\caption{\label{fig:supp_certifiedladder}Classical-accessibility diagnostics for the repaired regimes.  \textbf{a}, Certified weighted-MIS ladder scaling: median exact optimal-subspace probability $P_{\rm opt}$ (interquartile bands) over three transfer-DP-certified ladder instances and three optimizer seeds per point; depths $d=1,2$ are complete through $n=50$ and the more expensive $d=4$ pass through $n=30$.  These data are a limitation diagnostic: both QAOA and QAOA plus the local-field completion lose useful exact-optimum sampling as the certified system grows, and the completion is neutral or worse at several large-$n$ points.  \textbf{b}, Independent DMRG capstone on the two-row Ising strips, shown as the seed-median bond dimension needed to reach energy error $\le10^{-2}$.  TFIM needs $\chi^\star=4$ at every tested size and seed, and J1-J2 Ising stays at $\chi^\star=4$--$8$ with no monotone growth through $n=50$.}
\end{figure}

The certified weighted-MIS ladder in Fig.~\ref{fig:supp_certifiedladder} uses the transfer dynamic program to certify the exact optimum and optimal bitstring set, then evaluates the optimized QAOA and local-$Y$ completed distributions against that certificate.  Each completed point averages three certified ladder instances and three optimizer seeds.  The completed grid covers depths $d=1,2$ for $n=20,30,40,50$ and $d=4$ for $n=20,30$; the interrupted $d=4,n\ge40$ segments are excluded.  The archived completed runs use a bounded optimizer diagnostic with mean recorded target budget about 502 objective evaluations per trial, so the panel should be read as a fixed-depth sampling-accessibility stress test rather than as a fully tuned large-$n$ optimizer comparison.  The figure reports the exact optimal-subspace probability $P_{\rm opt}$ and the strict energy-success count for the completed grid.

\begin{lemma}[Shell locality]
\label{lem:shelllocal}
Let $P,Q$ be Pauli strings with supports $S_P,S_Q$.  Then $i[P,Q]$ is either zero or, up to a sign, a single Pauli string supported on $S_P\cup S_Q$.  Hence the stripped first shell $\K_1=\mathrm{strip}\{i[P,Q]:P\in\G_0,\,Q\in\Hterms\}$ consists of Pauli strings of weight at most $\max_{P,Q}(|S_P|+|S_Q|)$ and range at most the sum of the generator and Hamiltonian ranges.  For a QAOA host with local mixers $X_i$ and a diagonal cost with nearest-neighbor $Z_iZ_j$ and on-site $Z_i$ terms, $\K_1$ is at most two-local and nearest-neighbor: $\K_1\subseteq\{Y_iZ_j,Z_iY_j:(i,j)\in E\}\cup\{Y_i\}$.
\end{lemma}
\begin{proof}
The product of two Pauli strings is again a Pauli string, up to a phase in $\{\pm1,\pm i\}$, supported on the union of their supports.  If $P,Q$ commute then $[P,Q]=0$; if they anticommute then $[P,Q]=2PQ$, so $i[P,Q]$ is $\pm$ a Pauli string on $S_P\cup S_Q$.  The QAOA specialization is the explicit table $i[X_i,Z_iZ_j]=-2\,Y_iZ_j$, $i[X_j,Z_iZ_j]=-2\,Z_iY_j$, and $i[X_i,Z_i]=-2\,Y_i$, whose supports are contained in the corresponding lattice edges.
\end{proof}

\begin{lemma}[Iterated shell locality]
\label{lem:iteratedshelllocal}
Let the host generator set $\G_0$ and Hamiltonian term set $\Hterms$ consist of Pauli strings supported on balls of radius at most $r_G$ and $r_H$ on a bounded-degree geometric graph.  Define stripped commutator shells recursively by
\begin{align}
  \K_0&=\G_0,\\
  \K_{s+1}
  &=\mathrm{strip}\{i[P,Q]\;|\;
      P\in\K_s,\ Q\in\G_0\cup\Hterms\}.
\end{align}
If $R_s$ is the maximum geometric radius of a nonzero string in $\K_s$, then
\[
  R_s \le r_G+s\,r_\star,\qquad r_\star=\max(r_G,r_H).
\]
In particular, every bounded-order closure $\K_{\le k}=\cup_{s=1}^k\K_s$ has bounded geometric range whenever $k,r_G,r_H$ are fixed.
\end{lemma}
\begin{proof}
By Lemma~\ref{lem:shelllocal}, a nonzero commutator $i[P,Q]$ is supported on the union of the supports of $P$ and $Q$.  If the two Pauli strings are disjoint in a way that shares no qubit with anticommuting local factors, the commutator vanishes; every retained string therefore grows only through a local overlap between a radius-$R_s$ support and a radius-$r_\star$ support.  The radius increases by at most $r_\star$ at each shell, giving the induction $R_{s+1}\le R_s+r_\star$ and hence $R_s\le r_G+s r_\star$.
\end{proof}

\begin{theorem}[Bounded-order local closure has bounded light cones]
\label{thm:closurelightcone}
Let $H=\sum_a c_a h_a$ be a Hamiltonian on $n$ qubits of a bounded-degree geometric lattice (degree $\Delta_{\rm lat}$), with $O(n)$ terms, each $h_a$ a Pauli string supported within a ball of radius $r_H$.  Let $U_0(\theta)$ be any fixed-scheduled-depth local host circuit from a product initial state, generated by Pauli rotations of range at most $r_G$ and scheduled depth $D_0$.  Let $U_C(\phi)$ be a completion circuit generated by an implementable subset of the bounded-order closure $\K_{\le k}$ and scheduled into depth $D_C$, with overlapping rotations assigned to different local sublayers.  Let $U(\theta,\phi)=U_C(\phi)U_0(\theta)$.

For fixed $k,D_0,D_C,r_G,r_H,\Delta_{\rm lat}$, the loss
\[
  \langle 0|U^\dagger(\theta,\phi)HU(\theta,\phi)|0\rangle
\]
can be computed exactly by a classical light-cone contraction in time $O(n\,2^{\ell_k})$, where
\begin{align}
  \ell_k &\le B_{\Delta_{\rm lat}}(R_k^{\rm lc}),\\
  R_k^{\rm lc}
  &=O\!\left(D_0r_G+D_C(r_G+k r_\star)+r_H\right),
\end{align}
$r_\star=\max(r_G,r_H)$, and $B_{\Delta_{\rm lat}}(R)$ is the largest number of lattice sites in a radius-$R$ ball on a degree-$\Delta_{\rm lat}$ lattice.  Thus $\ell_k$ is independent of $n$ for fixed order, depth, and locality.  In one dimension, the prepared repaired state also has an MPS representation with bond dimension $\chi\le 2^{O(D_0r_G+D_C(r_G+k r_\star))}$; for a fixed-width strip the exponent is multiplied by a width-dependent constant.
\end{theorem}
\begin{proof}
By Lemma~\ref{lem:iteratedshelllocal}, every completion generator in $\K_{\le k}$ has range at most $R_k\le r_G+k r_\star$.  Conjugate a local Hamiltonian term $h_a$ backwards through $U_CU_0$ in the Heisenberg picture.  A scheduled local layer with generator range $\rho$ can enlarge the support of the back-propagated operator by at most $O(\rho)$ in graph distance.  After the host and completion layers, $U^\dagger h_a U$ is supported inside a ball whose radius is bounded by $O(D_0r_G+D_C R_k+r_H)$.  This ball contains at most $\ell_k$ qubits, independent of $n$ under the fixed-parameter hypotheses.  Therefore $\langle U^\dagger h_a U\rangle$ is obtained exactly by simulating only the restricted $\ell_k$-qubit light-cone subcircuit, at cost $O(2^{\ell_k})$.  Summing the $O(n)$ Hamiltonian terms gives $O(n2^{\ell_k})$ total time.

The MPS statement is the standard circuit-entanglement bound: a one-dimensional local circuit of depth $D$ and maximum gate range $R$ crossing any bipartition at most $O(DR)$ times can increase the Schmidt rank by at most a factor $2^{O(DR)}$ from a product state.  Here $R=O(r_G+k r_\star)$ and $D=O(D_0+D_C)$.  A fixed-width strip mapped to a one-dimensional snake ordering only multiplies the number of cut-crossing gates by a width-dependent constant.
\end{proof}

\begin{remark}[Companion axes]
Theorem~\ref{thm:closurelightcone} also covers the predeclared companion axes used in the numerical selector, provided those axes are bounded-range local Pauli rotations on the same geometric supports as the closure terms; this is the case for the local $Y$, edge $YZ/ZY$, and edge $XZ/ZX$ companions used in the reported local-lattice experiments.
\end{remark}

\begin{corollary}[No asymptotic loss-evaluation advantage under fixed-depth local scheduling]
\label{cor:noadv}
Under the hypotheses of Theorem~\ref{thm:closurelightcone}, the loss $\langle H\rangle(\theta,\phi)$ and its parameter-shift or finite-difference gradients---each a fixed number of shifted loss evaluations---are classically computable in time $O(n)$ per query.  Any polynomial-query optimization that accesses the ansatz only through these fixed-depth scheduled loss and gradient evaluations can therefore be reproduced classically in polynomial time on the same loss landscape, up to the same loss-query sequence.  This rules out an asymptotic loss-evaluation advantage for the geometrically local instances covered by the theorem; it does not address dense costs, different unscheduled circuit families, output sampling from the final circuit, or the global difficulty of finding the best parameters.
\end{corollary}

\begin{remark}[Dense or non-local costs]
\label{rem:dense}
The light-cone argument requires geometric locality, bounded interaction range, bounded degree, and fixed scheduled depth.  For dense or high-degree application QUBOs---for example MaxCut or maximum-independent-set instances on dense graphs---one cost layer can couple $\Theta(n)$ qubits, so the backwards light cone can grow with system size and Theorem~\ref{thm:closurelightcone} no longer gives an $O(n)$ fixed-light-cone contraction.  Bounded-degree but non-geometric graphs are a separate intermediate case: their graph-distance light cones may remain bounded at fixed depth, but the geometric ball-growth bound $B_{\Delta_{\rm lat}}(R)$ used above is no longer the right statement, and additional graph-specific analysis would be needed.  Whether such non-geometric or dense instances stay classically easy is a separate question that our exact argument does not settle.  One would turn to low-weight Pauli, LOWESA-type, or noiseless Pauli-propagation estimates \cite{fontana2023noisy,rudolph2023lowesa,angrisani2025classically}, whose additive-error cost is controlled by Pauli-weight decay conditions we do not establish here.  We therefore state the loss-evaluation accessibility theorem only for geometrically local hosts and Hamiltonians, and treat dense application-QUBO regimes as outside its scope.
\end{remark}
\end{appendices}

\clearpage
\bibliography{references}